\documentclass[iop]{emulateapj}

\usepackage{graphicx}
\usepackage{natbib}
\usepackage{soul}
\usepackage{longtable}

\def\dam{PTF\,12dam}
\def\dcc{iPTF\,13dcc}
\def\bdq{LSQ\,14bdq}
\def\des{DES\,14X3taz}
\def\ajg{iPTF\,13ajg}
\def\css{CSS\,121015}
\def\oz{SN~2006oz}
\def\lsim{\mathrel{\hbox{\rlap{\lower.55ex \hbox {$\sim$}}\kern-.0em
\raise.4ex \hbox{$<$}}}} 
\def\gsim{\mathrel{\hbox{\rlap{\lower.55ex \hbox {$\sim$}}\kern-.0em
\raise.4ex \hbox{$>$}}}} 

\def\ha{H$\alpha$}

\def\kms{km s$^{-1}$}
\def\a{$^{\rm a}$}
\def\b{$^{\rm b}$}
\def\c{$^{\rm c}$}

\newcommand{\Msun}{\mbox{M$_\odot$}}
\newcommand{\Rsun}{\mbox{R$_\odot$}}

\newcommand{\Msunyr}{\mbox{M$_\odot$ yr$^{-1}$}}

\newcommand{\ergsec}{\mbox{erg s$^{-1}$}}

\newcommand{\ergscmA}{\mbox{erg s$^{-1}$ cm$^{-2}$ \AA$^{-1}$}}

\submitted{}
\received{21 September 2016}
\revised{9 November 2016}
\accepted{29 November 2016}

\begin{document}

\title{On the early-time excess emission in hydrogen-poor
  superluminous supernovae}

\author{
  Paul M. Vreeswijk\altaffilmark{1,2},
  Giorgos Leloudas\altaffilmark{1,3},
  Avishay Gal-Yam\altaffilmark{1},
  Annalisa De Cia\altaffilmark{4,1},
  Daniel A. Perley\altaffilmark{3,5},
  Robert M. Quimby\altaffilmark{6,7},
  Roni Waldman\altaffilmark{8,1},
  Mark Sullivan\altaffilmark{9},
  Lin Yan\altaffilmark{10},
  Eran O. Ofek\altaffilmark{1},
  Christoffer Fremling\altaffilmark{11},
  Francesco Taddia\altaffilmark{11},
  Jesper Sollerman\altaffilmark{11},
  Stefano Valenti\altaffilmark{12},
  Iair Arcavi\altaffilmark{13,14,15}
  D. Andrew Howell\altaffilmark{14,13},
  Alexei V. Filippenko\altaffilmark{16},
  S. Bradley Cenko\altaffilmark{17,18},
  Ofer Yaron\altaffilmark{1},
  Mansi M. Kasliwal\altaffilmark{5},
  Yi Cao\altaffilmark{5},
  Sagi Ben-Ami\altaffilmark{19,1},
  Assaf Horesh\altaffilmark{1},
  Adam Rubin\altaffilmark{1},
  Ragnhild Lunnan\altaffilmark{5},
  Peter E. Nugent\altaffilmark{20,18},
  Russ Laher\altaffilmark{21},  
  Umaa D. Rebbapragada\altaffilmark{22},
  Przemys\l{}aw Wo\'zniak\altaffilmark{23},
  and Shrinivas R. Kulkarni\altaffilmark{5}
}

\altaffiltext{1}{Department of Particle Physics and Astrophysics,
  Weizmann Institute of Science, Rehovot 7610001, Israel}

\altaffiltext{2}{Benoziyo Fellow; email: paul.vreeswijk@weizmann.ac.il}

\altaffiltext{3}{Dark Cosmology Centre, Niels Bohr Institute,
  University of Copenhagen, Juliane Maries Vej 30, 2100 K{\o}benhavn
  {\O}, Denmark}

\altaffiltext{4}{European Southern Observatory, Karl-Schwarzschild-Strasse 2,
  85748 Garching bei M\"unchen, Germany}

\altaffiltext{5}{Cahill Center for Astrophysics, California Institute
  of Technology, Pasadena, CA 91125, USA}

\altaffiltext{6}{Department of Astronomy, San Diego State University,
  San Diego, CA 92182, USA}

\altaffiltext{7}{Kavli IPMU (WPI), UTIAS, The University of
  Tokyo, Kashiwa, Chiba 277-8583, Japan}

\altaffiltext{8}{Racah Institute of Physics, The Hebrew University,
  Jerusalem 91904, Israel}

\altaffiltext{9}{School of Physics and Astronomy, University of
  Southampton, Southampton SO17 1BJ, UK}

\altaffiltext{10}{Infrared Processing and Analysis Center, California
  Institute of Technology, Pasadena, CA 91125, USA}

\altaffiltext{11}{The Oskar Klein Centre, Department of Astronomy,
  Stockholm University, AlbaNova, 10691 Stockholm, Sweden}

\altaffiltext{12}{Department of Physics, University of California,
  Davis, CA 95616, USA}

\altaffiltext{13}{Department of Physics, University of California,
  Santa Barbara, CA 93106, USA}

\altaffiltext{14}{Las Cumbres Observatory Global Telescope, 6740
  Cortona Dr., Suite 102, Goleta, CA 93111, USA}

\altaffiltext{15}{Einstein Fellow}


\altaffiltext{16}{Department of Astronomy, University of California,
  Berkeley, CA 94720-3411, USA}

\altaffiltext{17}{Astrophysics Science Division, NASA Goddard Space
  Flight Center, Mail Code 661, Greenbelt, MD, 20771, USA}

\altaffiltext{18}{Joint Space-Science Institute, University of
  Maryland, College Park, MD 20742, USA}

\altaffiltext{19}{Smithsonian Astrophysical Observatory,
  Harvard-Smithsonian Center for Astrophysics, 60 Garden St.,
  Cambridge, MA 02138, USA}

\altaffiltext{20}{Computational Cosmology Center, Lawrence Berkeley
  National Laboratory, 1 Cyclotron Road, Berkeley, CA 94720, USA}

\altaffiltext{21}{Spitzer Science Center, MS 314-6, California
  Institute of Technology, Pasadena, CA 91125, USA}

\altaffiltext{22}{Jet Propulsion Laboratory, California Institute of
  Technology, Pasadena, CA 91109, USA}

\altaffiltext{23}{Los Alamos National Laboratory, MS D436, Los Alamos,
  NM 87545, USA}

\begin{abstract}
  We present the light curves of the hydrogen-poor superluminous
  supernovae (SLSNe-I) \dam\ and \dcc, discovered by the
  (intermediate) Palomar Transient Factory. Both show excess emission
  at early times and a slowly declining light curve at late times.
  The early bump in \dam\ is very similar in duration ($\sim$10~days)
  and brightness relative to the main peak (2--3~mag fainter) compared
  to those observed in other SLSNe-I. In contrast, the long-duration
  ($>$30~days) early excess emission in \dcc, whose brightness
  competes with that of the main peak, appears to be of a different
  nature. We construct bolometric light curves for both targets, and
  fit a variety of light-curve models to both the early bump and main
  peak in an attempt to understand the nature of these explosions.
  Even though the slope of the late-time light-curve decline in both
  SLSNe is suggestively close to that expected from the radioactive
  decay of $^{56}$Ni and $^{56}$Co, the amount of nickel required to
  power the full light curves is too large considering the estimated
  ejecta mass.  The magnetar model including an increasing escape
  fraction provides a reasonable description of the
  \dam\ observations.  However, neither the basic nor the
  double-peaked magnetar model is capable of reproducing the
  \dcc\ light curve. A model combining a shock breakout in an extended
  envelope with late-time magnetar energy injection provides a
  reasonable fit to the \dcc\ observations. Finally, we find that the
  light curves of both \dam\ and \dcc\ can be adequately fit with the
  circumstellar medium (CSM) interaction model.
\end{abstract}



\section{Introduction}
\label{sec:introduction}

Supernovae that reach an absolute magnitude brighter than the
(arbitrary) limit of $M=-21$ are labelled superluminous
\citep{2012Sci...337..927G}. Even though they are very rare, several
tens of them have been discovered over the past decade thanks to the
ever-increasing survey speed of optical telescopes. They are
observationally separated into two classes based on the detection of
hydrogen in their spectra, similar to classical supernovae
\citep[see][]{1997ARA&A..35..309F}: hydrogen-rich Type II SLSNe show
clear Balmer features
\citep[e.g.,][]{2007ApJ...659L..13O,2007ApJ...666.1116S}, while the
hydrogen-poor Type I SLSNe do not \citep{2012Sci...337..927G}. The
latter commonly exhibit a distinct W-shaped feature identified as
\ion{O}{2} at rest-frame wavelengths
4000--4500~\AA\ \citep{2011Natur.474..487Q,2016MNRAS.458.3455M}.  At
late times, the spectra of these SLSNe-I evolve to appear like those
of normal Type Ic SNe \citep{2010ApJ...724L..16P}, leading many
authors to refer to this class as SLSN-Ic. Sometimes this Type I/II
distinction is not so obvious; for example, spectra of
CSS121015:004244+132827 (hereafter referred to as \css) have
similarities to both Type II and I objects
\citep{2013arXiv1310.1311B}, while iPTF~13ehe, classified as Type I,
shows the emergence of broad \ha\ emission at late times
(\citealp{2015ApJ...814..108Y}; see also
\citealp{2015A&A...584L...5M,2016ApJ...828...87W}).

There is some evidence that the energy source powering the
hydrogen-rich SLSNe is interaction of the SN ejecta with
optically-thick material at a large distance ($\sim$10$^{15}$~cm), as
they typically reveal Balmer emission lines indicative of interaction
with a hydrogen-rich circumstellar medium \citep[CSM;
  e.g.,][]{1994ApJ...420..268C,1994MNRAS.268..173C,2014ApJ...788..154O}. Because
of the similarity with normal SNe of Type IIn, this class is also
referred to as SLSNe-IIn. However, some SLSNe-II do not exhibit narrow
emission lines, while they are of Type II as they reveal broad
hydrogen features during the photospheric phase
(\citealp{2016arXiv160401226I}; see also
\citealp{2012ApJ...747..118M}).

The energy source of the hydrogen-poor SLSNe is still under debate,
with the most promising candidates being (1) additional energy
input from a central engine, such as a spinning-down magnetar
\citep[e.g.,][]{2010ApJ...717..245K,2010ApJ...719L.204W,2013ApJ...770..128I}
or an accreting black hole \citep{2013ApJ...772...30D}; (2)
interaction of the ejecta with a hydrogen-poor shell expelled by the
progenitor star some time before the explosion
\citep{2010arXiv1009.4353B,2011ApJ...729L...6C}; and (3) the radioactive
decay of a large amount of nickel produced in the explosion,
potentially due to pair-instability conditions
\citep[see][]{2009Natur.462..624G} though this is still being debated
\citep{2010ApJ...717L..83M,2010A&A...512A..70Y,2013Natur.502..346N}. The
light curves of some hydrogen-poor SLSNe, such as the two PTF sources
presented in this paper, decay very slowly at late times, with a slope
similar to that expected from the decay of radioactive nickel and
cobalt.  These SLSNe are part of the hydrogen-poor class, but are
sometimes referred to as Type R \citep[``radioactive'';
  see][]{2012Sci...337..927G}.

The host galaxies of SLSNe are found to be irregular, compact,
low-mass galaxies with high specific star formation rates
\citep{2011ApJ...727...15N,2013arXiv1311.0026L,2015ApJ...804...90L,2015MNRAS.449..917L}.
A comprehensive study of 32 host galaxies of all PTF-discovered SLSNe
(until the end of 2012) found hydrogen-poor SLSNe to have a preference
for environments in hosts with a metallicity upper bound of about half
solar, while the hydrogen-rich SLSNe do not show such a preference
\citep{2016arXiv160408207P}.  A very similar conclusion was reached
independently by \citet{2016arXiv160504925C}.  In emission, the
galaxies hosting SLSNe have broadly similar characteristics as the
hosts of gamma-ray bursts (GRBs). \citet{2015MNRAS.449..917L} find the
host-galaxy emission-line strengths of redshift $z<1$ SLSNe-I to be
significantly stronger than in GRB hosts, but
\citet{2016arXiv160701045J} do not confirm this result over the range
$0.3<z<0.7$. In {\it absorption}, the environments of SLSNe-I appear
to be significantly poorer in their neutral gas content, as traced by
\ion{Mg}{1} and \ion{Mg}{2}, than those of GRBs
\citep{2014ApJ...797...24V}.

Focusing on the hydrogen-poor class, several of these have shown
evidence for an early-time light-curve ``bump'' or excess emission
before the onset of the main peak. Examples are
\oz\ \citep{2012A&A...541A.129L}, \bdq\ \citep{2015ApJ...807L..18N},
and \des\ \citep{2015arXiv151206043S}. In fact,
\citet{2016MNRAS.457L..79N} suggest that early bumps such as the ones
above may be ubiquitous in hydrogen-poor SLSNe. Early bumps have also
been observed in normal stripped-envelope SNe, and recently in a
normal SN Ic from a massive progenitor \citep{2016A&A...592A..89T}.
To date, such early excess emission has not been reported for any
hydrogen-rich SLSN-II. This early excess emission is of particular
interest, as it may provide a clue regarding what is powering these
explosions.

In the case of \oz, \citet{2012A&A...541A.129L} propose that the
precursor bolometric plateau might be related to a recombination wave
in a H-poor CSM. The study by \citet{2012ApJ...756L..22M} has shown
that a dip in the light curve is naturally expected when shock
breakout occurs within a dense CSM.  \citet{2015ApJ...807L..18N}
propose that the initial peak in \bdq\ may arise from the post-shock
cooling of extended stellar material, while reheating by a central
engine is driving the main peak. The high kinetic energy inferred from
fitting the \citet{2011ApJ...728...63R} model to the initial peak
($E_{\rm k}\sim 2\times10^{52}$~erg) of \bdq\ may favor a black hole
accretion engine \citep{2013ApJ...772...30D} rather than a magnetar.
The early-time excess emission in the case of \des\ shows rapid
cooling from 22,000~K to 8,000~K over the course of 15 rest-frame
days. The authors find that a shock-cooling model of CSM at a distance
of $\sim$400~\Rsun, followed by a magnetar causing the main peak of
the light curve can adequately explain the entire light curve.

\begin{deluxetable*}{lccccccccc}
  \tablecaption{Log of Spectroscopic Observations of \dcc\label{tab:13dcc_logspec}}
  \tablehead{
    \colhead{UTC Date} &
    \colhead{Telescope} &
    \colhead{Instrument} &
    \colhead{Exp.~Time} &
    \colhead{Grating/Grism/Filter} &
    \colhead{Slit~Width} &
    \colhead{$\lambda$ Coverage} &
    \colhead{Res.\tablenotemark{a}} &
    \colhead{I.Q.\tablenotemark{b}} &
    \colhead{Airm.} \\
    & & & (min.) & & \arcsec & (\AA) & (\AA) & \arcsec &
  }
  \startdata
  2013 Nov. 26 & P200    & DBSP   & 20       & 600/4000, 316/7500 & 1.5 &  3400--10,400 & 9.3 & 2.9 & 1.3 \\
  2013 Dec.  3 & Keck~I  & LRIS   & 21       & 400/3400, 400/8500 & 1.0 &  3200--10,240 & 6.0 & 1.6 & 1.4 \\ 
  2013 Dec.  4 & Keck~I  & LRIS   & 10       & 600/4000, 400/8500 & 1.0 &  3140--10,240 & 5.8 & 1.2 & 1.4 \\ 
  2013 Dec. 31 & Magellan Baade & IMACS & 25 & Gra-300-4.3        & 0.9 &  3700--9,700  & 6.1 & 0.9 & 1.3 \\
  2014 Jan.  6 & P200    & DBSP   & 60       & 600/4000, 316/7500 & 1.5 &  3300--10,400 & 8.5 & 1.8 & 1.6
  \enddata
  \tablenotetext{a}{The resolution of the spectra
    was determined from the width of the [O~I] $\lambda$5577 night-sky line.}
  \tablenotetext{b}{The image quality, or effective seeing, was
    measured directly from the width of the object's spatial profile
    around 6000~\AA.}
\end{deluxetable*}

%


In this paper, we present two SLSNe-I discovered by the (intermediate)
Palomar Transient Factory
\citep{2009PASP..121.1334R,2009PASP..121.1395L} that also show
evidence for early excess emission: \dam\ and \dcc. The modest bump in
\dam\ is very similar in duration and brightness relative to the main
peak to the cases discussed above. However, the long-duration early
excess emission in \dcc, whose brightness competes with that of
the main peak, appears to be of a different nature.

This paper is organized as follows. In Sec.~\ref{sec:observations} we
present the photometric observations that we obtained for \dam\ and
\dcc, as well as the spectroscopic sequence of \dcc.  We construct the
bolometric luminosity evolutions in Sec.~\ref{sec:photometry}, which
we confront with models in Sec.~\ref{sec:modelling}. The models
\citep[see][]{2012ApJ...746..121C,2013ApJ...773...76C,2015ApJ...808L..51P}
assume different energy sources (radioactive decay, magnetar heating,
and CSM interaction) and predict the ensuing light curve based on a
number of parameters; we infer estimates of the best-fit values by
fitting the semi-analytical models to the bolometric light curves. We
discuss our results and briefly conclude in Sec.~\ref{sec:discussion}.

Unless noted otherwise, the uncertainties listed in this paper are at
the 1$\sigma$ confidence level. We adopt the cosmological parameters
as derived by the Planck collaboration in 2015
\citep[H$_0=68$~km~s$^{-1}$~Mpc$^{-1}$, $\Omega_{\rm m}=0.31$,
  $\Omega_{\Lambda}=0.69$;][]{2015arXiv150201589P}.

\section{Identification, Observations, and Data Reduction}
\label{sec:observations}

\subsection{\dam}

\dam\ was flagged as a transient source as part of the regular PTF
operations on 2012 April 17 (UTC dates are used throughout this
paper); it was first detected on April 10. The source is located at
$\alpha = 14^{\rm h}24^{\rm m}46.20^{\rm s}$, $\delta = +46^\circ 13'
48.3''$ (J2000.0), with an uncertainty of 0.1\arcsec. At this location
the Galactic extinction is estimated to be low, $A_V=0.033$~mag
(\citealp{2011ApJ...737..103S}; see also
\citealp{1989ApJ...345..245C}). Spectroscopic follow-up observations
were performed with the Kast Spectrograph \citep{miller-stone93} at
the Lick 3~m Shane telescope, and the Low Resolution Imaging
Spectrograph \citep[LRIS;][]{1995PASP..107..375O} at the Keck~I 10~m
telescope (on Mauna Kea, Hawaii) on 2012 May 20, 21, and 22, showing
\dam\ to be an SLSN-I at $z=0.107$ \citep{2012ATel.4121....1Q}. The
full \dam\ spectroscopic sequence will be presented by Quimby et
al. (in prep.).

\dam\ was imaged with the Palomar Oschin 48~inch (P48) (i)PTF survey
telescope equipped with a 12k $\times$ 8k CCD mosaic camera
\citep{2008SPIE.7014E..4YR} in the Mould $R$ filter, the Palomar
60~inch (P60) and CCD camera \citep{2006PASP..118.1396C} in Johnson
$B$ and Sloan Digital Sky Survey (SDSS) $gri$, the Las Cumbres
Observatory Global Telescope Network
\citep[LCOGT;][]{2013PASP..125.1031B} in SDSS $r$, with LRIS mounted
on the Keck~I telescope in $R_s$.  Post-peak and late-time imaging
shows \dam\ to have a slowly declining light curve, i.e., of Type R
following the SLSN classification suggested by
\citet{2012Sci...337..927G}.

\dam, relatively nearby at $z=0.107$, has already received
considerable interest in the literature. \citet{2013Natur.502..346N}
use its light curve to argue against the claim that SN~2007bi, also an
SLSN-R with a similar late-time light curve as \dam, is produced by a
pair-instability explosion
\citep{2009Natur.462..624G}. \citet{2014arXiv1409.7728C} study both
the late-time SN decay and the host galaxy of \dam. They find that its
light curve can be fit with a magnetar model
\citep[see][]{2010ApJ...717..245K,2010ApJ...719L.204W} if an escape
fraction, increasing with time, of the magnetar energy input is
considered \citep[see also][]{2015ApJ...799..107W}. An even better fit
is reached with a model involving the interaction of the ejecta with
dense CSM. Using radiation-hydrodynamics calculations,
\citet{2015AstL...41...95B} show that PTF12dam and similar SLSNe can
be explained without a magnetar in a model with a radiative shock in a
dense circumstellar envelope.  The host galaxy of \dam\ is a compact,
low-mass ($3\times10^8$~\Msun), low-metallicity (12 + log[O/H] =
$8.05\pm0.09$) dwarf galaxy with a high star-formation rate
\citep[5~\Msunyr;][]{2014arXiv1409.7728C}. \citet{2014arXiv1411.1104T}
report on long-slit spectroscopy of the host to infer the presence of
a very young stellar population (even down to about 3~Myr).

\begin{figure*}
  \centering
  \includegraphics[width=\hsize]{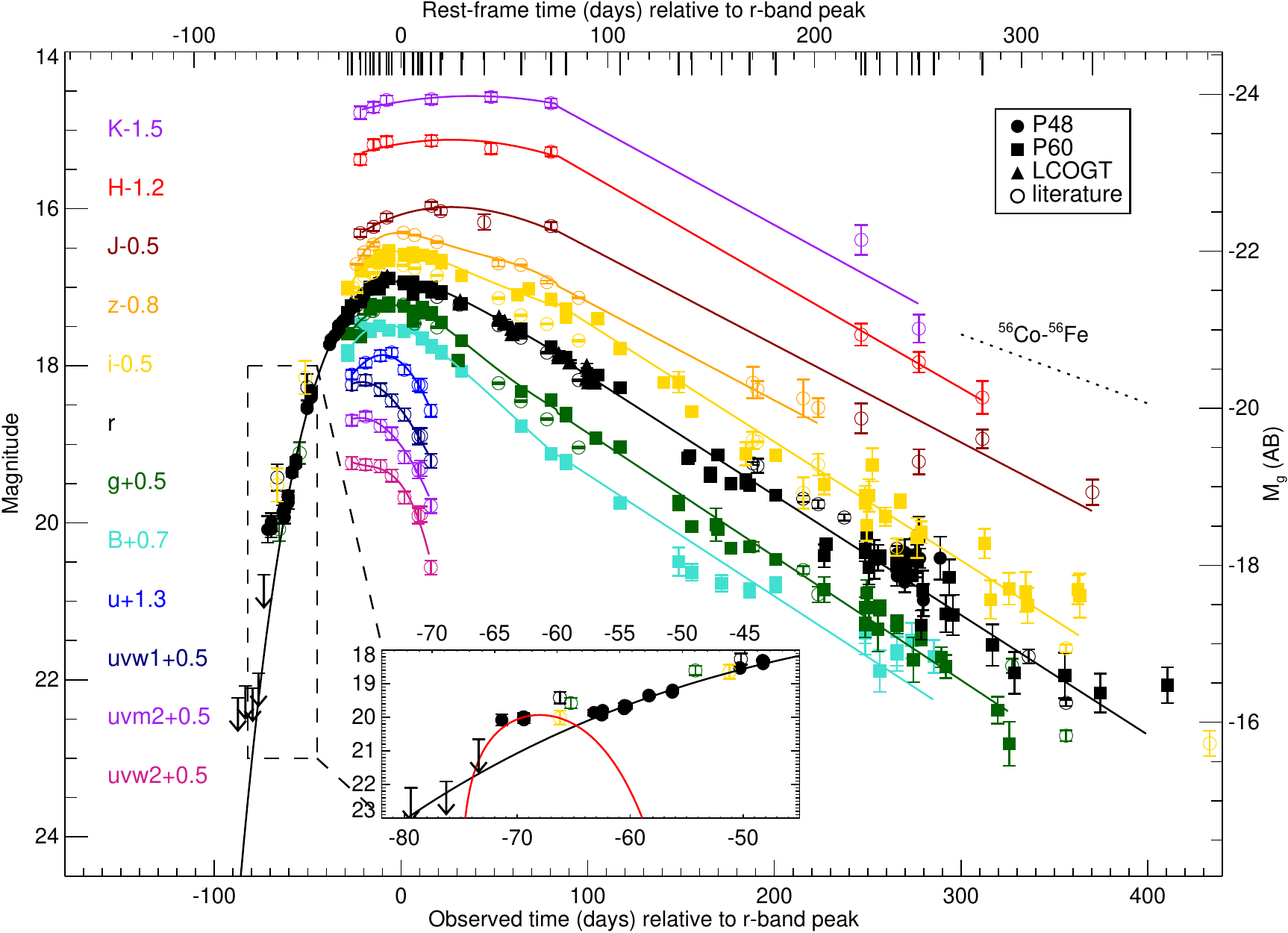}
  \caption{Light curves of \dam. Different colors and symbols denote
    different filters and telescopes, respectively. The solid lines
    show polynomial fits to the data points, where the early-time data
    (before $t_{\rm obs}=80$~days) are fit with a polynomial of order
    2--5, while the late-time data are fit with a straight line. The
    inset displays a zoom of the very early-time light curve, showing
    evidence for an initial plateau with a duration of about 5--10
    days. The solid red line in the inset shows a fit to the
    early-time $r$-band light curve using the
    \citet{2015ApJ...808L..51P} model, explained in more detail in
    Sec.~\ref{sec:modelling}. The late-time light curve is not that
    different from the decay expected from full trapping of gamma rays
    produced in the radioactive decay of $^{56}$Co to $^{56}$Fe, as
    indicated by the dotted line. The short vertical lines at the top
    show the epochs at which we constructed spectral energy
    distributions (SEDs) to derive the bolometric light curve. The
    absolute $g$-band AB magnitude corresponding to the observed
    $r$-band light curve around peak is indicated on the right-hand
    axis. \label{fig:12dam_lc}}
\end{figure*}

\subsection{\dcc}

\dcc\ has not had any exposure in the literature yet. It was flagged
as a transient source on 2013 August 29, which is the day that it was
also first detected, with a magnitude of $r\approx19.7$. It was
initially slowly fading, but it surprisingly rebrightened by almost a
magnitude, tens of days after discovery. iPTF had not observed this
field before this date for almost a year (2012 September 28). The
Catalina Sky Survey \citep[CSS;][]{2009ApJ...696..870D} monitored this
field in early January 2013\footnote{see
  http://nesssi.cacr.caltech.edu/DataRelease}, but there are no
additional detections or useful limits during the months preceding the
date of discovery. The images of the \dcc\ field from the 2010 and
2012 PTF observing campaigns only show upper limits at the
\dcc\ location. Its sky coordinates are $\alpha = 2^{\rm h}57^{\rm
  m}02.50^{\rm s}$, $\delta = -00^\circ 18' 44.0''$ (J2000.0), with an
uncertainty of 0.1\arcsec. It was independently discovered by the
Catalina Real Time Survey (CRTS) on 2013 September 12 and given the
name CSS130912:025702$-$001844 \citep{2013ATel.5437....1D}. At this
location the Galactic extinction is estimated to be moderate,
$A_V=0.18$~mag \citep{2011ApJ...737..103S,1989ApJ...345..245C}.
Spectroscopic follow-up observations were performed with the Double
Spectrograph (DBSP) at the Palomar 200~inch (P200), LRIS at Keck~I,
and the Inamori-Magellan Areal Camera \& Spectrograph (IMACS) at the
Magellan {\it Baade} telescope, showing \dcc\ to be a SLSN at
$z=0.4305$.

\dcc\ was imaged with the P48 Oschin (i)PTF survey telescope in the
Mould $R$ filter, the P60 in SDSS $gri$, the 4.3-m Discovery Channel
Telescope (DCT, at Lowell Observatory, Arizona) with the Large
Monolithic Imager (LMI) in SDSS $ri$, and finally with the {\it HST}
Advanced Camera for Surveys (ACS) Wide-Field Camera using filter F625W
(under program GO-13858; P.I. A. De Cia).

\subsection{Data Reduction}

All \dam\ and \dcc\ images were reduced in a standard fashion; for the
P48 images this was done using the IPAC pipeline
\citep{2014arXiv1404.1953L}. Image-subtraction point-spread function
(PSF) photometry is performed on all P48 and P60 images using a custom
routine written by one of us (M.S.). This pipeline is described by
\citet{2015MNRAS.446.3895F}; it constructs deep reference images --
from either before the SN explosion or after the SN has faded -- and
astrometrically aligns the images using the Automated Astrometry
described by \citet{2008ASPC..394...27H} and the Naval Observatory
Merged Astrometric Dataset \citep[NOMAD;][]{2004AAS...205.4815Z}. The
image PSFs are matched in order to perform image subtraction, and PSF
photometry is extracted from the difference image at the SN
location. The fluxes are calibrated against the SDSS Data Release 10
\citep{2014ApJS..211...17A} when available, and otherwise using the
photometric catalog of \citet{2012PASP..124...62O}.  The LCOGT data
have been reduced using a custom pipeline developed by one of us
(S.V.). This pipeline is described in the Appendix of
\citet{2016MNRAS.459.3939V}, and employs standard procedures (PYRAF,
DAOPHOT) in a PYTHON framework. Host-galaxy flux was removed using
image-subtraction technique HOTPANTS\footnote{ High Order Transform of
  PSF ANd Template Subtraction;
  http://www.astro.washington.edu/users/becker/v2.0/hotpants.html}).
PSF magnitudes were computed on the subtracted images and transformed
to the standard SDSS filter system via standard-star observations
taken during clear nights. The logs of the photometric observations of
\dam\ and \dcc\ are presented in Tables~\ref{tab:12dam_logphotometry}
and \ref{tab:13dcc_logphotometry}, respectively.

The \dcc\ spectra were reduced using standard pipelines and will be
made digitally available via WISeREP \citep{2012PASP..124..668Y}.  The
log of these spectroscopic observations is given in
Table~\ref{tab:13dcc_logspec}.

\begin{figure*}[t]
  \includegraphics[width=0.5\hsize]{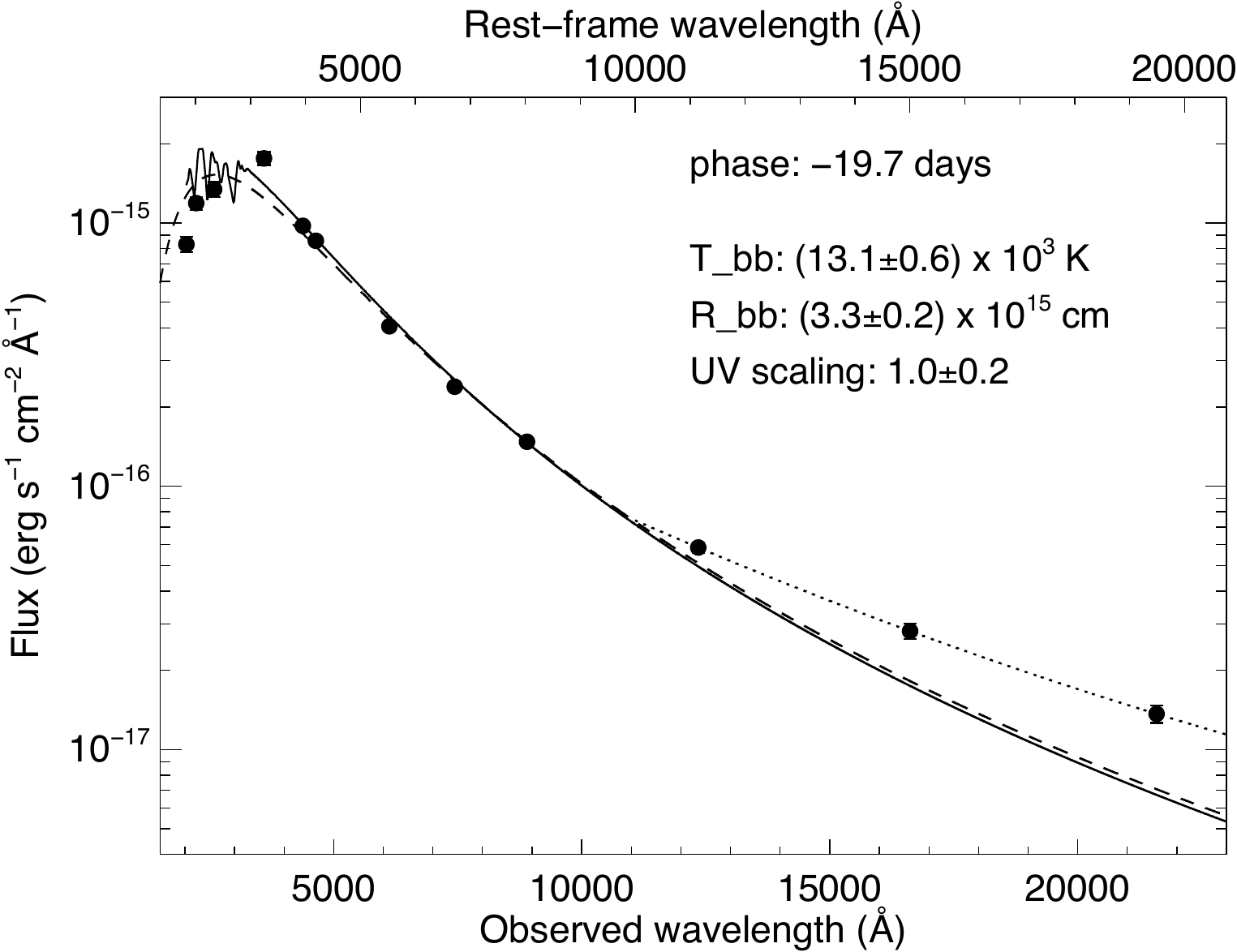}
  \includegraphics[width=0.5\hsize]{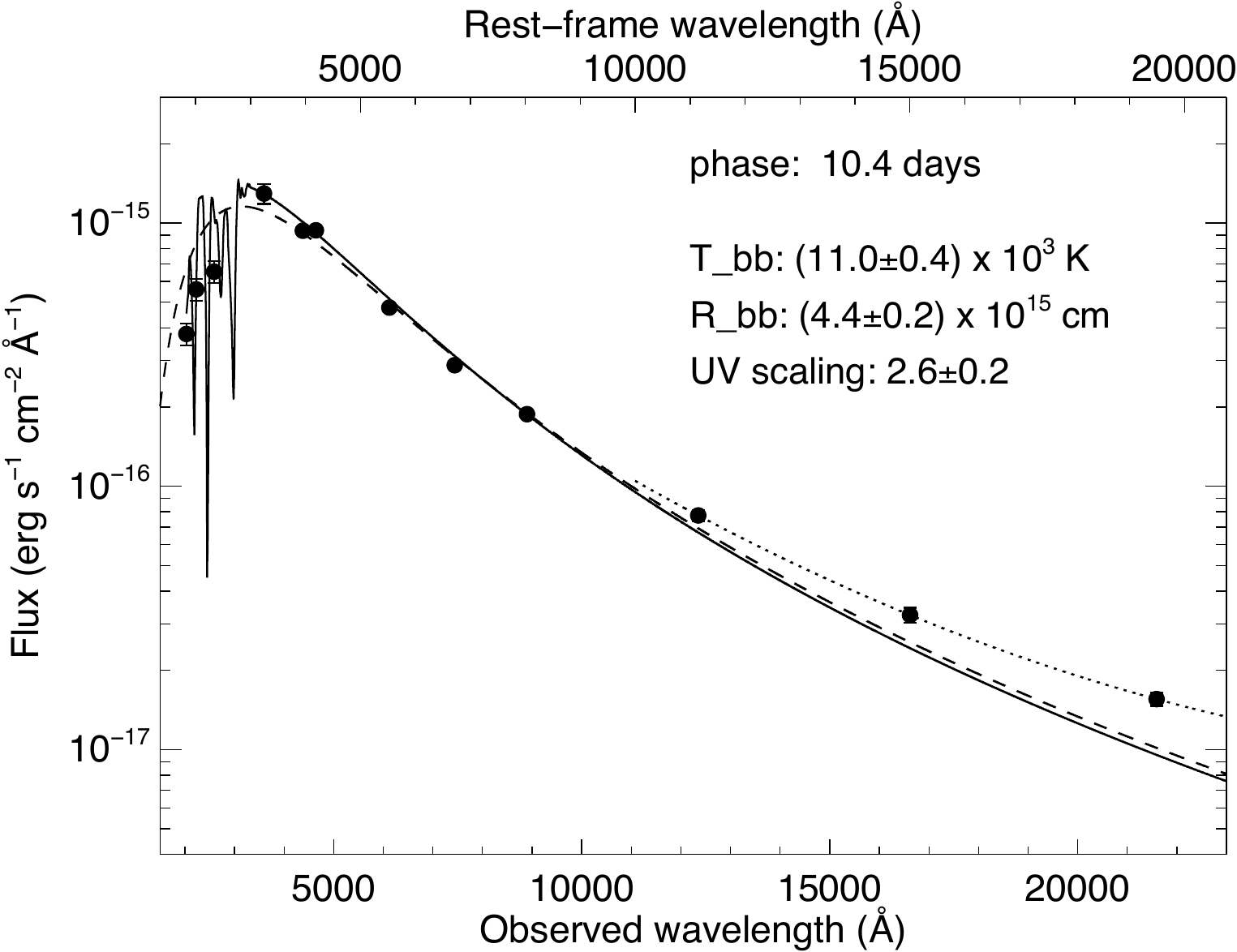}
  \caption{Two examples of Planck-function fits to the SEDs
    constructed from the photometry. The dashed lines show the
    best-fitting Planck function to the data, while the solid line
    depicts the best-fitting Planck function that has been modified
    below 3000~\AA\ using an observed near-UV spectrum taken with
    {\it HST}. Simply comparing the near-UV magnitudes (the three data
    points on the left) with the $uBg$ magnitudes suggests that the UV
    absorption is increasing with time. The dotted lines display a
    low-order polynomial fit to the NIR data, which is used to
    estimate the amount of NIR excess above the Planck fit. This NIR
    excess emission might be produced by re-emission of photons absorbed in
    the near-UV. The blackbody temperature, radius, and UV scaling
    factor derived from these fits are also listed (see the text for
    details).
    \label{fig:planckfits}}
\end{figure*}

\section{Photometric Evolution}
\label{sec:photometry}

\subsection{\dam}
\label{sec:phot12dam}

Figure~\ref{fig:12dam_lc} shows the resulting light curve of \dam,
combining data from our PTF campaign (filled circles, squares, and
triangles for P48, P60, and LCOGT data, respectively) with
measurements published in the literature (open circles) from
\citet{2013Natur.502..346N} and \citet{2014arXiv1409.7728C}, which
include ultraviolet (UV) observations by {\it Swift}
\citep{2004ApJ...611.1005G}. All UV and optical filters are in the AB
system \citep{oke83}, while the near-infrared (NIR) filters $JHK$ are
in the Vega system. The magnitudes shown have been corrected for the
Galactic extinction along this sightline
\citep[$E_{B-V}=0.01$~mag;][]{2011ApJ...737..103S}. We performed
polynomial fits to the light curves (magnitudes versus linear time)
with the order ranging from two (for filters with a sparsely sampled
light curve, such as $JHK$) to five (for the well-sampled $r$-band
data) at phases between $-60$ and $+75$ days (here and throughout the
paper, phase refers to the time in days relative to the main peak in
the $r$-band filter in the {\it rest} frame). For the filters with
measurements beyond $+75$ days, the data are adequately described by a
straight line -- that is, an exponential decay of the flux in
time. The combined early- and late-time fits are shown by the solid
lines. For the filters $Bgri$ these fits were performed only to PTF
data. The observed late-time slopes are in the range
$0.012-0.016$~mag~day$^{-1}$ for all filters. The inset shows a
close-up view of the early-time light curve, which includes a
\citet{2015ApJ...808L..51P} model fit (solid red line) discussed in
Sec.~\ref{sec:modelling}.

\begin{figure*}
  \centering
  \includegraphics[width=\hsize]{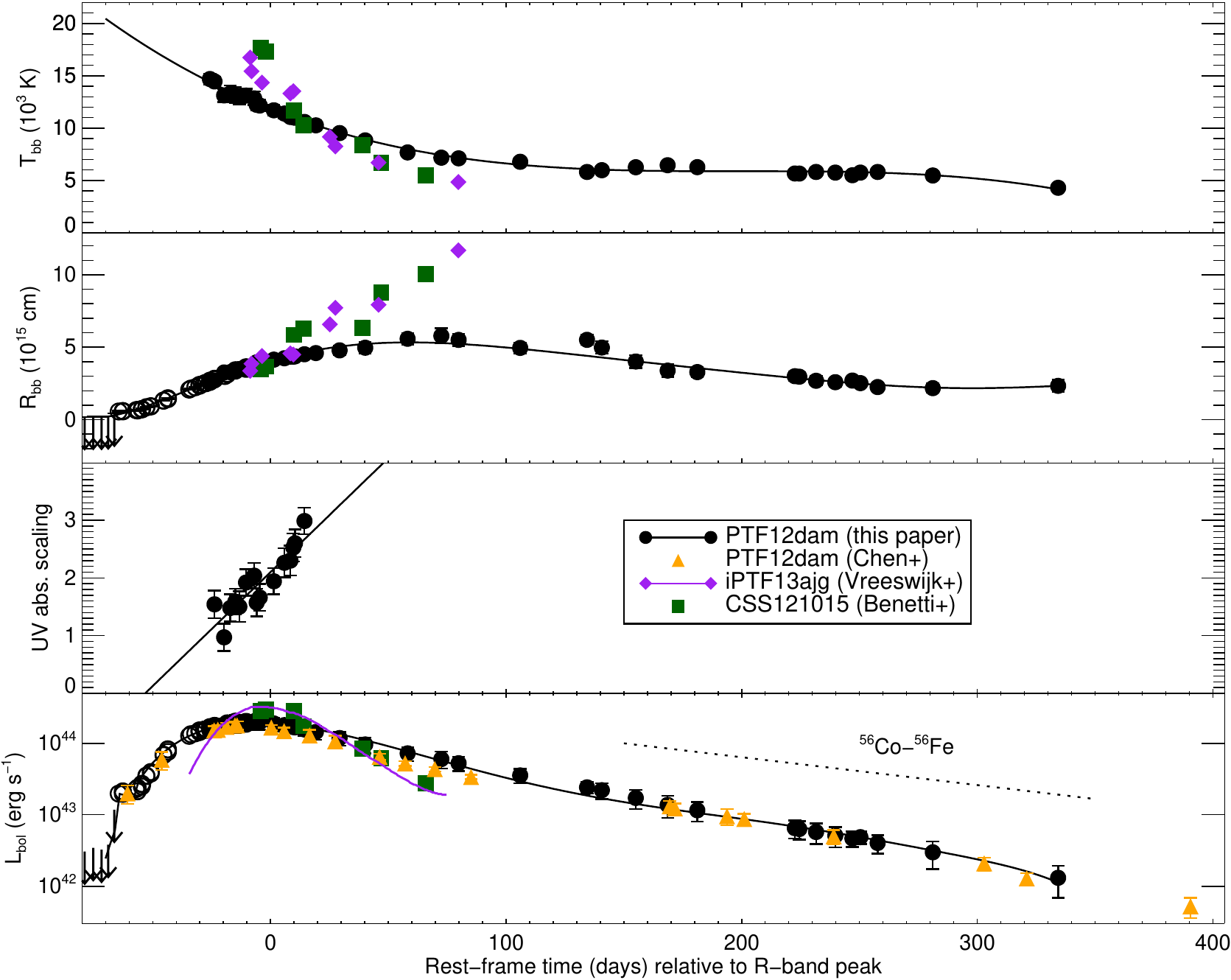}
  \caption{Time evolution of (from top to bottom) the blackbody
    effective temperature, blackbody radius, UV absorption strength
    scaling of observed {\it HST} spectrum, and bolometric luminosity
    of \dam\ compared to a few other SLSNe.  The filled black circles
    correspond to the best-fit values inferred from the
    Planck-function fits to the SEDs constructed from the light curve,
    with fit parameters $T_{\rm bb}$, $R_{\rm bb}$, and UV absorption
    scale factor. The solid lines in the top three panels are
    polynomial fits to the best-fit values, and the solid black line
    in the lower panel is derived from these. The radii inferred from
    the early $r$-band data points alone (instead of the multifilter
    SEDs), adopting the temperature evolution shown in the top panel,
    are depicted by the open circles, as are the corresponding
    bolometric luminosity values.  The filled circles in the bottom
    panel show the \dam\ bolometric light curve inferred from the
    $T_{\rm bb}$ and $R_{\rm bb}$ values and modified by the
    polynomial fit to the UV scaling.  We also show the
    \dam\ bolometric light curve as inferred by
    \citet{2014arXiv1409.7728C} and updated by Chen et al. (2016,
    Erratum, in prep.).  The temperature, radius, and bolometric
    luminosity evolution of \dam\ are compared to a few other SLSNe:
    iPTF~13ajg \citep[filled purple diamonds;][]{2014ApJ...797...24V}
    and \css\ \citep[filled green squares;][]{2013arXiv1310.1311B}.
    \label{fig:bol12dam}}
\end{figure*}

The peak brightness in the observed $r$ band of $m_r=16.9$~mag is
reached at a modified Julian date (MJD) of 56096.7, corresponding to
UTC 2012 June 18.7. At this epoch, we determine a K-correction
\citep[see][]{2002astro.ph.10394H} from observed $r$ to rest-frame
SDSS $g$ of $-0.01$, resulting in an absolute $g$-band peak magnitude
of $M_g,{\rm peak}=-21.6$ (the distance modulus of \dam\ is
38.55~mag). The K$_{rg}$-correction is determined assuming that
\dam\ is radiating like a blackbody, adopting the temperature
evolution as inferred below (see Fig.~\ref{fig:bol12dam}). The
absolute $g$-band magnitude is plotted on the right-hand axis of
Fig.~\ref{fig:12dam_lc}. We note that as the K-correction evolves in
time (K$_{rg}$ develops from $+0.2$ at a phase of $-70$~days, to
$-1.0$ at $+334$~days), the scale on this axis is only valid for the
$r$-band light curve (in black) around peak magnitude. Our earliest
detections of the SN, between $-60$ and $-55$ rest-frame days before
the $r$-band peak, show evidence for significant emission in excess of
that expected from extrapolation of the polynomial fit -- see the
inset in Fig.~\ref{fig:12dam_lc}.  What appears to be a single data
point in the light curve, at $-70$~days in the observer's frame,
actually consists of three independent measurements, making the excess
above the polynomial fit highly significant. This excess emission is
only clear from the PTF data, and has not been inferred from imaging
data sets published to date \citep[e.g.,][]{2013Natur.502..346N}.  The
$r$-band limiting magnitudes shown have been determined by coadding
all the images (typically three) taken during a single night.

In order to derive the bolometric light curve of \dam, we select a
number of epochs at which we construct spectral energy distributions
(SEDs). These correspond to the epochs at which observations were
performed in the near-UV by {\it Swift}, the $B$- or $K$-band filters;
they are indicated as short solid lines at the top of
Fig.~\ref{fig:12dam_lc}.  We adopt the magnitude as estimated by the
polynomial fits at a particular epoch, and the magnitude error from
the measurement in the same filter closest in time to the SED epoch.

Two example SEDs, taken from epochs with phases $-19.7$ and $+10.4$
days, are shown in Fig.~\ref{fig:planckfits}. The overall shape of the
SEDs for all epochs is very similar to that of a Planck function. We
first fit Planck functions to the observed SEDs (which have been
corrected for Galactic extinction) by converting the observed
magnitude to flux in the rest frame at the central wavelengths of the
filters. These fits provide an estimate of the blackbody effective
temperature ($T_{\rm bb}$) and radius ($R_{\rm bb}$); they are shown
by the dashed lines in the example SEDs in
Fig.~\ref{fig:planckfits}. It is clear that both the near-UV and the
NIR parts of the SED are poorly fit.

We account for the additional UV absorption by utilizing the spectrum
of \dam\ as observed by {\it HST} on 2012 May 26, at a phase of $-20$
days (Quimby et al., in prep.). This spectrum is well approximated by
a Planck function redward of 3000~\AA, while strong absorption
features are present blueward of that. These near-UV features are at
very similar wavelengths as the features observed in other SLSNe, such
as SCP~06F6 \citep{2009ApJ...690.1358B}, SNLS~06D4eu
\citep{2013ApJ...779...98H}, and \ajg\ \citep{2014ApJ...797...24V}. We
normalize the {\it HST} spectrum blueward of 3000~\AA\ (rest frame),
so it can be used to modify the Planck function to be fit, as
explained below. The flux below a rest-frame wavelength of
1800~\AA\ in this template spectrum is assumed to be zero, which is a
fair assumption as the SLSNe with spectra down to this rest-frame
wavelength show very low flux levels in that spectral region and the
flux is rapidly declining toward the blue.

We again perform Planck-function fitting, but rather than converting
magnitudes to rest-frame flux, we instead first construct a Planck
spectrum at a distance of 10~pc in the rest frame, based on the two
fit parameters $T_{\rm bb}$ and $R_{\rm bb}$. This spectrum is then
modified in the region 1800--3000~\AA\ with the normalized {\it HST} spectrum
(in the rest frame), by simply multiplying the {\it HST} and the blackbody
spectra over this wavelength region. Below 1800~\AA\ it is assumed to
be zero, beyond 3000~\AA\ it is a pure blackbody. This template
spectrum is then converted to the observer's frame [with the spectrum
in units of \ergscmA; the spectrum is multiplied by $(10/{\rm d}_{\rm
  l,pc})^2/(1+z)$, where $d_{\rm l,pc}$ is the luminosity distance of
\dam\ in pc], and magnitudes are extracted from this spectrum by
performing synthetic photometry (using the filter transmission curves)
to obtain model magnitudes for each filter. These model magnitudes are
fit to the observed magnitudes to obtain the best-fit $T_{\rm bb}$ and
$R_{\rm bb}$. To allow for the absorption features in the {\it HST} spectrum
to vary with time, we introduce a fit parameter which simply scales
the strength of these absorption features. In particular, the
normalized, scaled {\it HST} spectrum is calculated as flux$_{\rm
  HST,norm,scaled} = 1 - (1 - {\rm flux}_{\rm HST,norm}$) $\times$
scale factor. For example, in Fig.~\ref{fig:planckfits}, the solid lines show
the best-fitting modified blackbody spectrum.  In the left panel, at a
phase within a day of the epoch that the {\it HST} spectrum was taken, the
UV absorption scaling was found to be around unity, as expected. On
the right, the best-fit scale factor is much stronger, almost a factor
of three. The third panel of Fig.~\ref{fig:bol12dam} shows the
best-fitting UV absorption scale factor for all epochs for which UV
photometry is available; it is clearly increasing in time.

Interestingly, an increase in the UV absorption lines is also observed
for \ajg\ \citep{2014ApJ...797...24V}, one of the few SLSNe-I with
good temporal coverage of the near-UV region where this can be
measured. The increase is clear from Fig.~4 of
\citet{2014ApJ...797...24V}, where the spectra with UV coverage have
been normalized by the blackbody fits. Measuring the combined
equivalent width of the \ajg\ model spectra in the rest-wavelength
range of 2000--2800~\AA\ for epochs 2--5, we obtain a linear increase
in rest-frame equivalent width of 35\% between phases $-8$ and
$+10$~days. The corresponding increase in equivalent width (or
increase in UV scale factor) for \dam\ over the same phase is very
similar, 39\%. However, the rest-frame equivalent width over the range
2000--2800~\AA\ at phase $-8$~days is about a factor of three smaller
for \dam\ compared to \ajg: 200~\AA\ for \dam\ versus 600~\AA\ for
\ajg. An increase in the near-UV absorption is not unexpected: as the
ejecta cool down from initially being very hot, the fraction of ions
in the lower-ionization states increases, and these are thought to be
responsible for the near-UV absorption lines (\ion{Mg}{2}, \ion{C}{2},
\ion{C}{3}, \ion{Fe}{3}).

\begin{deluxetable}{rrrr}
  \tablecaption{Blackbody temperature, radius, and bolometric
    luminosity of \dam. \label{tab:lbol12dam}}
  \tablehead{
    \colhead{Phase} &
    \colhead{$T_{\rm bb}$} &
    \colhead{$R_{\rm bb}$} &
    \colhead{log $L_{\rm bol}$}\\
    (days) & (10$^3$~K) & (10$^{15}$~cm) & (\ergsec)
  }
  \startdata
$-$78.86 &                  &  $<$ 0.19       &  $<$ 42.49 \\ 
$-$75.24 &                  &  $<$ 0.21       &  $<$ 42.54 \\ 
$-$71.68 &                  &  $<$ 0.21       &  $<$ 42.51 \\ 
$-$68.86 &                  &  $<$ 0.23       &  $<$ 42.58 \\ 
$-$66.28 &                  &  $<$ 0.42       &  $<$ 43.07 \\ 
$-$25.71 & 14.69 $\pm$ 0.52 & 2.57 $\pm$ 0.11 & 44.21 $\pm$ 0.07 \\ 
$-$23.76 & 14.46 $\pm$ 0.47 & 2.71 $\pm$ 0.11 & 44.24 $\pm$ 0.06 \\ 
$-$19.69 & 13.13 $\pm$ 0.64 & 3.26 $\pm$ 0.20 & 44.25 $\pm$ 0.09 \\ 
$-$17.06 & 13.32 $\pm$ 0.77 & 3.28 $\pm$ 0.24 & 44.28 $\pm$ 0.10 \\ 
$-$14.89 & 13.17 $\pm$ 0.77 & 3.39 $\pm$ 0.25 & 44.29 $\pm$ 0.11 \\ 
$-$13.22 & 12.99 $\pm$ 0.71 & 3.49 $\pm$ 0.24 & 44.29 $\pm$ 0.10 \\ 
$-$10.37 & 13.11 $\pm$ 0.66 & 3.50 $\pm$ 0.23 & 44.31 $\pm$ 0.09 \\ 
 $-$6.92 & 12.85 $\pm$ 0.65 & 3.64 $\pm$ 0.24 & 44.31 $\pm$ 0.09 \\ 
 $-$5.87 & 12.23 $\pm$ 0.56 & 3.91 $\pm$ 0.24 & 44.30 $\pm$ 0.09 \\ 
 $-$4.54 & 12.16 $\pm$ 0.55 & 3.94 $\pm$ 0.24 & 44.30 $\pm$ 0.08 \\ 
   1.44 &  11.71 $\pm$ 0.51 & 4.14 $\pm$ 0.24 & 44.28 $\pm$ 0.08 \\ 
   5.89 &  11.39 $\pm$ 0.45 & 4.24 $\pm$ 0.24 & 44.25 $\pm$ 0.08 \\ 
   8.45 &  11.07 $\pm$ 0.43 & 4.35 $\pm$ 0.23 & 44.23 $\pm$ 0.08 \\ 
   9.69 &  11.04 $\pm$ 0.43 & 4.35 $\pm$ 0.24 & 44.23 $\pm$ 0.08 \\ 
  10.39 &  10.99 $\pm$ 0.43 & 4.37 $\pm$ 0.24 & 44.22 $\pm$ 0.08 \\ 
  14.42 &  10.59 $\pm$ 0.40 & 4.52 $\pm$ 0.24 & 44.19 $\pm$ 0.07 \\ 
  19.26 &  10.27 $\pm$ 0.45 & 4.60 $\pm$ 0.30 & 44.16 $\pm$ 0.09 \\ 
  29.34 &   9.53 $\pm$ 0.41 & 4.78 $\pm$ 0.31 & 44.07 $\pm$ 0.08 \\ 
  40.21 &   8.83 $\pm$ 0.42 & 4.97 $\pm$ 0.40 & 43.98 $\pm$ 0.10 \\ 
  58.20 &   7.68 $\pm$ 0.33 & 5.58 $\pm$ 0.40 & 43.86 $\pm$ 0.09 \\ 
  72.52 &   7.18 $\pm$ 0.35 & 5.79 $\pm$ 0.53 & 43.78 $\pm$ 0.10 \\ 
  79.85 &   7.12 $\pm$ 0.30 & 5.50 $\pm$ 0.41 & 43.72 $\pm$ 0.09 \\ 
 105.98 &   6.79 $\pm$ 0.29 & 4.95 $\pm$ 0.37 & 43.55 $\pm$ 0.09 \\ 
 134.35 &   5.81 $\pm$ 0.20 & 5.51 $\pm$ 0.34 & 43.38 $\pm$ 0.07 \\ 
 140.65 &   5.98 $\pm$ 0.26 & 4.98 $\pm$ 0.44 & 43.34 $\pm$ 0.10 \\ 
 155.05 &   6.27 $\pm$ 0.32 & 4.01 $\pm$ 0.45 & 43.24 $\pm$ 0.11 \\ 
 168.56 &   6.46 $\pm$ 0.37 & 3.39 $\pm$ 0.43 & 43.14 $\pm$ 0.13 \\ 
 181.16 &   6.27 $\pm$ 0.32 & 3.28 $\pm$ 0.36 & 43.06 $\pm$ 0.11 \\ 
 222.56 &   5.67 $\pm$ 0.28 & 3.00 $\pm$ 0.31 & 42.81 $\pm$ 0.11 \\ 
 224.44 &   5.66 $\pm$ 0.28 & 2.96 $\pm$ 0.31 & 42.80 $\pm$ 0.11 \\ 
 231.57 &   5.81 $\pm$ 0.31 & 2.68 $\pm$ 0.32 & 42.76 $\pm$ 0.12 \\ 
 239.73 &   5.75 $\pm$ 0.30 & 2.58 $\pm$ 0.30 & 42.71 $\pm$ 0.12 \\ 
 246.99 &   5.50 $\pm$ 0.24 & 2.70 $\pm$ 0.23 & 42.67 $\pm$ 0.09 \\ 
 250.37 &   5.75 $\pm$ 0.19 & 2.52 $\pm$ 0.19 & 42.69 $\pm$ 0.08 \\ 
 257.69 &   5.81 $\pm$ 0.29 & 2.25 $\pm$ 0.24 & 42.61 $\pm$ 0.11 \\ 
 281.11 &   5.48 $\pm$ 0.41 & 2.17 $\pm$ 0.32 & 42.47 $\pm$ 0.15 \\ 
 334.29 &   4.30 $\pm$ 0.35 & 2.33 $\pm$ 0.41 & 42.12 $\pm$ 0.17    
  \enddata
\end{deluxetable}

The blackbody fits to the \dam\ photometry are consistently below the
$JHK$ flux measurements. The absorption in the UV could potentially be
related to this excess emission in the NIR, as the absorbed UV
emission is expected to be reradiated at longer wavelengths. This
effect is observed in Type Ia SNe: the flux absorbed at blue
wavelengths due to \ion{Fe}{2} and \ion{Co}{2} is redistributed to the
red, eventually leading to a NIR secondary maximum \citep[e.g., see
  Figs.~4 and 5 of][]{2000ApJ...530..757P,2007ApJ...656..661K}.

\begin{figure*}
  \centering
  \includegraphics[width=\hsize]{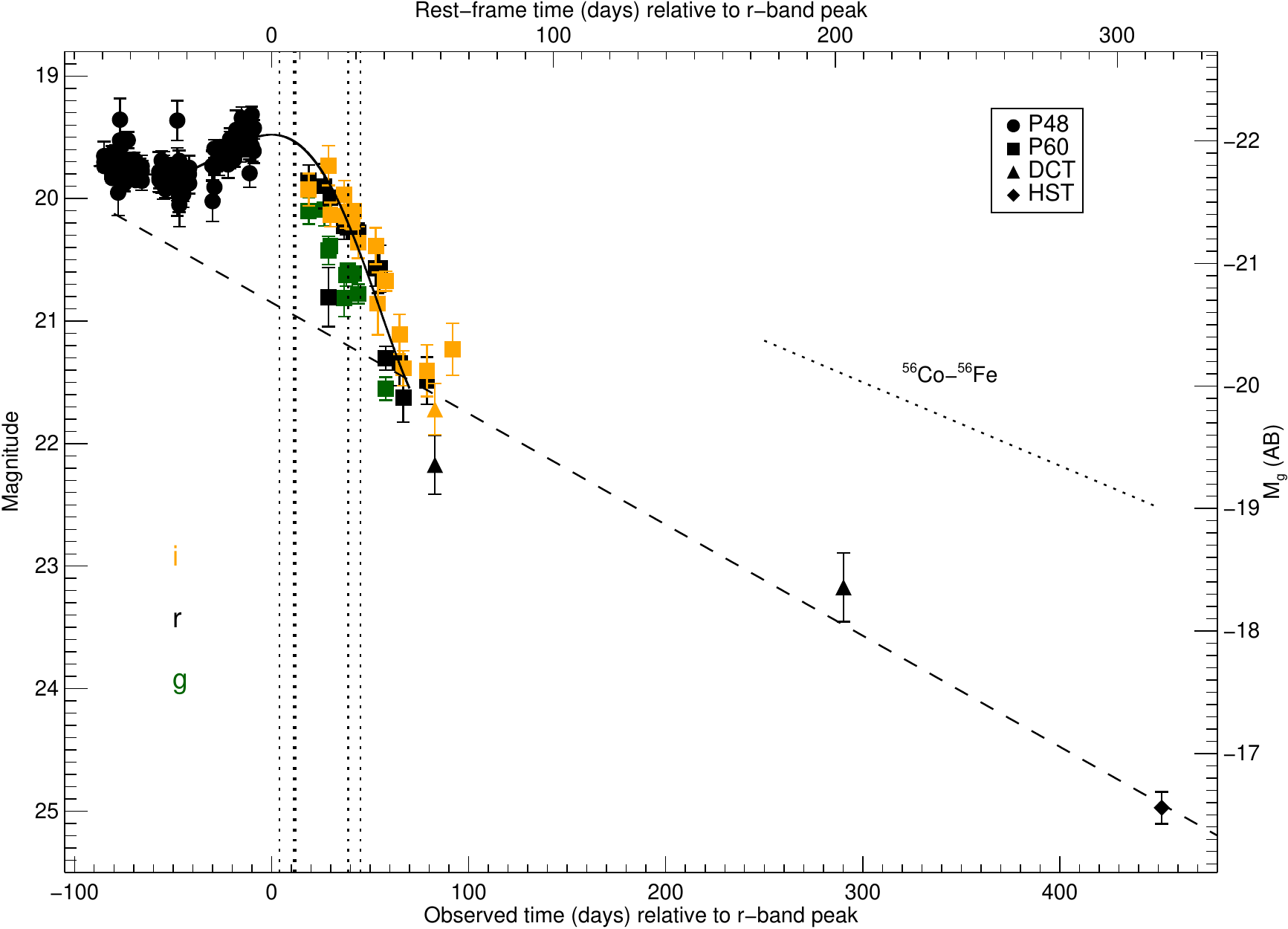}
  \caption{Light curves of \dcc. Different colors and symbols denote
    different filters and telescopes, respectively. The solid line
    shows a polynomial fit to the $r$-band data up to about t$_{\rm
      obs}=70$~days. The dashed line displays a straight-line fit
    (i.e., exponential decay of the flux vs. linear time) to the last
    seven $r$-band data points. The late-time light curve is similar
    to the decay expected from radioactive decay of $^{56}$Co to
    $^{56}$Fe, as indicated by the slanted dotted line. The vertical
    dotted lines show the epochs at which the \dcc\ spectra were
    obtained. \label{fig:13dcc_lc}}
\end{figure*}

We investigate if the NIR excess is related to the UV absorption in
\dam\ by determining the absorbed UV flux over the rest-frame range
2000--3000~\AA, and the excess NIR emission over the rest-frame range
1--2.5~$\mu$m.  For the NIR excess determination, we approximate the
observed spectrum with a low-order polynomial shown as dotted lines in
Fig.~\ref{fig:planckfits}, and integrate over the flux difference
between this spectrum and the best-fitting blackbody (solid line). The
inferred ratio of near-UV absorption over NIR excess shows a smooth
evolution in time, starting at a factor of two at a phase of
$-20$~days and reaching a factor of four at around $-5$~days, after
which it very slowly decreases to a factor between 3.5 and 4 at
$+15$~days. The wavelength upper limit of integration for the NIR
excess of 2.5~$\mu$m is arbitrary; if we extend it to a longer
wavelength the above factor of roughly four would decrease. We note
that the NIR excess is apparent until a phase of $+130$~days, after
which the blackbody fits are consistent with the optical through NIR
flux measurements. The apparent additional NIR emission above the
Planck fits could potentially be caused by photons being emitted at
longer wavelengths following absorption in the UV part of the
spectrum, but it is difficult to confirm this. The lack of NIR excess
beyond $+130$~days can be explained by the SN temperature having
decreased to around 5000~K at this point, resulting in very few
available UV photons to be re-emitted in the NIR.

The resulting blackbody temperature, radius, UV absorption scale
factor, and bolometric luminosity evolution are listed in
Table~\ref{tab:lbol12dam} and shown in Fig.~\ref{fig:bol12dam}. The
estimate of the bolometric luminosity includes a correction for the UV
absorption as described by the straight-line fit to the scale factor
shown in the third panel from the top. Adopting the temperature
evolution estimated by the polynomial fit at early times, we also
infer the blackbody radius from each $r$-band measurement before
maximum light, which provides a rough estimate of the radius evolution
at early times.

Integrating the bolometric light curve that we derive over the phase
interval from $-70$ to $+334$~days results in an estimated total
radiated energy of $E_{\rm rad}=1.8\times10^{51}$~erg for \dam.  This
value is similar to that found for other SLSNe, such as the Type I
SNLS~06D4eu \citep{2013ApJ...779...98H} and
\ajg\ \citep{2014ApJ...797...24V} and the Type II
\css\ \citep{2013arXiv1310.1311B}.

\begin{figure}[t!]
  \centering
  \includegraphics[width=\hsize]{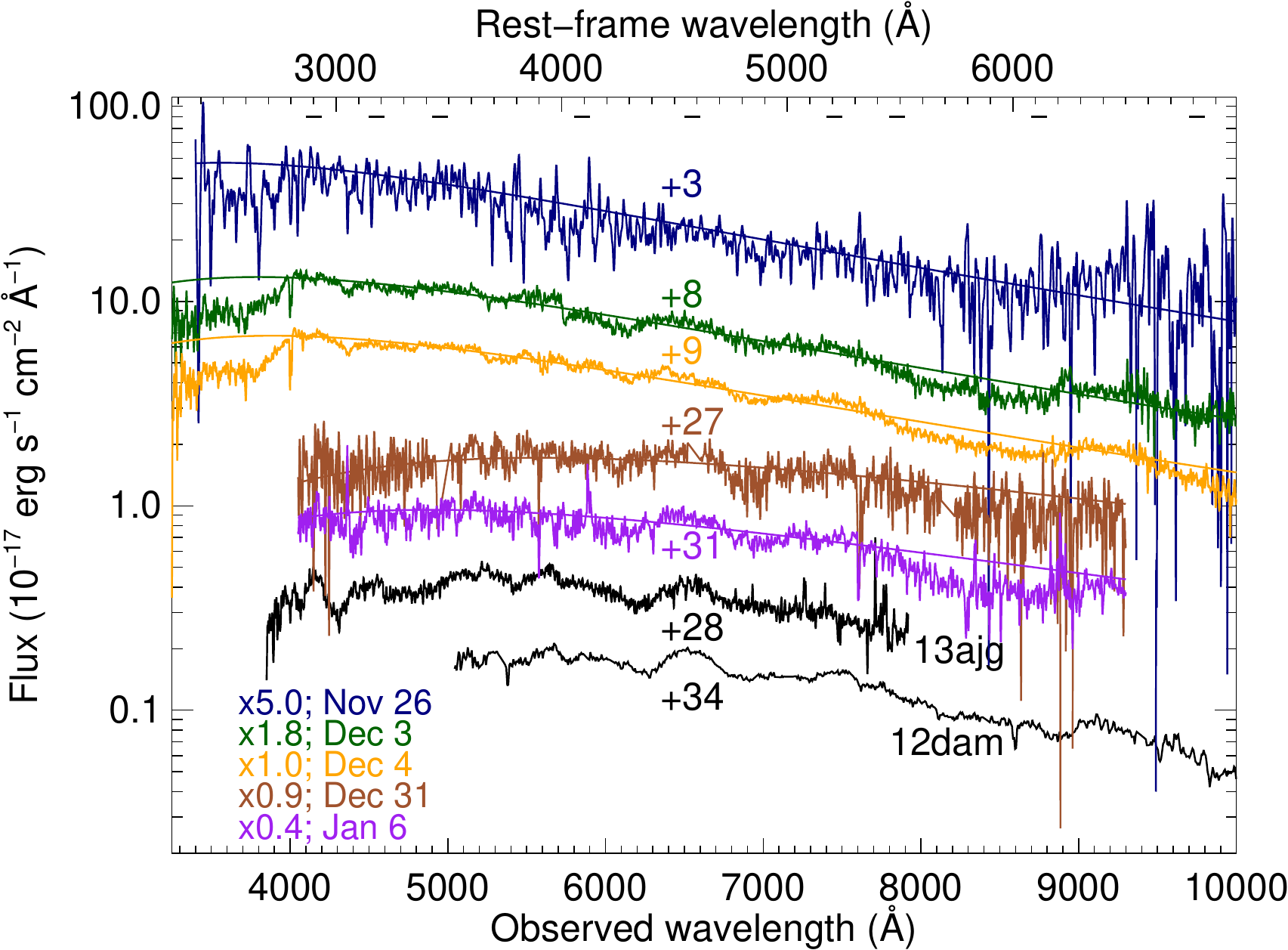}
  \caption{Time series of spectra of \dcc\ (see
    Table~\ref{tab:13dcc_logspec}). The spectra were corrected for
    Galactic extinction \citep{2011ApJ...737..103S} and scaled to the
    polynomial fit to the $r$-band photometry depicted in
    Fig.~\ref{fig:13dcc_lc} to ensure a proper absolute calibration.
    To avoid the spectra overlapping each other, an additional scaling
    was applied as indicated at the bottom left, along with the date
    of observation (Nov./Dec. 2013 and Jan. 2014); the phase
    (rest-frame days relative to the $r$-band maximum) is shown next
    to each spectrum. The \dcc\ spectra were smoothed with a Gaussian
    filter having a full width at half-maximum intensity (FWHM) of
    5~\AA.  The spectra were fit with a Planck function to selected
    50~\AA-wide wavelength regions (the same for all spectra; these
    regions are indicated with dashes at the top of the plot). For
    comparison, we show the spectra of two known SLSNe-I at the bottom
    in black: \ajg\ \citep{2014ApJ...797...24V} and
    \dam\ \citep{2013Natur.502..346N} at phases of $+28$ and
    $+34$~days, respectively. \label{fig:13dcc_spectra}}
\end{figure}

\subsection{\dcc}
\label{sec:phot13dcc}

Figure~\ref{fig:13dcc_lc} shows the light curve of \dcc. The $r$-band
evolution displays an initial slow decline extending over at least 30
rest-frame days, after which the SLSN is rebrightening to reach its
peak brightness around 60 rest-frame days after the initial detection.
The main peak brightness of the polynomial fit through the $r$-band
data (shown by the solid line in Fig.~\ref{fig:13dcc_lc}) of
$r=19.5$~mag is reached at a modified Julian date of MJD = 56618.3,
corresponding to UTC 2013 November 22.3.  Although there is a gap in
the data around the time of the peak, we adopt this date and magnitude
as the main peak time and $r$-band magnitude of \dcc. At this epoch,
we determine a K-correction \citep[see][]{2002astro.ph.10394H} from
observed $r$ to rest-frame SDSS $g$ of $-0.4$ mag, leading to an
absolute $g$-band peak magnitude of $M_g,{\rm peak}=-22.0$ (the
distance modulus of \dcc\ is 41.93~mag). The absolute $g$-band
magnitude is plotted on the right-hand axis of
Fig.~\ref{fig:13dcc_lc}. At the redshift of \dcc\ ($z=0.431$), the
effective wavelengths of observed $r$ and rest-frame $g$ match very
well, leading to a K-correction that is almost independent of the
color evolution of the SN with a magnitude depending mainly on the
redshift: K$_{rg} = -2.5\, {\rm log} (1+z) = -0.4$.

\begin{deluxetable}{rlrr}
  \tablecaption{Blackbody temperature (adopted from \dam), radius, and bolometric
    luminosity of \dcc. \label{tab:lbol13dcc}}
  \tablehead{
    \colhead{Phase} &
    \colhead{$T_{\rm bb}$} &
    \colhead{$R_{\rm bb}$} &
    \colhead{log $L_{\rm bol}$}\\
    (days) & ($10^3$~K) & ($10^{15}$~cm) & ([\ergsec])
  }
  \startdata
$-$59.33 & 18.80   & 2.50 $\pm$ 0.04 & 44.75 $\pm$ 0.07 \\
$-$56.54 & 18.39   & 2.52 $\pm$ 0.03 & 44.71 $\pm$ 0.06 \\
$-$55.82 & 18.29   & 2.51 $\pm$ 0.05 & 44.70 $\pm$ 0.08 \\
$-$55.17 & 18.19   & 2.58 $\pm$ 0.05 & 44.71 $\pm$ 0.08 \\
$-$54.43 & 18.08   & 2.34 $\pm$ 0.20 & 44.62 $\pm$ 0.15 \\
$-$53.79 & 17.99   & 2.71 $\pm$ 0.09 & 44.74 $\pm$ 0.10 \\
$-$53.03 & 17.89   & 2.67 $\pm$ 0.05 & 44.72 $\pm$ 0.08 \\
$-$52.32 & 17.79   & 2.56 $\pm$ 0.04 & 44.67 $\pm$ 0.07 \\
$-$51.01 & 17.60   & 2.94 $\pm$ 0.09 & 44.77 $\pm$ 0.10 \\
$-$50.24 & 17.49   & 2.58 $\pm$ 0.04 & 44.65 $\pm$ 0.07 \\
$-$49.54 & 17.40   & 2.66 $\pm$ 0.06 & 44.66 $\pm$ 0.08 \\
$-$48.83 & 17.30   & 2.72 $\pm$ 0.04 & 44.67 $\pm$ 0.07 \\
$-$48.13 & 17.20   & 2.67 $\pm$ 0.03 & 44.65 $\pm$ 0.06 \\
$-$47.40 & 17.10   & 2.69 $\pm$ 0.03 & 44.64 $\pm$ 0.06 \\
$-$46.71 & 17.01   & 2.68 $\pm$ 0.04 & 44.63 $\pm$ 0.07 \\
$-$46.04 & 16.92   & 2.70 $\pm$ 0.06 & 44.63 $\pm$ 0.09 \\
$-$39.80 & 16.10   & 2.82 $\pm$ 0.08 & 44.58 $\pm$ 0.10 \\
$-$39.02 & 16.00   & 2.89 $\pm$ 0.06 & 44.59 $\pm$ 0.08 \\
$-$38.32 & 15.92   & 2.95 $\pm$ 0.07 & 44.60 $\pm$ 0.09 \\
$-$37.63 & 15.83   & 2.84 $\pm$ 0.06 & 44.56 $\pm$ 0.08 \\
$-$36.94 & 15.74   & 2.91 $\pm$ 0.05 & 44.57 $\pm$ 0.08 \\
$-$36.25 & 15.66   & 2.97 $\pm$ 0.05 & 44.58 $\pm$ 0.07 \\
$-$35.56 & 15.57   & 3.01 $\pm$ 0.04 & 44.58 $\pm$ 0.07 \\
$-$34.86 & 15.49   & 2.93 $\pm$ 0.05 & 44.55 $\pm$ 0.08 \\
$-$34.15 & 15.40   & 3.05 $\pm$ 0.05 & 44.57 $\pm$ 0.08 \\
$-$33.45 & 15.31   & 3.17 $\pm$ 0.17 & 44.60 $\pm$ 0.12 \\
$-$32.75 & 15.23   & 2.99 $\pm$ 0.09 & 44.54 $\pm$ 0.10 \\
$-$32.06 & 15.15   & 3.04 $\pm$ 0.08 & 44.54 $\pm$ 0.09 \\
$-$31.37 & 15.06   & 2.95 $\pm$ 0.06 & 44.50 $\pm$ 0.08 \\
$-$29.26 & 14.82   & 3.13 $\pm$ 0.06 & 44.53 $\pm$ 0.08 \\
$-$20.96 & 13.88   & 3.39 $\pm$ 0.12 & 44.48 $\pm$ 0.10 \\
$-$20.26 & 13.81   & 3.55 $\pm$ 0.07 & 44.51 $\pm$ 0.08 \\
$-$19.53 & 13.73   & 3.62 $\pm$ 0.09 & 44.52 $\pm$ 0.09 \\
$-$18.85 & 13.66   & 3.74 $\pm$ 0.05 & 44.54 $\pm$ 0.07 \\
$-$18.13 & 13.58   & 3.80 $\pm$ 0.04 & 44.54 $\pm$ 0.06 \\
$-$15.39 & 13.29   & 3.86 $\pm$ 0.09 & 44.52 $\pm$ 0.08 \\
$-$14.68 & 13.22   & 4.03 $\pm$ 0.09 & 44.55 $\pm$ 0.08 \\
$-$13.98 & 13.15   & 4.00 $\pm$ 0.08 & 44.53 $\pm$ 0.08 \\
$-$13.30 & 13.08   & 4.03 $\pm$ 0.07 & 44.53 $\pm$ 0.07 \\
$-$12.59 & 13.01   & 4.19 $\pm$ 0.07 & 44.55 $\pm$ 0.08 \\
$-$11.24 & 12.88   & 4.28 $\pm$ 0.15 & 44.56 $\pm$ 0.10 \\
$-$10.52 & 12.80   & 4.49 $\pm$ 0.08 & 44.59 $\pm$ 0.08 \\
 $-$9.74 & 12.73   & 4.28 $\pm$ 0.05 & 44.53 $\pm$ 0.06 \\
 $-$9.03 & 12.66   & 4.35 $\pm$ 0.07 & 44.54 $\pm$ 0.07 \\
 $-$7.71 & 12.53   & 4.39 $\pm$ 0.09 & 44.53 $\pm$ 0.08 \\
 $-$7.02 & 12.47   & 4.73 $\pm$ 0.11 & 44.59 $\pm$ 0.08 \\
 $-$6.30 & 12.40   & 4.57 $\pm$ 0.12 & 44.55 $\pm$ 0.09 \\
{\bf 2.81} & {\bf 11.65 $\pm$ 2.46} & {\bf 5.28 $\pm$ 0.44} & {\bf 44.56 $\pm$ 0.27} \\
{\bf 7.81} & {\bf 10.91 $\pm$ 0.40} & {\bf 5.44 $\pm$ 0.20} & {\bf 44.48 $\pm$ 0.07} \\
{\bf 8.50} & {\bf 10.66 $\pm$ 0.34} & {\bf 5.55 $\pm$ 0.19} & {\bf 44.45 $\pm$ 0.06} \\
   13.16 & 10.73   & 4.79 $\pm$ 0.29 & 44.33 $\pm$ 0.13 \\
   18.75 & 10.31   & 5.00 $\pm$ 0.42 & 44.30 $\pm$ 0.15 \\
   20.16 & 10.20   & 3.35 $\pm$ 0.37 & 43.94 $\pm$ 0.17 \\
   20.84 & 10.15   & 4.90 $\pm$ 0.10 & 44.26 $\pm$ 0.08 \\
   25.80 &  9.81   & 4.67 $\pm$ 0.23 & 44.16 $\pm$ 0.12 \\
   26.47 &  9.77   & 4.71 $\pm$ 0.16 & 44.16 $\pm$ 0.10 \\
   27.15 &  9.72   & 4.83 $\pm$ 0.07 & 44.17 $\pm$ 0.07 \\
{\bf 27.17} & {\bf 7.42 $\pm$ 0.76} & {\bf 7.29 $\pm$ 0.51} & {\bf 44.06 $\pm$ 0.16} \\
   27.84 &  9.67   & 4.78 $\pm$ 0.09 & 44.15 $\pm$ 0.08 \\
   29.26 &  9.58   & 4.77 $\pm$ 0.11 & 44.13 $\pm$ 0.08 \\
   30.70 &  9.49   & 4.82 $\pm$ 0.17 & 44.13 $\pm$ 0.10 \\
{\bf 31.48} & {\bf 8.45 $\pm$ 0.74} & {\bf 5.56 $\pm$ 0.34} & {\bf 44.05 $\pm$ 0.14} \\
   36.98 &  9.10   & 4.53 $\pm$ 0.30 & 44.00 $\pm$ 0.13 \\
   37.68 &  9.06   & 4.58 $\pm$ 0.43 & 44.00 $\pm$ 0.16 \\
   38.37 &  9.02   & 4.60 $\pm$ 0.40 & 44.00 $\pm$ 0.15 \\
   40.43 &  8.90   & 3.36 $\pm$ 0.15 & 43.70 $\pm$ 0.11 \\
   45.41 &  8.63   & 3.50 $\pm$ 0.30 & 43.69 $\pm$ 0.15 \\
   46.72 &  8.56   & 3.12 $\pm$ 0.28 & 43.57 $\pm$ 0.15 \\
   55.12 &  8.15   & 3.66 $\pm$ 0.33 & 43.62 $\pm$ 0.15 \\
   57.95 &  8.02   & 2.76 $\pm$ 0.31 & 43.35 $\pm$ 0.17 \\
  202.87 &  5.88   & 3.58 $\pm$ 0.46 & 43.04 $\pm$ 0.18 \\
  315.76 &  4.78   & 2.90 $\pm$ 0.17 & 42.49 $\pm$ 0.13   
  \enddata
  \tablecomments{The bold-faced values are inferred from
    the \dcc\ spectra, while for the others the \dam\ temperature
    evolution was adopted with the radii inferred using
    the $r$-band magnitude.}
\end{deluxetable}

\begin{figure*}
  \centering
  \includegraphics[width=\hsize]{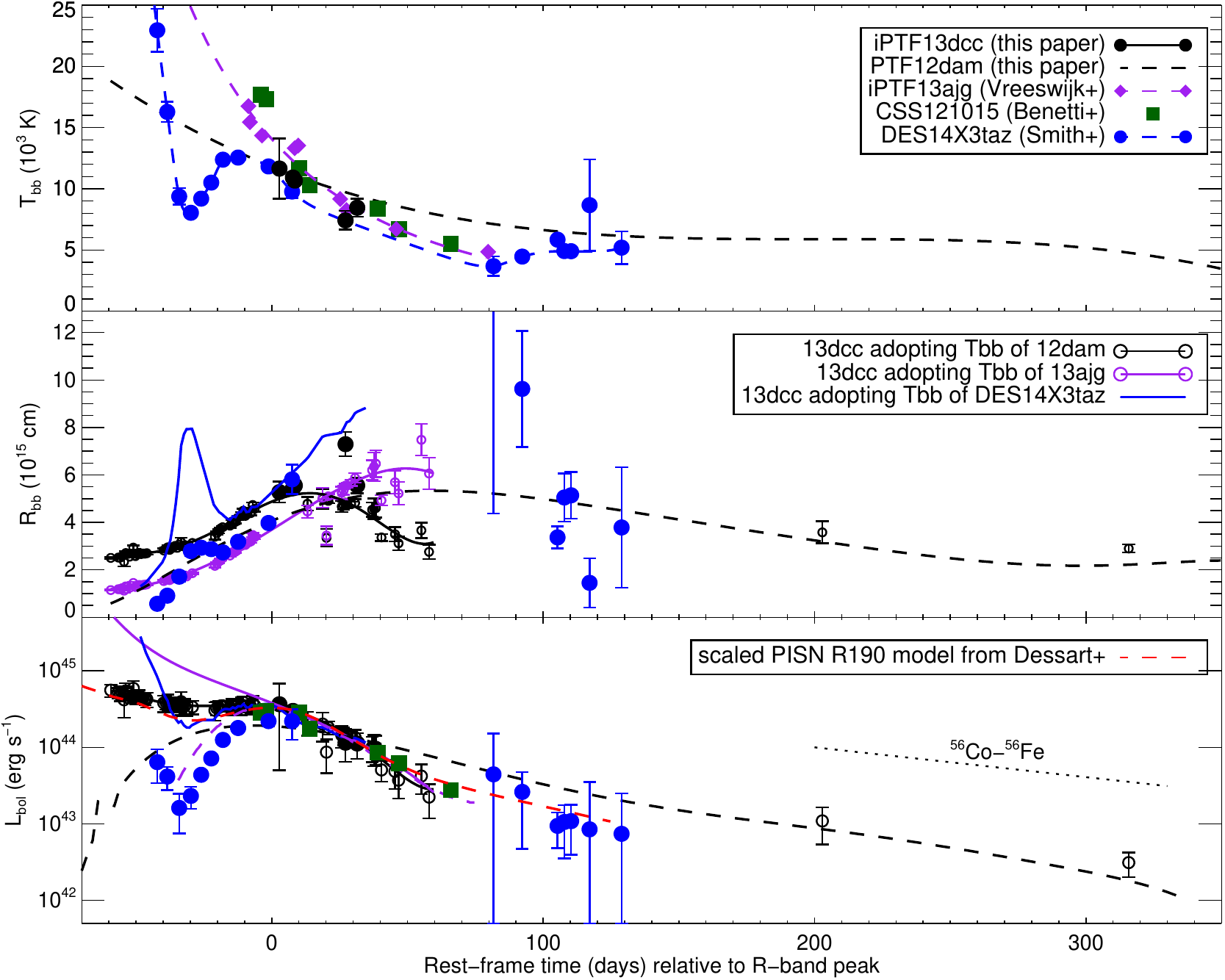}
  \caption{Time evolution of (from top to bottom) the blackbody
    temperature, radius, and bolometric luminosity of \dcc.  The
    filled black circles in all panels correspond to the best-fit
    values inferred from the Planck-function fits to the spectra, with
    fit parameters $T_{\rm bb}$ and $R_{\rm bb}$. {\bf Top panel:} the
    temperature evolution of \dcc\ is compared to that inferred for
    \dam\ (open black triangles and dashed line), \css\ \citep[green
      squares;][]{2013arXiv1310.1311B}, \ajg\ \citep[purple
      diamonds;][]{2014ApJ...797...24V}, and \des\ \citep[filled blue
      circles;][]{2015arXiv151206043S}. {\bf Middle panel:} the radius
    evolution of \dcc\ inferred from our $r$-band observations,
    adopting the temperature evolution of \dam\ (small open black
    circles and solid line), \ajg\ (small open purple circles and
    solid line), and \des\ (solid blue line). The filled blue circles
    show the radius evolution inferred for \des\ by
    \citet{2015arXiv151206043S}. The black dashed line shows the
    radius evolution that we inferred for \dam\ (see
    Fig.~\ref{fig:bol12dam}). {\bf Bottom panel:} the \dcc\ bolometric
    light curve inferred from the polynomial fits to the radius and
    assuming the temperature evolution of \dam\ (open black circles),
    \ajg\ (solid purple line), and \des\ (solid blue line). These are
    compared with the evolution inferred for \dam\ (dashed black
    line), \css\ (filled green squares), \ajg\ (dashed purple line),
    \des\ (filled blue circles), and the pair-instability supernova
    (PISN) R190 model from \citet{2012MNRAS.426L..76D}, scaled and
    shifted to match the peak of \dcc\ (red dashed
    line). \label{fig:bol13dcc}}
\end{figure*}

The apparent $r$-band brightness of the early excess emission is only
a few tenths of a magnitude fainter than the brightness at peak, and
the early decline suggests it probably was even brighter before. iPTF
or other surveys did not observe this sky location during several
months prior to our first detection, so we do not have any constraints
on the early brightness evolution and the time of explosion. After the
main peak, the light curve is declining rapidly until about
50~days after peak, followed by a very slow late-time decay with a
slope that is only slightly steeper than that expected from the decay
of radioactive $^{56}$Co, assuming full trapping of the gamma-ray
photons produced.

Since we do not have any information on the \dcc\ light-curve
evolution before our first detection, it is possible that the main
peak occurred earlier and that the peak we consider to be the main
peak is a late-time rebrightening.  Such a rebrightening has been
detected in other SLSNe, such as SN2015bn \citep{2016arXiv160304748N}.
However, in the case of \dcc, the current main peak reaches an
absolute $g$-band magnitude of $M_g=-22.0$. If this is a late-time
rebrightening similar to that of SN2015bn, the main peak would have
reached around $M_g \approx -23$, which would be stretching the
brightness budget of SLSNe. Also, below we show that the estimated
temperature of \dcc\ at the presumed peak is very similar to that of
various other SLSNe-I at peak. If indeed the actual main peak occurred
before our first detection, we would have expected the temperature at
the time of this supposed rebrightening to be well below 10,000~K. For
these reasons, we consider it unlikely that the actual main peak
occurred before our first detection.

The light curve of \dcc\ is rather limited in multifilter
observations, making it difficult to derive a bolometric light curve
from the photometry as we did for \dam. Instead, we turn to the
spectroscopic sequence that was secured for this source as part of the
iPTF follow-up campaign; see Table~\ref{tab:13dcc_logspec} and
Fig.~\ref{fig:13dcc_spectra}. The times at which the five
\dcc\ spectra were taken, ranging from around the main peak until
$+30$~days in the rest frame, are indicated by the vertical dotted
lines in Fig.~\ref{fig:13dcc_lc}. For each spectrum, we fit a Planck
function to selected wavelength regions of the continuum free from
obvious features (these regions are indicated with the short dashes at
the top of the figure). This results in a best-fit blackbody
temperature, radius, and corresponding bolometric luminosity at each
epoch, shown by the filled black circles in the top, middle, and
bottom panels of Fig.~\ref{fig:bol13dcc}, respectively. These values
correspond to the bold-faced entries in Table~\ref{tab:lbol13dcc}. The
\dcc\ spectra and the blackbody fits are shown in
Fig.~\ref{fig:13dcc_spectra}.

The temporal range spanned by the spectra is limited, from the time of
the main peak to $+30$ rest-frame days. Therefore, to estimate the
full bolometric light curve we have to make an assumption about the
\dcc\ temperature evolution outside of this range. In the top panel of
Fig.~\ref{fig:bol13dcc}, we also show the inferred temperature
evolution of \dam\ (derived in Sec.~\ref{sec:phot12dam}),
\css\ \citep{2013arXiv1310.1311B}, \ajg\ \citep{2014ApJ...797...24V},
and \des\ \citep{2015arXiv151206043S}. The temperature derived from
the \dcc\ spectra appears to match the evolution of \dam\ quite
well. However, it is also not too far from the temperature evolution
of the other SLSNe-I that are shown.

Assuming a particular temperature evolution and that \dcc\ is
radiating as a blackbody, our $r$-band photometry can be used to infer
the radius evolution. This is done assuming the temperature evolution
of different SLSNe: \dam, \ajg, and \des. The resulting radius
evolution for each of these three cases is shown by the solid lines
in the middle panel of Fig.~\ref{fig:bol13dcc}. The radius evolution
adopting the \des\ temperature appears contrived; if the temperature
evolved that dramatically for \dcc\ as well, we would have detected
it in our $r$-band photometry. In principle, it is possible that the
\dcc\ temperature evolution contained a similar dramatic drop and rise
as \des, but if it did that must have occurred before our first $r$-band
detection. The middle panel also shows the radii inferred directly
from the \dcc\ spectra (filled black circles), the radius evolution
inferred for \dam\ (dashed black line, see Sec.~\ref{sec:phot12dam}),
and that inferred for \des\ by \citet{2015arXiv151206043S}.

The bottom panel of Fig.~\ref{fig:bol13dcc} shows the bolometric
luminosity evolution of \dcc\ adopting the temperature evolution of
\dam, \ajg, and \des. As mentioned above, the latter appears contrived
and we consider it irrelevant for \dcc, whereas adopting the
temperature evolution of \ajg\ leads to unreasonably large values for
the early-time bolometric luminosity of \dcc. We therefore adopt the
temperature evolution of \dam\ to derive the bolometric light curve of
\dcc, listed in Table~\ref{tab:lbol13dcc} and shown by the open
circles in the bottom panel of Fig.~\ref{fig:bol13dcc}. We will adopt
this bolometric evolution of \dcc\ when performing the modelling in
Sec.~\ref{sec:modelling}. Integrating this bolometric light curve over
the phase interval from $-59$ to $+60$~days ($-59$ to $+315$~days)
results in an estimated total radiated energy of $E_{\rm
  rad}=2.8\times10^{51}$~erg ($3.0\times10^{51}$~erg) for \dcc, which
is about 60\% larger than the value we derived for \dam\ (see
Sec.~\ref{sec:phot12dam}). For comparison, we also show the
bolometric evolution of \dam, \css, \ajg, and \des.

Finally, we also plot the pair-instability supernova (PISN) model
denoted R190 from \citet{2012MNRAS.426L..76D}, scaled up in luminosity
($\times 15$) and shifted and contracted in time ($\times 0.45$) to
best match the bolometric behavior of \dcc. The time of the main peak
(i.e., at phase = 0 in the figure) of the unscaled R190 model is about
220~days after explosion, or 100 days in the contracted model shown.
\citet{2012MNRAS.426L..76D} focus on three types of PISN, with the
stars exploding being a red supergiant (RSG), a blue supergiant (BSG),
or a Wolf-Rayet (WR) star \citep[see also][]{2013MNRAS.428.3227D}. The
R190 model refers to a 190~\Msun\ main-sequence star dying as a
164~\Msun\ RSG with a surface radius of about 4000~\Rsun\ and an
extended hydrogen envelope. Despite the R190 model being scaled both
in luminosity and time, the similarity in the general brightness
evolution with \dcc\ is intriguing. However, this model with a
hydrogen envelope is not the most natural for explaining the
observations of a hydrogen-poor SLSN. In addition,
\citet{2016MNRAS.455.3207J} compare their PISN models with the nebular
spectra of SN~2007bi and \dam, finding discrepancies for several key
observables and thus no support for a PISN interpretation.

\section{Modelling}
\label{sec:modelling}

In this section, we explore different light-curve models that could
explain the bolometric light-curve evolution of \dam\ and
\dcc\ derived in the previous section. We start by applying the
semi-analytical light-curve models described by
\citet{2012ApJ...746..121C} and \citet{2013ApJ...773...76C}. These
include the following three independent power inputs: (1) radioactive
decay of $^{56}$Ni and $^{56}$Co, (2) magnetar spin-down, and (3) forward-
and reverse-shock heating due to SN ejecta interacting with CSM. 
The radioactive decay and interaction power inputs
have also been combined by \citet{2012ApJ...746..121C} into a hybrid
model in which both processes can be modelled simultaneously.  These
models are described in more detail below, but we also refer the
reader to the above-mentioned papers.

\begin{figure*}
  \centering
  \includegraphics[width=\hsize]{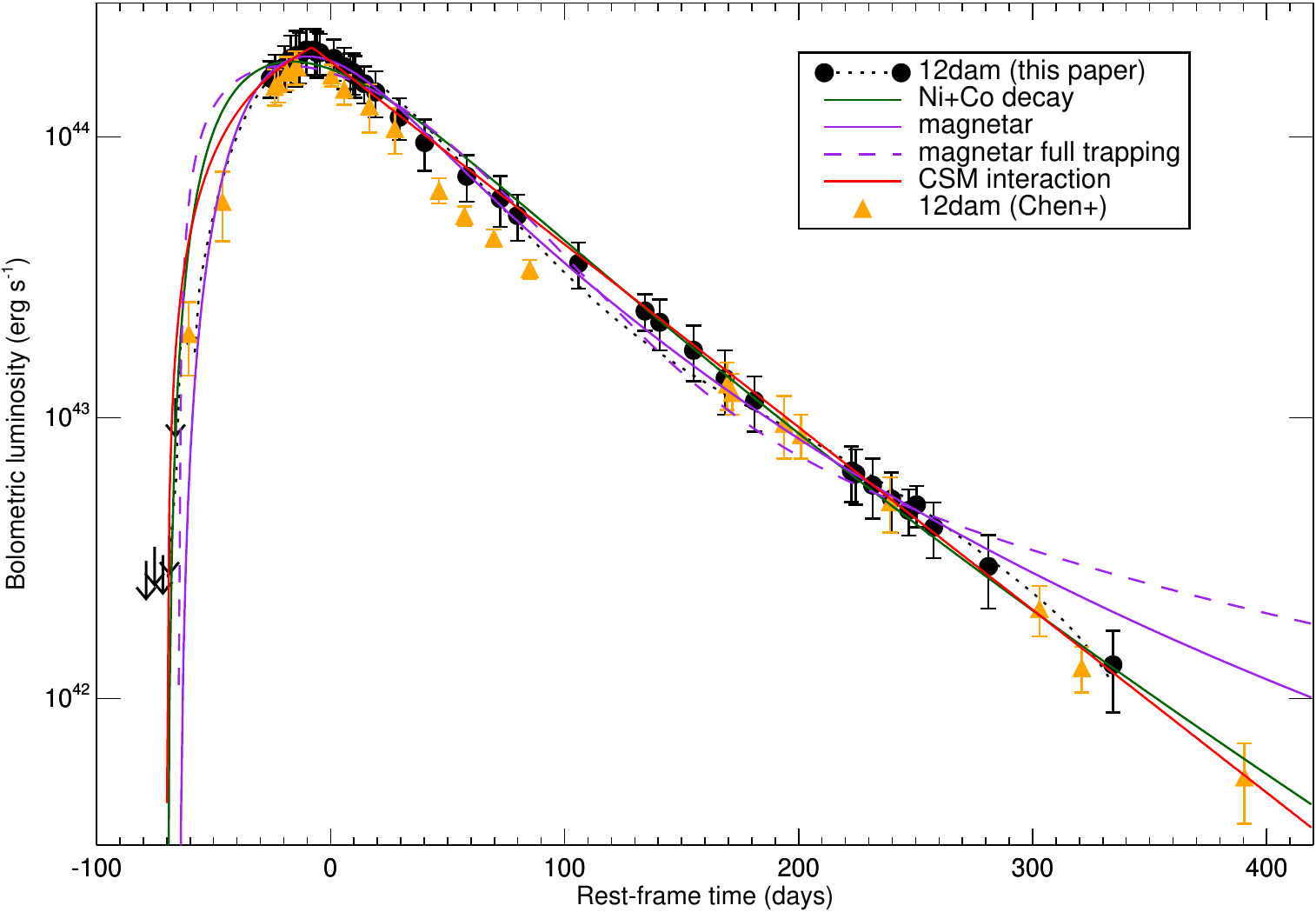}
  \caption{Model fits of radioactive decay (solid green line),
    magnetar (solid and dashed purple lines), and CSM interaction
    (solid red line) to the bolometric light curve of \dam\ that we
    derived in Sec.~\ref{sec:photometry} (filled black circles and
    dotted line).  See Tables~\ref{tab:nico}, \ref{tab:magnetar}, and
    \ref{tab:csm} for the corresponding parameter best-fit values.
    All three models provide a good description of the \dam\ data. The
    magnetar fits including increased leakage of energy as a function
    of time and assuming full trapping are shown with the solid and
    dashed purple lines, respectively. The filled orange triangles
    show the \dam\ bolometric light curve as inferred by
    \citet{2014arXiv1409.7728C} and updated by Chen et al. (2016,
    Erratum, in prep.). \label{fig:model12dam}}
\end{figure*}

In addition, we investigate if the early-time bump or plateau observed
in \dam\ and \dcc\ could be produced by cooling emission from material
heated by the shock produced in the explosion. For standard SNe, the
expected cooling emission has been modelled with the material being
the outer part of the stellar envelope \citep{2011ApJ...728...63R} and
also for the case in which an extended low-mass envelope is
surrounding a compact core \citep{2014ApJ...788..193N}.

\citet{2014ApJ...788..193N} show that a standard progenitor cannot
explain the fast drop in bolometric luminosity after the first peak,
which is supported by comparison of the Rabinak \& Waxman (2011) model
to observations of SLSNe with double-peaked light curves:
\bdq\ \citep{2015ApJ...807L..18N} and
\des\ \citep{2015arXiv151206043S}. Although
\citet{2016arXiv160703700S} extend the Rabinak \& Waxman analytic
expressions to later times and do infer a drop in luminosity for
normal SNe-II, their model is strictly valid only at times when the
photospheric (i.e. effective) temperature is above the recombination
temperature for the opacity in the relevant outer shells of the
envelope to be time and space independent. For hydrogen-poor envelopes
this latter temperature is of the order of the recombination
temperature of doubly-ionized helium: 2$-$2.5~eV or 23$-$29~kK, and so
this model is not applicable to \dam\ or \dcc\ where the early excess
emission peak occurs at around 10~days when the effective temperature
has already dropped well below this temperature limit.
For both \bdq\ and \des, shock cooling of an
extended envelope appears to describe the initial peak rather well
\citep{2015ApJ...808L..51P,2015arXiv151206043S}.  We therefore focus
on the extended envelope case, following the prescription of
\citet{2015ApJ...808L..51P}. We note that this model also successfully
reproduces the early bump of iPTF~15dtg, a spectroscopically normal SN
Ic from a massive progenitor \citep{2016A&A...592A..89T}.

Finally, we combine the Piro model with the Chatzopoulos magnetar
and CSM interaction models to fit the entire bolometric light curve of
\dcc. In the CSM case, this has the advantage that two fit parameters
are the same in these models: the total SN energy, and the sum of
Piro's core and extended mass is the same as the ejecta mass in the
CSM interaction model.  For \dam\ the early excess emission has too
few data points to attempt this combined modelling. As a comparison
for the best-fit values inferred for \dam\ and \dcc, we also fit all
the above-mentioned models, including the combined ones, to the
bolometric light curve of \des\ presented by
\citet{2015arXiv151206043S}.

\subsection{\dam}

Figure~\ref{fig:model12dam} features the bolometric light curve of
\dam\ as derived in Sec.~\ref{sec:photometry}. Around peak, the
bolometric luminosity that we find is up to 50\% higher than the
\dam\ bolometric luminosity evolution derived by
\citet{2013Natur.502..346N} and \citet{2014arXiv1409.7728C}.  In
private communication with the first authors of these papers to try to
understand this discrepancy, it was found that they used the wrong
photometric zeropoints to convert the {\it Swift} UV magnitudes to
fluxes (the zeropoint for Vega was used while the magnitudes published
by Nicholl et al. and Chen et al. are in the AB system), leading to an
underestimate of the true bolometric luminosity (see the Errata of
Chen et al. 2016, in prep. and
\citealt{2013Natur.502..346N}\footnote{published online at
  http://www.nature.com/nature/journal/\\v539/n7630/full/nature19850.html}). Correcting
the Chen et al. bolometric light curve for this calibration error
brings it up to a similar luminosity level as our bolometric light
curve. In Fig.~\ref{fig:model12dam} we show the {\it corrected}
bolometric light curve from the Erratum of Chen et al. 2016 (in
prep.), which is still slightly below our light curve around peak and
up to a phase of 100~days.

We fit the Chatzopoulos models to our light-curve data (the filled
circles shown in Fig.~\ref{fig:model12dam}) using a custom IDL
$\chi^2$ minimization program, in which we make grateful use of the
MPFIT procedures\footnote{see http://purl.com/net/mpfit} written by
Craig Markwardt \citep{2009ASPC..411..251M,mpfit_minpack}. Where
relevant, we adopt a Thomson electron scattering opacity of
$\kappa=0.2$~cm$^2$\,g$^{-1}$, appropriate for material without
hydrogen, and also fit for the time of explosion which is assumed to
be within the phase range of $-70$ to $-65$~days as constrained by the
observations. We loosely limit the progenitor radius, $R_0$, to be
within the range 0.1--500~\Rsun, and use a fiducial value for the
expansion velocity of $v_{\rm exp}=10,000$~\kms\ for the radioactive
decay and magnetar models (for the CSM interaction case this velocity
is not assumed but is calculated in the model).

\begin{deluxetable}{lccc}
  \tablecaption{Ni+Co decay best-fit parameters. \label{tab:nico}}
  \tablehead{
    \colhead{} &
    \colhead{\dam} &
    \colhead{\dcc} &
    \colhead{\des\tablenotemark{a}}
  }
  \startdata
  $t_{\rm expl, rest}$ [d] & $-70$\b          & $-89$           & $-43$\b \\
  $M_{\rm Ni}$ [\Msun]    & $21.9\pm0.5$     & $115\pm209$     & $26\pm39$\\             
  $t_{\rm d}$ [d]         & $46\pm2$         & $64\pm25$       & $61\pm48$\\
  $A_{\rm leakage}$ [$10^3~{\rm d}^2$] & $28\pm1$ & $3\pm8$         & $4\pm7$\\
  $R_0$ [\Rsun]          & 0.1$^{\rm b}$     & 0.1$^{\rm b}$    & 0.1$^{\rm b}$ \\
  $\chi^2$/DOF           & 5.9/32           & 119/51          & 73/10
  \enddata
  \tablenotetext{a}{For \des\ the first two detections have been
    discarded in the fit, focussing on fitting the main peak.}
  \tablenotetext{b}{Parameter reached limit of allowed range.}
\end{deluxetable}

The power input in the first model that we fit is radioactive decay of
$^{56}$Ni and $^{56}$Co, which produces energetic gamma rays that are
thermalized in the expanding ejecta. The fit parameters are the time
of explosion ($t_{\rm expl, rest}$), the nickel mass produced ($M_{\rm
  Ni}$), the mean of the hydrodynamical and diffusion timescales or
effective light-curve timescale ($t_{\rm d}$), a factor determining
the fraction of gamma rays that escape as a function of time ($A_{\rm
  leakage}$, modifying the luminosity with the factor $1-e^{-A
  t^{-2}}$; i.e., a large $A$ corresponds to full trapping), and the
progenitor radius ($R_0$). The goodness-of-fit $\chi^2$ value divided
by the number of degrees of freedom (DOF) and best-fit values for the
Ni+Co decay-model parameters are listed in Table~\ref{tab:nico}, and
the corresponding best fit to the \dam\ luminosity evolution is shown
by the solid green line in Fig.~\ref{fig:model12dam}.

This model describes the data very well. The very low reduced $\chi^2$
indicates that the uncertainties in the luminosity are likely
overestimated. However, the best-fit nickel mass, $M_{\rm
  Ni}=22$~\Msun, is much greater than the estimated ejecta mass,
$M_{\rm ej}\approx 8$~\Msun, which is calculated using Eq. 1 of
\citet{2015MNRAS.450.1295W}, repeated here:
\begin{equation}
M_{\rm ej}\approx \frac{1}{2} \frac{\beta c}{\kappa} v_{\rm exp} t_{d}^{2},
\end{equation}
where we have substituted the rise time to maximum light ($t_r$) with
$t_d$, $\beta$ is a constant of integration (13.8), and $c$ is the
speed of light. We note that this estimate of the ejecta mass is
uncertain due to various assumptions made in its derivation, amongst
others regarding the effective opacity and homologous
expansion. Nonetheless, this model can probably be rejected on the
basis that the required nickel mass to power the peak luminosity of
\dam\ is unrealistically large \citep[see
  also][]{2013Natur.502..346N,2014arXiv1409.7728C}.  We note that
\citet{2013Natur.502..346N} inferred a lower nickel mass of
14--16~\Msun\ (see their extended data Fig.~6) than we do, because
they underestimated the true bolometric luminosity around peak of
\dam, as we discussed above.

To check if the light curve tail can be powered by radioactive decay,
we fit our late-time data beyond 100 rest-frame days. The nickel mass
required (12~\Msun) is still large for the estimated ejecta mass
(22~\Msun, which is larger than before because the best-fit effective
diffusion time scale increased for this late-time fit). For the light
curve beyond 200~days, the best nickel mass (10~\Msun) and estimated
ejecta mass (140~\Msun) are compatible. Therefore, the radioactive
decay of $^{56}$Ni and $^{56}$Co has difficulty accounting for the
full bolometric light curve of \dam, but it could be that it is
powering the late-time light curve.

In the magnetar model, the energy input is produced by the spin-down
of a rapidly spinning, highly magnetic neutron star, the remnant of
the explosion. Although it is not yet clear if and how a large
fraction of the rotational and/or magnetic energy can be
converted to radiation, in this model it is assumed that the energy
from the magnetar is thermalized in the expanding ejecta. The
parameters of the magnetar model are the time of explosion ($t_{\rm
  expl, rest}$), the initial spin period of the neutron star in
milliseconds ($P_{\rm sp,init}$), its magnetic field strength in units
of $10^{14}$~Gauss ($B_{14}$), the effective light-curve timescale
($t_{\rm d}$), a constant $A_{\rm leakage}$, and the progenitor radius
($R_0$). The parameters' best-fit values are listed in
Table~\ref{tab:magnetar}, and the corresponding best-fit model to the
bolometric light curve of \dam\ is shown by the solid purple line in
Fig.~\ref{fig:model12dam}.

\begin{deluxetable}{lccc}
  \tablecaption{Magnetar best-fit parameters. \label{tab:magnetar}}
  \tablehead{
    \colhead{} &
    \colhead{\dam} &
    \colhead{\dcc} &
    \colhead{\des\tablenotemark{a}}
  }
  \startdata
  $t_{\rm expl, rest}$ [d]         & $-65$\b        & $-70$\b        & $-43$\b \\
  $P_{\rm sp,init}$ [ms]           & $2.30\pm0.03$  & $1.69\pm0.38$  & $1.8\pm2.8$ \\
  $B_{14}$ [$10^{14}$~G]    & $0.7\pm0.7$    & $0.33\pm0.18$  & $0.39\pm0.68$\\ 
  $t_{\rm d}$ [d]                 & $50\pm27$      & 10$^{\rm b}$    & $49\pm202$ \\
  $A_{\rm leakage}$ [$10^3~{\rm d}^2$] & $32\pm51$      & $7\pm4$        & $3\pm11$\\
  $R_0$ [\Rsun]                  & 0.1\b          & 0.1\b          & 0.1\b \\
  $\chi^2$/DOF                   & 5.9/31         & 127/50         & 33.7/9
  \enddata
  \tablenotetext{a}{For \des\ the first two detections have been
    discarded in the fit, focussing on fitting the main peak.}
  \tablenotetext{b}{Parameter reached limit of allowed range.}
\end{deluxetable}

Although it is not present in Eq.~2 of \citet{2013ApJ...773...76C}, we
include a leakage factor in exactly the same way as is done for the
radioactive decay model \citep[see also][]{2015ApJ...799..107W}. We
stress that when assuming full trapping (i.e., a large value for
$A_{\rm leakage}$), the magnetar model predicts a late-time luminosity
brighter than the observations, as already found by
\citet{2014arXiv1409.7728C}. We show this full-trapping model fit with
the dashed purple line in Fig.~\ref{fig:model12dam}, which has
best-fit values $P_{\rm sp,init}=1$~ms (reaching the lower limit of
the allowed range) and $B_{14}=1.63\pm0.03$~G
($\chi^2$/DOF=$\chi^2_{\rm \nu}=0.7$). The magnetar model including
the leakage factor fits our data well ($\chi^2_{\rm \nu}=0.2$).  In
the radioactive decay model, the leakage is understood: the opacity of
the gamma rays produced in the radioactive decay of nickel and cobalt
is decreasing with time, allowing an increasing fraction of the energy
to escape.  In the magnetar model, however, it is not clear how the
rotational and/or magnetic energy is transferred to the expanding
ejecta. It might also be in the form of high-energy photons, in which
case the leakage would occur in a similar way as in the radioactive
decay model.  We note that the initial spin period that we find
($P_{\rm sp,init}=2.30\pm0.03$~ms) is lower than that inferred by
\citet[][$P_{\rm sp,init}=2.72$~ms]{2014arXiv1409.7728C}, which moves
the location of \dam\ in the spin period vs. host-metallicity plot
shown in Fig.~7 of \citet{2016arXiv160504925C} slightly away from the
suggested relation.

\begin{deluxetable}{lccc}
  \tablecaption{CSM interaction best-fit parameters.$^{\rm a}$ \label{tab:csm}}
  \tablehead{
    \colhead{} &
    \colhead{\dam} &
    \colhead{\dcc} &
    \colhead{\des$^{\rm b}$}
  }
  \startdata
  $t_{\rm expl, rest}$ [d]              & $-70$\c       & $-79$      & $-43$\c \\
  $\delta$                            & 0             & 2          & 2 \\
  $n$                                 & 8             & 6          & 14 \\
  $s$                                 & 0             & 0          & 0  \\
  $R_{\rm prog}$ [\Rsun]               & 0.1\c         & 0.1\c      & 0.1\c \\
  $\rho_{\rm CSM}$ [$10^{-13}$~g~cm$^{-3}$] &  5  & 0.3        & 4 \\
  $M_{\rm CSM}$ [\Msun]                & 15            & 14         & 6 \\
  $M_{\rm ejecta}$ [\Msun]              & 13            & 3          & 7 \\
  $E_{\rm SN}$ [$10^{51}$~erg]          & 3             & 3          & 1\\
  $\chi^2$/DOF                        & 1.8/31        & 41.9/49    & 16.4/8\\
  & & & \\
  $R_{\rm CSM}^{\rm d}$ [$10^{15}$~cm]  &  2  &   6  & 2 \\
  $v_{\rm exp}^{\rm d}$ [$10^{3}$~\kms] & 15  &  25  & 15
  \enddata
  \tablenotetext{a}{Owing to the complexity of the model, these best-fit
    values are very uncertain and should be considered as rough
    approximations.}
  \tablenotetext{b}{For \des\ the first two detections have been
    discarded in the fit, focussing on fitting the main peak.}
  \tablenotetext{c}{Parameter reached limit of allowed range.}
  \tablenotetext{d}{The lower two parameters are not fit, but are
    calculated within the model.}
\end{deluxetable}

The energy input in the CSM interaction model
originates from the SN ejecta interacting with dense CSM 
(e.g., a dense progenitor wind or circumstellar shell previously
cast off by the progenitor), resulting in a forward/circumstellar and a
reverse/ejecta shock. \citet{2013ApJ...773...76C} use
\citet{1982ApJ...258..790C} and \citet{1994ApJ...420..268C} to derive
an expression for the time-dependent luminosity produced by the
forward and reverse shocks, which are depositing kinetic energy into
the CSM and SN ejecta, respectively. This is combined with the
prescription of radiative diffusion developed by
\citet{1980ApJ...237..541A,1982ApJ...253..785A} to derive an analytic
expression for the output bolometric light curve \citep[see Eq.~20
  of][]{2013ApJ...773...76C}. The model has many free parameters: the
time of the explosion ($t_{\rm expl, rest}$), the power-law index for
the inner and outer ejecta density profile ($\delta$ and $n$,
respectively), the power-law index for the CSM density profile ($s$),
the progenitor radius ($R_{\rm prog}$), the density immediately
outside the progenitor envelope ($\rho_{\rm CSM}$), the CSM mass
($M_{\rm CSM}$), the ejecta mass ($M_{\rm ejecta}$), and the total SN
energy ($E_{\rm SN}$). In our fitting routine, we loop over the
following typical values for the parameters: $\delta=[0,2]$, $n=6$--14,
and $s=[0,2]$ ($s=0$ signifies a constant CSM density profile, such as
that of a shell of material, while $s=2$ corresponds to that produced by
a progenitor wind), and find the best-fit values for the remaining
parameters. The best-fit values are listed in Table~\ref{tab:csm}, and
the corresponding best-fit model for \dam\ is shown by the solid red
line in Fig.~\ref{fig:model12dam}.

The CSM interaction model provides an excellent fit to the data
($\chi^2_{\rm \nu}=0.1$), but considering the number of free
parameters, this may not be so surprising \citep[see
  also][]{2013ApJ...773...76C}. This CSM interaction model can be
combined with the model for radioactive decay of nickel and cobalt,
adding another free parameter, the nickel mass ($M_{\rm Ni}$), but we
fixed this to zero in the CSM fit.

We also fit the analytic shock-cooling model of an extended envelope
surrounding the SN developed by \citet{2014ApJ...788..193N} and
\citet{2015ApJ...808L..51P} to the observed early-time $r$-band light
curve of \dam. The parameters are the time of the explosion ($t_{\rm
  expl}$), the core mass ($M_{\rm core}$), the energy in the SN shock
(which is transferred to the extended material, $E_{\rm SN}$), and the
mass, radius, and opacity of the extended material ($M_{\rm ext}$,
$R_{\rm ext}$, and $\kappa$, respectively). The term $M_{\rm core}$ is
somewhat misleading; it refers to the part of the original stellar
core that is ejected, i.e. the mass of the core minus the mass of the
stellar remnant \citep[see][]{2014ApJ...788..193N}. We fix
$\kappa=0.2$~cm$^2$\,g$^{-1}$ as for all the other models. We limit
the SN energy and extended radius to be at most $E_{\rm
  SN}\leq10^{53}$~erg and $R_{\rm ext}\leq10^{14}$~cm, respectively.
The extended mass is constrained to be at most half the core mass, and
the latter is required to be at least 1~\Msun.  Given these
parameters, the time evolution of the bolometric luminosity and radius
of the extended material is provided by the model.

\begin{deluxetable}{lccc}
  \tablecaption{\citet{2015ApJ...808L..51P} best-fit parameters.\a\ \label{tab:piro}}
  \tablehead{
    \colhead{} &
    \colhead{\dam} &
    \colhead{\dcc} &
    \colhead{\des}
  }
  \startdata
  $t_{\rm expl, rest}$ [d]      & $-75$\b     & $-70$\b       & $-45$ \\
  $M_{\rm core}$ [\Msun]       & 1\b         & 35            & 4 \\
  $E_{\rm SN}$ [$10^{51}$~erg] & 3\c         & 100\b         & 1\c \\
  $M_{\rm ext}$ [\Msun]        & 0.5\b       & 17.5\b        & 0.4 \\
  $R_{\rm ext}$ [$10^{13}$~cm]  & 0.6         & 10\b          & 10\c \\
  $\chi^2$/DOF                & 3.0/2       & 52.7/17       & 0.004/1
  \enddata
  \tablenotetext{a}{Owing to the degeneracy of the parameters in this
    model, these best-fit values are very uncertain and should be
    considered as rough approximations.}
  \tablenotetext{b}{Parameter reached limit of allowed range.}
  \tablenotetext{c}{Parameter was held fixed; in the case of \dam\ and
    \des\ the explosion energy was fixed to the same value as the
    best-fit value of the CSM interaction model (see
    Table~\ref{tab:csm}).}
\end{deluxetable}

\begin{figure*}
  \includegraphics[width=0.5\hsize]{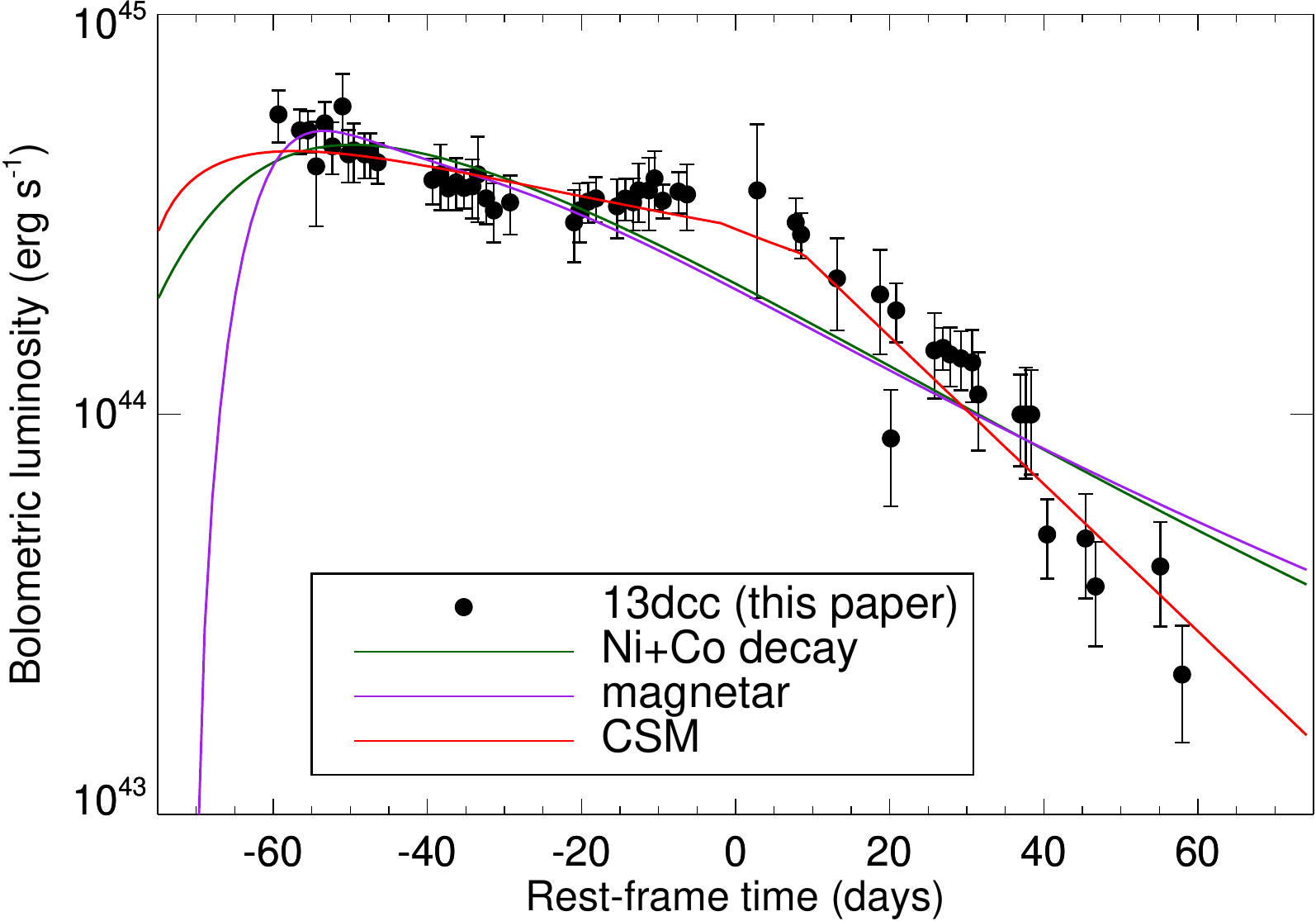}
  \includegraphics[width=0.5\hsize]{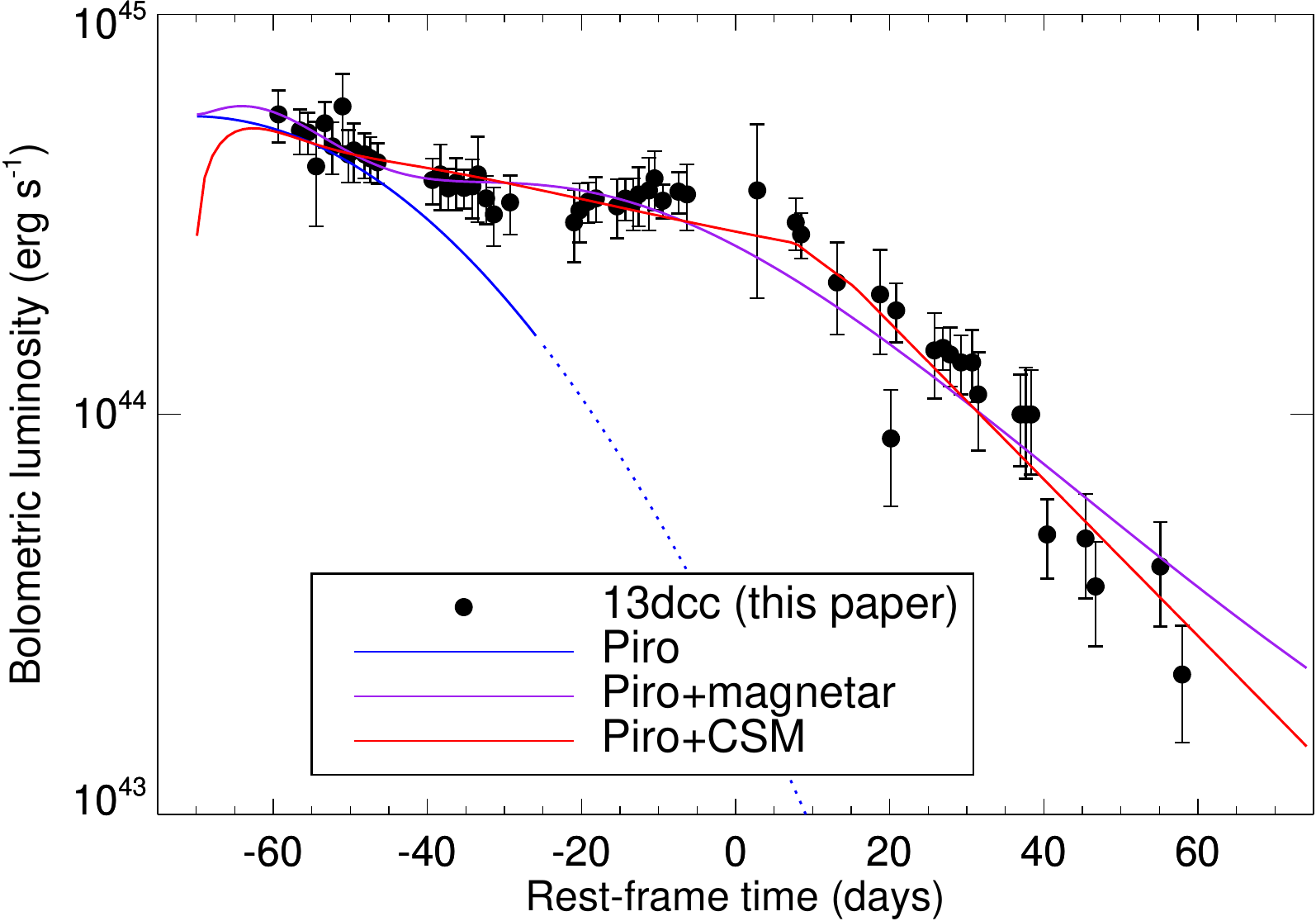}
  \caption{Left panel: model fits of radioactive decay, magnetar, and
    CSM interaction to the bolometric light curve of \dcc\ (up to a
    phase of $+80$~days). While the radioactive decay and magnetar
    models fail to fit both the early excess emission and main peak,
    the CSM interaction model describes the observations rather
    well. Right panel: a Piro model fit to the early-time light curve
    (phase $<-25$~days), and a combination of the Piro plus magnetar
    and Piro plus CSM interaction models to the \dcc\ light curve.
    The two breaks in the CSM interaction model light curve, seen in
    both panels, correspond to the times of termination of the forward
    and reverse shocks. See Tables~\ref{tab:nico}-\ref{tab:pirocsm}
    for the corresponding parameter best-fit values for all these
    model fits.
    \label{fig:model13dcc}}
\end{figure*}

To fit the model to the observed magnitudes, we determine the
effective temperature evolution based on the model parameters and
transform the resulting rest-frame spectral blackbody flux to the
observer's frame. We then extract observed model magnitudes by
performing synthetic photometry on the spectrum using the P48 $r$-band
transmission curve. To make use of the $r$-band upper limits nearest
in time to the first detections, we converted the limits at an
observed time of about $-73$ and $-76$~days to expectation values of
half the limit, with an uncertainty (1$\sigma$) the size of one sixth
of the limit, so these added data points are consistent with the
3$\sigma$ upper limit. We did the same for the measurement at an
observed time around $-63$~days, as by this time the luminosity is
already dominated by whatever is powering the main peak. Although this
is not clear in Fig.~\ref{fig:12dam_lc}, the $r$-band measurement at
an observed time of $-70$~days is in fact composed of three different
data points; hence, in the fit we use a total of seven
measurements. Even when making use of the limits in the fit, the
number of data points is limited. We therefore decided to fix the
explosion energy to the best-fit value found for the CSM interaction
fit. The resulting best-fit values are listed in Table~\ref{tab:piro}.
This fit is shown by the red solid line in the inset of
Fig.~\ref{fig:12dam_lc}. As already noted, the number of degrees of
freedom is very low (two), and so the parameters are not well
constrained. However, this is just to show that the Piro model is
capable of explaining the early-time evolution of the
\dam\ light-curve with reasonable values for the parameters.  For
\bdq, \citet{2015ApJ...808L..51P} find acceptable fits with the
following ranges of values for the fit parameters: $M_{\rm
  ext}=0.3$--0.5~\Msun, $R_{\rm
  ext}=500$--5000~\Rsun\ ($=3.5\times10^{13-14}$~cm), and $E_{\rm
  SN}=$ (9--55) $\times10^{51}$~erg.

\begin{deluxetable}{lcc}
  \tablecaption{Piro+magnetar best-fit parameters.\a \label{tab:piromag}}
  \tablehead{
    \colhead{} &
    \colhead{\dcc} &
    \colhead{\des}
  }
  \startdata
  $t_{\rm expl, rest}$ [d]           & $-70$\b       & $-43$\b \\
  $M_{\rm core}$ [\Msun]            & 12            & 1.2 \\
  $M_{\rm ext}$ [\Msun]             & 6             & 0.6\b \\
  $R_{\rm ext}$ [$10^{13}$~cm] & 4             & 0.2 \\
  $E_{\rm SN}$ [$10^{51}$~erg] & 100\b         & 20 \\
  $P_{\rm sp,init}$ [ms]             & $1.2\pm1.7$   & $1.8\pm2.8$ \\
  $B_{14}$ [$10^{14}$~G]      & $1.1\pm0.8$   & $0.4\pm0.7$\\ 
  $t_{\rm d}$ [d]                   & $66\pm72$     & $49\pm209$ \\
  $A_{\rm leakage}$ [$10^{3}~{\rm d}^2$] & $11\pm37$     & $3\pm11$\\
  $R_0$ [\Rsun]                    & 0.1\b         & 0.1\b \\
  $\chi^2$/DOF                     & 48.9/46       & 34.2/7
  \enddata
  \tablenotetext{a}{Owing to the degeneracy of the Piro parameters in
    this model, these Piro-model best-fit values are very uncertain
    and should be considered as rough approximations.}
  \tablenotetext{b}{Parameter reached limit of allowed range.}
\end{deluxetable}

\begin{deluxetable}{lcc}
  \tablecaption{Piro+CSM interaction best-fit parameters.\a \label{tab:pirocsm}}
  \tablehead{
    \colhead{} &
    \colhead{\dcc\tablenotemark{a}} &
    \colhead{\des}
  }
  \startdata
  $t_{\rm expl, rest}$ [d]              & $-70$\b        & $-43$\b \\
  $M_{\rm core}$ [\Msun]               & 2.4            & 7.9 \\
  $M_{\rm ext}$ [\Msun]                & 1.2\b          & 0.1  \\
  $R_{\rm ext}$ [$10^{13}$~cm]    & 10\b           & 7  \\
  $\delta$                            & 2              & 0     \\
  $n$                                 & 6              & 14    \\
  $s$                                 & 0              & 0     \\
  $R_{\rm prog}$ [\Rsun]               & 0.4            & 0.1\b \\
  $\rho_{\rm CSM}$\c [$10^{-13}$~g~cm$^{-3}$] & 0.3 & 5    \\
  $M_{\rm CSM}$ [\Msun]                 & 12             & 6    \\
  $M_{\rm ejecta}$ [\Msun]              & 3.6            & 8.0  \\
  $E_{\rm SN}$ [$10^{51}$~erg]          & 3              & 2     \\
  $\chi^2$/DOF                        & 36.6/47         & 16.6/8\\
  & & \\
  $R_{\rm CSM}$\c [$10^{15}$~cm]  & 6              & 2 \\
  $v_{\rm exp}$\c [$10^{3}$~\kms] & 23             & 18
  \enddata
  \tablenotetext{a}{Owing to the complexity of the model, these best-fit
    values are very uncertain and should be considered as rough
    approximations.}
  \tablenotetext{b}{Parameter reached limit of allowed range.}
  \tablenotetext{c}{The lower two parameters are not fit, but are
    calculated within the CSM interaction model.}
\end{deluxetable}

\subsection{\dcc}

Figure~\ref{fig:model13dcc} shows the bolometric light curve of
\dcc\ as derived in Sec.~\ref{sec:photometry}. As we did for \dam, we
fit the Chatzopoulos and Piro models to our light-curve data. Again,
we adopt a Thomson electron scattering opacity of
$\kappa=0.2$~cm$^2$\,g$^{-1}$ and for \dcc\ we assume that the time of
explosion is in the phase range of $-100$ to $-70$~days. We loosely
limit the radius of the progenitor star, $R_0$, to be within the range
0.1--500~\Rsun, and use a fiducial value for the expansion velocity of
10,000~\kms\ for the radioactive decay and magnetar models.  The
resulting best-fit values for the model parameters of the radioactive
decay, magnetar, and CSM interaction models are listed in
Tables~\ref{tab:nico}, \ref{tab:magnetar}, and \ref{tab:csm},
respectively; the corresponding model fits are shown in the left panel
of Fig.~\ref{fig:model13dcc}. The amount of nickel required to power
\dcc\ is very large (albeit with an even larger formal uncertainty),
much larger than the calculated mass in the ejecta ($M_{\rm ej}\approx
15$~\Msun), rendering this model unphysical. Also, the magnetar model
is unable to describe both the early emission and ``main'' peak.

\begin{figure*}
  \includegraphics[width=0.5\hsize]{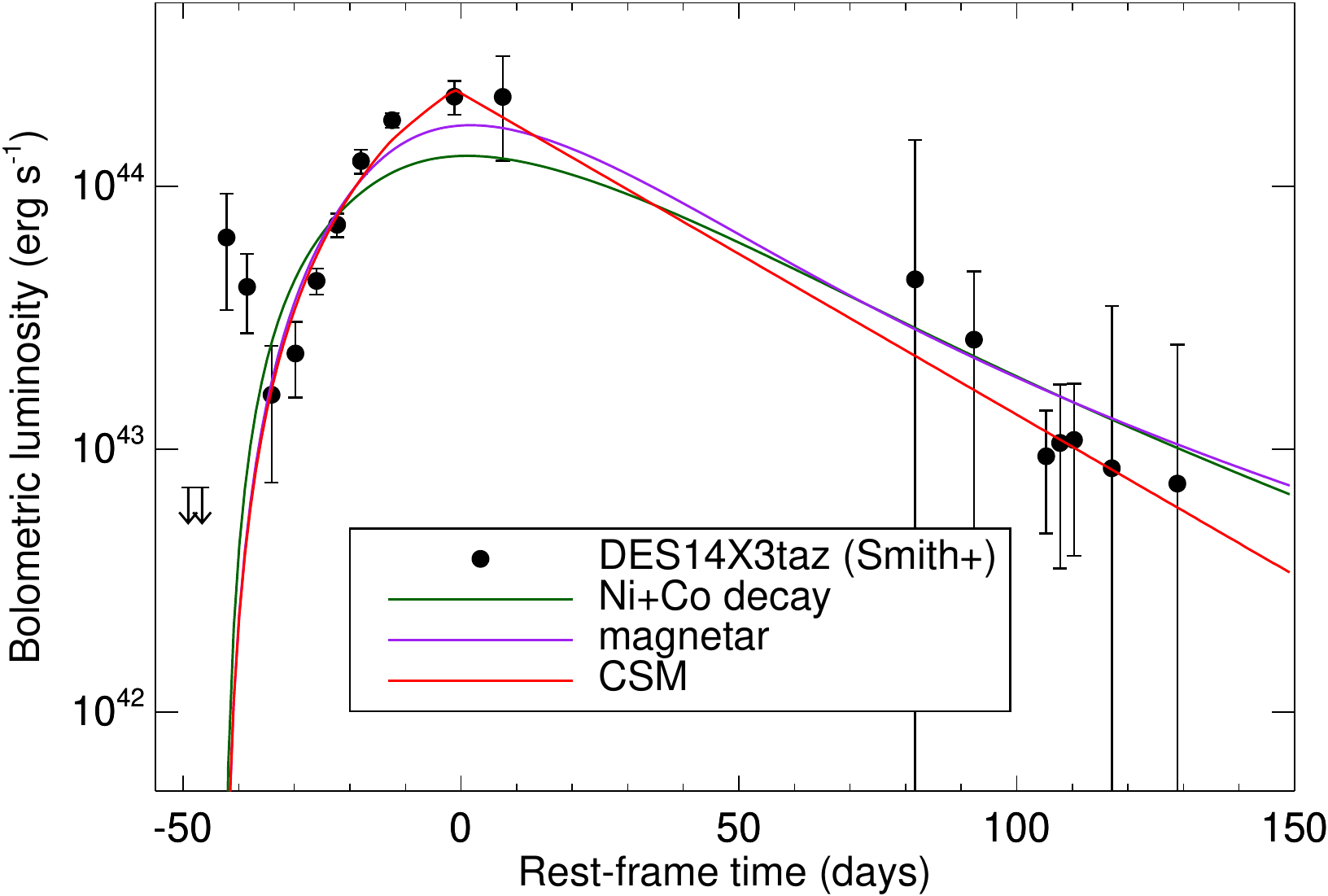}
  \includegraphics[width=0.5\hsize]{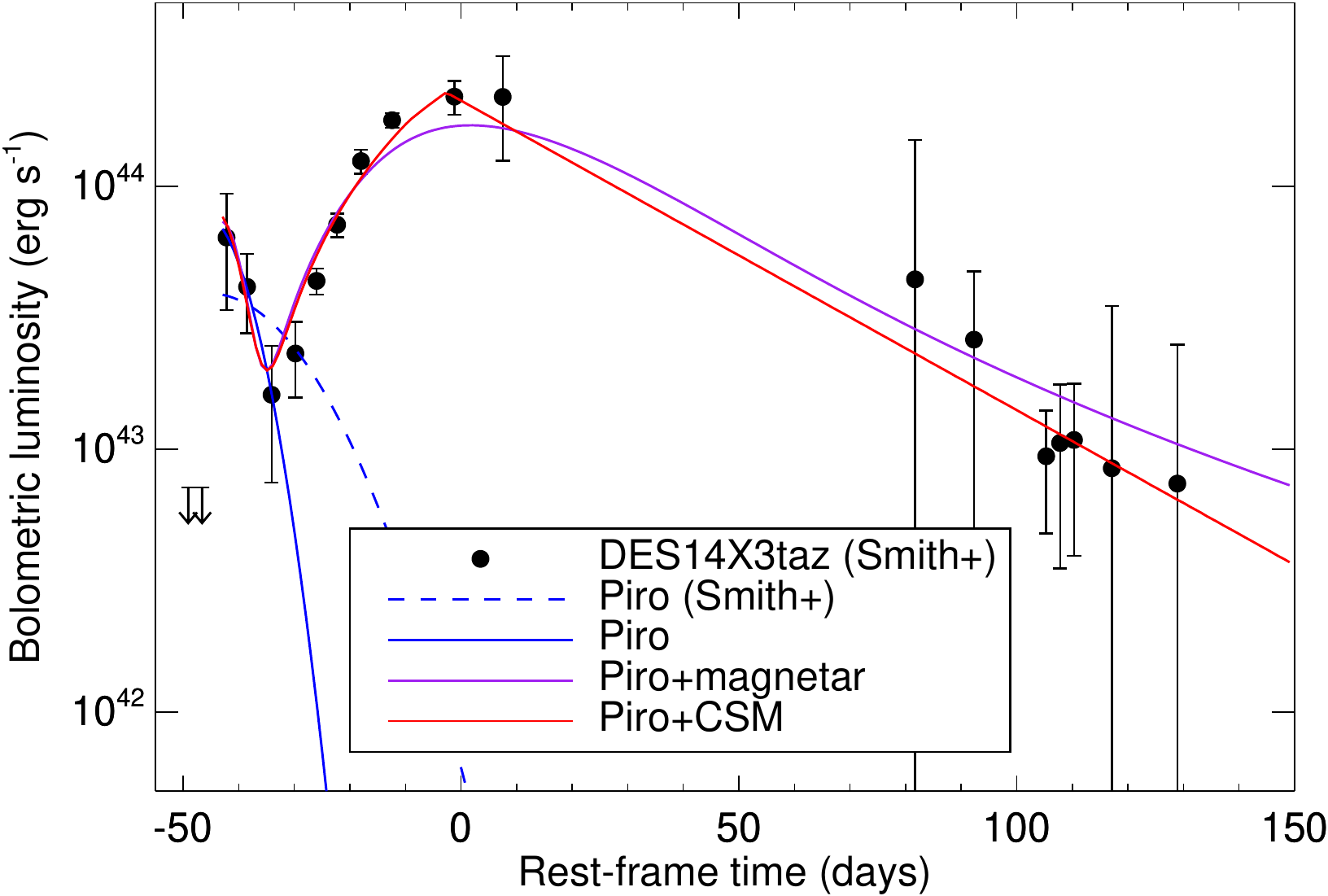}
  \caption{Model fits of radioactive decay, magnetar, CSM interaction,
    and the Piro model to the bolometric light curve of
    \des\ \citep{2015arXiv151206043S}.  In the left panel, we only fit
    the main peak, and the first two detections have been excluded in
    the fits. The radioactive decay model provides a poor fit, while
    the magnetar model fits the data reasonably well. The CSM
    interaction model provides the best match with the
    observations. On the right, we fit the \citet{2015ApJ...808L..51P}
    model to the first three detections, indicated with the solid blue
    line. For reference, we also show the model parameters found by
    \citet{2015arXiv151206043S} from fitting the $griz$ magnitudes
    rather than the bolometric luminosity as we do in this paper. We
    also perform combined Piro+magnetar and Piro+CSM interaction fits,
    shown by the solid lines in the right panel.  See
    Tables~\ref{tab:nico}--\ref{tab:pirocsm} for the corresponding
    parameter best-fit values for all these model
    fits. \label{fig:model14des}}
\end{figure*}

We also fit the analytic shock-cooling model of
\citet{2015ApJ...808L..51P} to the bolometric light curve of \dcc.  As
for \dam, we limit the SN energy and extended radius to be at most
$E_{\rm SN}\leq10^{53}$~erg and $R_{\rm ext}\leq10^{14}$~cm,
respectively, and the extended mass is constrained to be at most half
the core mass. The best-fit values are given in Table~\ref{tab:piro}
and the best fit is shown in the right panel of
Fig.~\ref{fig:model13dcc}.  The line turns from solid to dotted at a
phase of $-25$~days, beyond which the data points were excluded in
this fit. The fit is good at early times, but undershoots the data
points between a phase of $-45$ and $-25$~days. A larger explosion
energy of $1.5\times10^{53}$~erg would provide a very good fit, but
this is unreasonably high. The inferred extended mass, which is the
main parameter determining the light-curve width, is much larger in
the case of \dcc\ ($\sim18$~\Msun) than that needed to explain the
excess emission in \dam\ ($\sim0.5$~\Msun).

Finally, we fit a combination of the Piro plus magnetar and
also Piro plus CSM interaction to the \dcc\ observations. In the latter
combined model, the explosion energy is set to the same value in both
the Piro and CSM interaction models, and the ejecta mass (of the CSM
interaction model) set to the sum of the core mass and the extended
mass (both of the Piro model): $M_{\rm core}+M_{\rm ext} = M_{\rm
  ejecta}$. This is because the ejected core mass is sweeping up the
extended mass, which is at a lower distance than the CSM mass, and
thus the sum of these masses equals the ejecta mass in the CSM
interaction model. The best-fit parameters of these combination models
are listed in Tables~\ref{tab:piromag} and \ref{tab:pirocsm}, and the
corresponding light curves are shown in the right panel of
Fig.~\ref{fig:model13dcc}.

\subsection{\des}

\citet{2015arXiv151206043S} present observations of the hydrogen-poor
SLSN \des, which also shows evidence for early-time excess emission.
These authors were able to catch this early emission in multiple
filters, for the first time, allowing for a quite accurate
determination of the temperature, radius, and bolometric light-curve
evolution for the early-time emission. Since the bolometric evolution
of this SLSN is so well constrained from a very early epoch, we also
fit \des\ with the same models as we have done for \dam\ and
\dcc\ above. The filled black circles in Fig.~\ref{fig:model14des}
show the bolometric light curve of \des\ as presented by
\citet{2015arXiv151206043S}. The time of explosion for \des\ is very
well constrained by observations to be in the range $-50$ to
$-43$~days. As for \dam\ and \dcc, we loosely limit the radius of the
progenitor star, $R_0$, to be within the range 0.1--500~\Rsun, and use
a fiducial value for the expansion velocity of 10,000~\kms. We exclude
the first two detections and focus on fitting the models to the main
peak, but constraining the time of explosion to be before the first
detection. The resulting best-fit values for the model parameters of
the radioactive decay, magnetar, and CSM interaction models are listed
in Tables~\ref{tab:nico}, \ref{tab:magnetar}, and \ref{tab:csm},
respectively. The corresponding bolometric-luminosity evolution is
shown in the left panel of Fig.~\ref{fig:model14des}.  As we did for
\dcc\ in the previous section, a combination of the Piro plus magnetar
and Piro plus CSM interaction models is fit to the full light curve of
\des. The best-fit parameters of these combination models are listed
alongside those of \dcc\ in Tables~\ref{tab:piromag} and
\ref{tab:pirocsm}, and the corresponding bolometric model light curves
are shown in the right panel of Fig.~\ref{fig:model14des}.

\begin{figure*}
  \centering
  \includegraphics[width=\hsize]{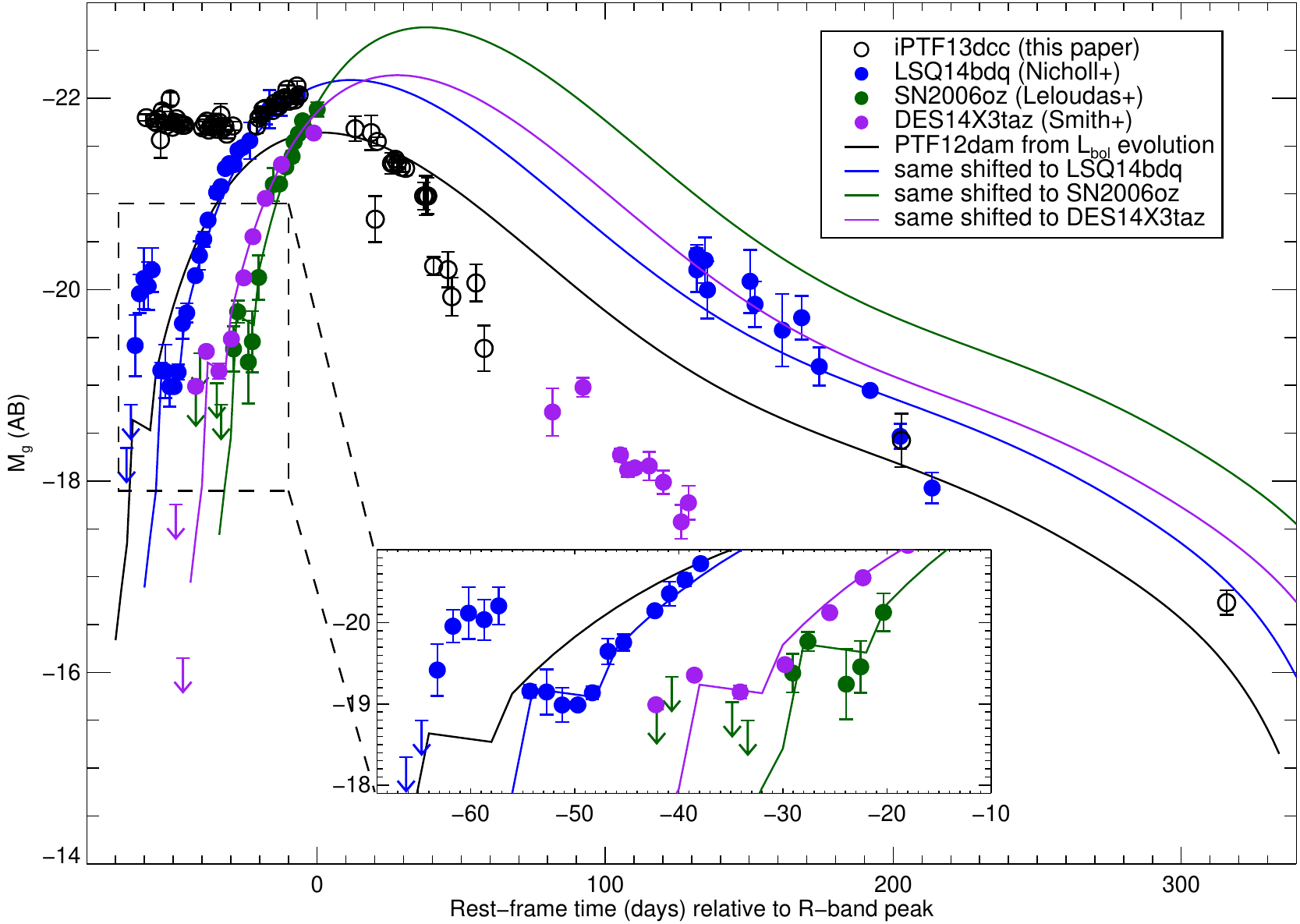}
  \caption{Absolute $g$-band light curves of \dcc\ (open black
    circles), LSQ14bdq \citep[filled blue
      circles;][]{2015ApJ...807L..18N}, \oz\ \citep[filled green
      circles;][]{2012A&A...541A.129L}, and \des\ \citep[filled purple
      circles;][]{2015arXiv151206043S}. The solid black line shows the
    \dam\ evolution assuming the temperature and radius evolution as
    inferred from the photometry and as shown in
    Fig.~\ref{fig:bol12dam}.  The blue, green, and purple lines
    correspond to the \dam\ $M_g$ light curve simply shifted in time
    and magnitude to roughly match the data points of LSQ14bdq,
    \oz, and \des, respectively.  The open black circles are the
    \dcc\ $M_g$ values inferred from the observed P48 and P60 $r$-band
    magnitudes, applying a K-correction that is based on the
    temperature evolution of \dam\ (see
    Fig.~\ref{fig:bol13dcc}). \label{fig:Mg}}
\end{figure*}

\section{Discussion}
\label{sec:discussion}

Figure~\ref{fig:Mg} shows a comparison of the absolute $g$-band light
curves of \dcc, \bdq, \oz, \des, and \dam. The \dcc\ data were derived
by converting the observed $r$-band measurements to rest-frame $g$,
where we computed the K-corrections adopting the temperature evolution
of \dam\ (see Fig.~\ref{fig:bol13dcc}). Since the effective
wavelengths of the observed $r$-band and rest-frame $g$-band filters
match well at the redshift of \dcc\ ($z=0.431$), these K-corrections
are very close to $-$2.5~log~(1+$z$) and do not depend very much on
the assumed temperature evolution. For \dam\ ($z=0.107$), we instead
computed a continuous rest-frame $g$-band light curve by adopting the
temperature and bolometric luminosity evolution as derived in
Fig.~\ref{fig:bol12dam}. For each SLSN shown in Fig.~\ref{fig:Mg},
except for \dcc, we shifted the \dam\ light curve, in both time and
magnitude, to match the data points. This is simply to show how the
observed early excess emission of each SLSN compares in duration and
magnitude with that of the others.  For both \oz\ and \des\ the match
with \dam\ at early epochs is quite good, whereas at late times the
\des\ light curve is dropping much more rapidly than that of \dam. For
\bdq\ the early excess emission is longer and brighter than that of
\dam, but at late times the two match surprisingly well. The obvious
outlier is the light curve of \dcc: its early emission is of a much
longer duration and a few magnitudes brighter than those of the
others.

\citet{2016ApJ...821...36K} suggest that the early excess emission
observed for \bdq\ could be caused by a magnetar central engine whose
energy input drives a shock through the pre-exploded SN ejecta,
resulting in a burst of shock-breakout emission several days after the
explosion. The radiation is expected to be released in the optical/UV
wavelengths and to have a duration of several days, with the emission
being dimmer than the main light-curve peak resulting from continued
magnetar heating.  As can be seen in Fig.~7 of
\citet{2016ApJ...821...36K}, for standard magnetar and ejecta
parameters, the shock-breakout emission produces only a kink in the
overall light curve. In principle, this model is capable of explaining
the early excess emission in \dam. However, the early emission that we
observe for \dcc\ is too bright and too extended to be accommodated in
this model. Even when pushing the ejecta mass and kinetic energy to
large values \citep[see Fig.~8 of][]{2016ApJ...821...36K}, this model
is not able to reproduce the \dcc\ observations.

Based on the model fits presented in Sec.~\ref{sec:modelling}, the CSM
interaction model seems to provide better fits to the data.  Moreover,
the magnetar model is unable by itself to fit the precursor bumps seen
in SLSNe but requires the addition of the Piro model (or similar) and
the presence of extended material at large radii to explain the first
peak. The combination of Piro and magnetar provides a reasonable
description of the \dcc\ light curve, even though it does not fit the
data well around peak and appears to overshoot the data beyond
$+40$~days. Moreover, the explosion energy for this model reaches the
upper end of the range that we allowed: $E_{\rm SN}=10^{53}$~erg.  We
find that the CSM interaction model, which in fact contains one free
parameter less than the number used in the combined Piro and magnetar
model, can provide a good fit to the \dcc\ light curve, even without
the addition of the Piro model (see Table~\ref{tab:csm} and
Fig.~\ref{fig:model13dcc}).  The first minimum in the light curve
cannot be accurately reproduced (see the left panel of
Fig.~\ref{fig:model13dcc}), but a time-dependent opacity (we have
assumed a constant opacity throughout this paper) could be invoked to
explain the early light-curve shape. Also, this \dcc\ light-curve dip
is reminiscent of the picture proposed by \citet{2012ApJ...756L..22M},
where a drop in the light curve arises naturally. These authors
suggest that such a dip is a solid prediction from the strong
interaction scenario regardless of the power source for the early
emission.

The ability of the CSM interaction model to reproduce the light curves
of many SLSNe can be explained by the fact that it includes many free
parameters. In addition, the CSM model by
\citet{2012ApJ...746..121C,2013ApJ...773...76C} includes a number of
simplifying assumptions of which the most important are those of a
central power source and a stationary photosphere, allowing for the
use of Arnett-style diffusion by using analytical equations. However,
it is not clear if these assumptions hold in real CSM
interaction. Chatzopoulos et al. verify their model against a more
sophisticated hydrodynamical model developed for the H-rich SN~2006gy
\citep{2013MNRAS.428.1020M} and obtain a reassuringly similar light
curve for similar parameters. Extending the use of the model to H-poor
SLSNe, such as \dam\ and \dcc, is not trivial, however, owing to the
different treatment of the opacity when hydrogen is
absent. Nevertheless, models that successfully reproduce the light
curves of H-poor SLSNe by using CSM interaction have now been
reproduced in radiation hydrodynamics simulations by
\citet{2015arXiv151000834S}, although these authors do report
discrepancies in some parameters of their models in comparison to
those obtained with the Chatzopoulos models for the same SLSNe. The
CSM interaction semi-analytical models remain a valuable tool, and it
is clear that H-poor CSM interaction can reproduce the light curves of
SLSNe, more naturally explaining features such as premaximum bumps in
a self-consistent way. However, these caveats show that caution should
be used in interpreting the best-fit model parameters based on
$\chi^2$ minimization of the semi-analytical CSM interaction models.

On a final note, an argument often invoked against the CSM interaction
model for hydrogen-poor SLSNe is the lack of narrow emission lines, as
observed for Type IIn SNe. However, to date, the spectroscopic
signature(s) of interaction with a H-deficient dense CSM has not been
investigated through spectral synthesis modelling, due to its
complexity.

\begin{deluxetable}{crccc}
  \tablecaption{Log of observations of \dam.\label{tab:12dam_logphotometry}}
  \tablehead{
    \colhead{MJD} &
    \colhead{Phase\a} &
    \colhead{Telescope} &
    \colhead{Filter} &
    \colhead{Magnitude\b} \\
    (days) & (days) & & & AB
  }
  \startdata
    56009.34  &$-$78.92  &       P48  &   R  & $>$ 22.26 \\
    56013.35  &$-$75.30  &       P48  &   R  & $>$ 22.10 \\
    56017.29  &$-$71.73  &       P48  &   R  & $>$ 22.13 \\
    56020.41  &$-$68.91  &       P48  &   R  & $>$ 21.94 \\
    56023.28  &$-$66.33  &       P48  &   R  & $>$ 20.69 \\
    56025.33  &$-$64.47  &       P48  &   R  &     20.11 $\pm$ 0.17 \\
    56027.26  &$-$62.73  &       P48  &   R  &     20.02 $\pm$ 0.12 \\
    56027.29  &$-$62.70  &       P48  &   R  &     20.11 $\pm$ 0.11 \\
    56027.32  &$-$62.68  &       P48  &   R  &     20.04 $\pm$ 0.10 \\
    56033.49  &$-$57.10  &       P48  &   R  &     19.89 $\pm$ 0.08 \\
    56034.16  &$-$56.49  &       P48  &   R  &     19.96 $\pm$ 0.08 \\
    56034.19  &$-$56.46  &       P48  &   R  &     19.96 $\pm$ 0.08 \\
    56034.26  &$-$56.41  &       P48  &   R  &     19.83 $\pm$ 0.07 \\
    56036.17  &$-$54.68  &       P48  &   R  &     19.77 $\pm$ 0.08 \\
    56036.21  &$-$54.64  &       P48  &   R  &     19.68 $\pm$ 0.07 \\
    56036.37  &$-$54.50  &       P48  &   R  &     19.70 $\pm$ 0.07 \\
    56038.35  &$-$52.71  &       P48  &   R  &     19.37 $\pm$ 0.04 \\
    56038.38  &$-$52.69  &       P48  &   R  &     19.39 $\pm$ 0.04 \\
    56040.43  &$-$50.83  &       P48  &   R  &     19.28 $\pm$ 0.04 \\
    56040.46  &$-$50.81  &       P48  &   R  &     19.22 $\pm$ 0.04 \\
    56046.40  &$-$45.44  &       P48  &   R  &     18.55 $\pm$ 0.03 \\
    56046.43  &$-$45.41  &       P48  &   R  &     18.57 $\pm$ 0.03 \\
    56048.43  &$-$43.60  &       P48  &   R  &     18.34 $\pm$ 0.03 \\
    56048.47  &$-$43.57  &       P48  &   R  &     18.43 $\pm$ 0.02 \\
    56048.50  &$-$43.54  &       P48  &   R  &     18.36 $\pm$ 0.03 \\
    56058.47  &$-$34.53  &       P48  &   R  &     17.76 $\pm$ 0.02 \\
    56059.28  &$-$33.81  &       P48  &   R  &     17.69 $\pm$ 0.01 \\
    56061.33  &$-$31.95  &       P48  &   R  &     17.64 $\pm$ 0.01 \\
    56063.29  &$-$30.18  &       P48  &   R  &     17.53 $\pm$ 0.01 \\
    56063.32  &$-$30.15  &       P48  &   R  &     17.51 $\pm$ 0.01 \\
    56063.35  &$-$30.12  &       P48  &   R  &     17.57 $\pm$ 0.01 \\
    56065.44  &$-$28.24  &       P48  &   R  &     17.46 $\pm$ 0.02 \\
    56066.38  &$-$27.39  &       P48  &   R  &     17.44 $\pm$ 0.01 \\
    56066.41  &$-$27.36  &       P48  &   R  &     17.46 $\pm$ 0.01 \\
    56066.45  &$-$27.33  &       P48  &   R  &     17.44 $\pm$ 0.01 \\
    56068.48  &$-$25.50  &       P48  &   R  &     17.38 $\pm$ 0.01 \\
    56070.40  &$-$23.76  &       P48  &   R  &     17.28 $\pm$ 0.01 \\
    56070.44  &$-$23.73  &       P48  &   R  &     17.28 $\pm$ 0.01 \\
    56070.47  &$-$23.70  &       P48  &   R  &     17.29 $\pm$ 0.01 \\
    56076.37  &$-$18.36  &       P48  &   R  &     17.12 $\pm$ 0.01 \\
    56076.40  &$-$18.33  &       P48  &   R  &     17.16 $\pm$ 0.01 \\
    56362.48  &  240.09  &       P48  &   R  &     20.70 $\pm$ 0.13 \\
    56362.51  &  240.12  &       P48  &   R  &     20.42 $\pm$ 0.10 \\
    56362.54  &  240.14  &       P48  &   R  &     20.59 $\pm$ 0.13 \\
    56366.49  &  243.71  &       P48  &   R  &     20.40 $\pm$ 0.17 \\
    56366.50  &  243.72  &       P48  &   R  &     20.56 $\pm$ 0.10 \\
    56366.51  &  243.73  &       P48  &   R  &     20.78 $\pm$ 0.13 \\
    56366.52  &  243.74  &       P48  &   R  &     20.60 $\pm$ 0.11 \\
    56369.54  &  246.47  &       P48  &   R  &     20.51 $\pm$ 0.13 \\
    56374.51  &  250.96  &       P48  &   R  &     20.48 $\pm$ 0.12 \\
    56376.51  &  252.76  &       P48  &   R  &     21.01 $\pm$ 0.20 \\
    56385.51  &  260.90  &       P48  &   R  &     20.48 $\pm$ 0.27 \\
    56068.22  &$-$25.73  &       P60  &   i  &     17.52 $\pm$ 0.01 \\
    56068.22  &$-$25.73  &       P60  &   r  &     17.42 $\pm$ 0.01 \\
    56068.22  &$-$25.73  &       P60  &   B  &     17.14 $\pm$ 0.01 \\
    56068.22  &$-$25.72  &       P60  &   g  &     17.13 $\pm$ 0.01 \\
    56068.23  &$-$25.72  &       P60  &   i  &     17.54 $\pm$ 0.01 \\
    56068.23  &$-$25.72  &       P60  &   r  &     17.34 $\pm$ 0.01 \\
    56068.23  &$-$25.72  &       P60  &   B  &     17.21 $\pm$ 0.01 \\
    56068.23  &$-$25.72  &       P60  &   g  &     17.13 $\pm$ 0.01 \\
    56075.23  &$-$19.39  &       P60  &   i  &     17.31 $\pm$ 0.01 \\
    56075.24  &$-$19.39  &       P60  &   r  &     17.22 $\pm$ 0.01 \\
    56075.24  &$-$19.39  &       P60  &   B  &     16.80 $\pm$ 0.01 \\
    56075.24  &$-$19.39  &       P60  &   g  &     17.17 $\pm$ 0.01 \\
    56080.20  &$-$14.90  &       P60  &   i  &     17.19 $\pm$ 0.01 \\
    56080.20  &$-$14.90  &       P60  &   r  &     17.02 $\pm$ 0.01 \\
    56080.20  &$-$14.90  &       P60  &   B  &     16.87 $\pm$ 0.01 \\
    56080.21  &$-$14.90  &       P60  &   g  &     16.82 $\pm$ 0.01 \\
    56080.21  &$-$14.90  &       P60  &   i  &     17.19 $\pm$ 0.01 \\
    56080.21  &$-$14.90  &       P60  &   r  &     17.05 $\pm$ 0.01 \\
    56080.21  &$-$14.90  &       P60  &   B  &     16.90 $\pm$ 0.01 \\
    56080.21  &$-$14.89  &       P60  &   g  &     16.76 $\pm$ 0.01
  \enddata
  \tablenotetext{a}{Calculated using MJD$_{r,\rm peak} = 56096.7$ and $z=0.107$.}
  \tablenotetext{b}{The magnitudes have {\it not} been corrected for
    Galactic extinction.}
\end{deluxetable}

\addtocounter{table}{-1}
\begin{deluxetable}{crccc}
  \tablecaption{(continued) Log of observations of \dam.}
  \tablehead{
    \colhead{MJD} &
    \colhead{Phase\a} &
    \colhead{Telescope} &
    \colhead{Filter} &
    \colhead{Magnitude\b} \\
    (days) & (days) & & & AB
  }
  \startdata
    56085.21  &$-$10.38  &       P60  &   i  &     17.11 $\pm$ 0.01 \\
    56085.21  &$-$10.38  &       P60  &   r  &     16.98 $\pm$ 0.01 \\
    56085.21  &$-$10.38  &       P60  &   B  &     16.83 $\pm$ 0.01 \\
    56085.21  &$-$10.38  &       P60  &   i  &     17.21 $\pm$ 0.01 \\
    56085.21  &$-$10.38  &       P60  &   r  &     17.05 $\pm$ 0.01 \\
    56085.21  &$-$10.37  &       P60  &   B  &     16.83 $\pm$ 0.01 \\
    56085.22  &$-$10.37  &       P60  &   g  &     16.78 $\pm$ 0.01 \\
    56090.20  & $-$5.87  &       P60  &   i  &     17.05 $\pm$ 0.01 \\
    56090.20  & $-$5.87  &       P60  &   r  &     16.92 $\pm$ 0.01 \\
    56090.20  & $-$5.87  &       P60  &   g  &     16.74 $\pm$ 0.01 \\
    56090.20  & $-$5.87  &       P60  &   i  &     17.19 $\pm$ 0.01 \\
    56090.21  & $-$5.87  &       P60  &   r  &     16.91 $\pm$ 0.01 \\
    56090.21  & $-$5.87  &       P60  &   B  &     16.89 $\pm$ 0.01 \\
    56090.22  & $-$5.85  &       P60  &   g  &     16.78 $\pm$ 0.01 \\
    56098.29  &    1.44  &       P60  &   i  &     17.10 $\pm$ 0.01 \\
    56098.29  &    1.44  &       P60  &   r  &     16.98 $\pm$ 0.01 \\
    56098.29  &    1.44  &       P60  &   B  &     16.90 $\pm$ 0.01 \\
    56098.30  &    1.44  &       P60  &   i  &     17.12 $\pm$ 0.01 \\
    56098.30  &    1.44  &       P60  &   r  &     16.95 $\pm$ 0.01 \\
    56098.30  &    1.44  &       P60  &   B  &     16.89 $\pm$ 0.01 \\
    56098.30  &    1.44  &       P60  &   g  &     16.78 $\pm$ 0.01 \\
    56103.21  &    5.88  &       P60  &   i  &     17.08 $\pm$ 0.01 \\
    56103.21  &    5.88  &       P60  &   r  &     16.96 $\pm$ 0.01 \\
    56103.22  &    5.89  &       P60  &   g  &     16.97 $\pm$ 0.01 \\
    56103.22  &    5.89  &       P60  &   i  &     17.13 $\pm$ 0.01 \\
    56103.23  &    5.90  &       P60  &   r  &     17.12 $\pm$ 0.01 \\
    56103.23  &    5.90  &       P60  &   B  &     16.86 $\pm$ 0.01 \\
    56103.23  &    5.90  &       P60  &   g  &     16.84 $\pm$ 0.01 \\
    56108.21  &   10.40  &       P60  &   i  &     17.11 $\pm$ 0.01 \\
    56108.21  &   10.40  &       P60  &   r  &     17.02 $\pm$ 0.01 \\
    56108.21  &   10.40  &       P60  &   B  &     17.00 $\pm$ 0.01 \\
    56108.21  &   10.40  &       P60  &   g  &     16.80 $\pm$ 0.01 \\
    56108.21  &   10.40  &       P60  &   i  &     17.12 $\pm$ 0.01 \\
    56108.21  &   10.40  &       P60  &   r  &     17.02 $\pm$ 0.01 \\
    56108.21  &   10.40  &       P60  &   B  &     16.99 $\pm$ 0.01 \\
    56108.21  &   10.40  &       P60  &   g  &     16.87 $\pm$ 0.01 \\
    56113.20  &   14.91  &       P60  &   i  &     17.13 $\pm$ 0.01 \\
    56113.20  &   14.91  &       P60  &   r  &     17.09 $\pm$ 0.01 \\
    56113.20  &   14.91  &       P60  &   B  &     17.08 $\pm$ 0.02 \\
    56113.21  &   14.91  &       P60  &   g  &     16.86 $\pm$ 0.01 \\
    56113.21  &   14.91  &       P60  &   i  &     17.15 $\pm$ 0.01 \\
    56113.21  &   14.91  &       P60  &   r  &     17.02 $\pm$ 0.01 \\
    56113.21  &   14.91  &       P60  &   B  &     17.10 $\pm$ 0.02 \\
    56113.21  &   14.91  &       P60  &   g  &     16.89 $\pm$ 0.01 \\
    56118.25  &   19.47  &       P60  &   i  &     17.17 $\pm$ 0.01 \\
    56118.25  &   19.47  &       P60  &   r  &     17.11 $\pm$ 0.01 \\
    56118.25  &   19.47  &       P60  &   B  &     17.18 $\pm$ 0.01 \\
    56118.25  &   19.47  &       P60  &   g  &     16.99 $\pm$ 0.01 \\
    56118.26  &   19.47  &       P60  &   i  &     17.21 $\pm$ 0.01 \\
    56118.26  &   19.47  &       P60  &   r  &     17.08 $\pm$ 0.01 \\
    56118.26  &   19.47  &       P60  &   B  &     17.16 $\pm$ 0.01 \\
    56118.26  &   19.48  &       P60  &   g  &     16.99 $\pm$ 0.01 \\
    56127.28  &   27.62  &       P60  &   g  &     17.47 $\pm$ 0.02 \\
    56129.20  &   29.36  &       P60  &   i  &     17.37 $\pm$ 0.01 \\
    56129.20  &   29.36  &       P60  &   r  &     17.24 $\pm$ 0.01 \\
    56129.20  &   29.36  &       P60  &   B  &     17.41 $\pm$ 0.01 \\
    56129.20  &   29.36  &       P60  &   g  &     17.22 $\pm$ 0.01 \\
    56159.20  &   56.46  &       P60  &   i  &     17.62 $\pm$ 0.02 \\
    56161.18  &   58.24  &       P60  &   r  &     17.56 $\pm$ 0.01 \\
    56161.18  &   58.24  &       P60  &   B  &     18.11 $\pm$ 0.02 \\
    56161.18  &   58.25  &       P60  &   g  &     17.87 $\pm$ 0.01 \\
    56165.17  &   61.85  &       P60  &   i  &     17.54 $\pm$ 0.02 \\
    56177.15  &   72.67  &       P60  &   i  &     17.67 $\pm$ 0.02 \\
    56177.15  &   72.67  &       P60  &   r  &     17.78 $\pm$ 0.01 \\
    56177.15  &   72.67  &       P60  &   B  &     18.46 $\pm$ 0.03 \\
    56177.15  &   72.67  &       P60  &   g  &     17.98 $\pm$ 0.01 \\
    56185.16  &   79.91  &       P60  &   i  &     17.90 $\pm$ 0.02 \\
    56185.16  &   79.91  &       P60  &   r  &     17.92 $\pm$ 0.02 \\
    56185.16  &   79.91  &       P60  &   B  &     18.54 $\pm$ 0.03 \\
    56185.16  &   79.91  &       P60  &   g  &     18.18 $\pm$ 0.02 \\
    56185.17  &   79.91  &       P60  &   i  &     17.80 $\pm$ 0.02 \\
    56185.17  &   79.92  &       P60  &   r  &     17.92 $\pm$ 0.02
  \enddata
  \tablenotetext{a}{Calculated using MJD$_{r,\rm peak} = 56096.7$ and $z=0.107$.}
  \tablenotetext{b}{The magnitudes have {\it not} been corrected for
    Galactic extinction.}
\end{deluxetable}

\addtocounter{table}{-1}
\begin{deluxetable}{crccc}
  \tablecaption{(continued) Log of observations of \dam.}
  \tablehead{
    \colhead{MJD} &
    \colhead{Phase\a} &
    \colhead{Telescope} &
    \colhead{Filter} &
    \colhead{Magnitude\b} \\
    (days) & (days) & & & AB
  }
  \startdata 
    56185.17  &   79.92  &       P60  &   B  &     18.59 $\pm$ 0.04 \\
    56185.17  &   79.92  &       P60  &   g  &     18.15 $\pm$ 0.02 \\
    56201.11  &   94.32  &       P60  &   g  &     18.46 $\pm$ 0.03 \\
    56202.11  &   95.22  &       P60  &   i  &     17.92 $\pm$ 0.02 \\
    56202.11  &   95.22  &       P60  &   r  &     18.15 $\pm$ 0.02 \\
    56214.10  &  106.05  &       P60  &   i  &     18.30 $\pm$ 0.03 \\
    56214.10  &  106.06  &       P60  &   r  &     18.31 $\pm$ 0.02 \\
    56214.10  &  106.06  &       P60  &   B  &     19.09 $\pm$ 0.04 \\
    56214.11  &  106.06  &       P60  &   g  &     18.58 $\pm$ 0.02 \\
    56237.55  &  127.23  &       P60  &   i  &     18.73 $\pm$ 0.06 \\
    56245.54  &  134.45  &       P60  &   i  &     18.73 $\pm$ 0.13 \\
    56245.54  &  134.45  &       P60  &   B  &     19.84 $\pm$ 0.18 \\
    56245.54  &  134.45  &       P60  &   g  &     19.29 $\pm$ 0.09 \\
    56250.51  &  138.95  &       P60  &   r  &     19.21 $\pm$ 0.04 \\
    56251.51  &  139.85  &       P60  &   r  &     19.17 $\pm$ 0.03 \\
    56252.51  &  140.75  &       P60  &   i  &     19.10 $\pm$ 0.05 \\
    56252.51  &  140.75  &       P60  &   B  &     19.97 $\pm$ 0.11 \\
    56252.51  &  140.75  &       P60  &   g  &     19.59 $\pm$ 0.05 \\
    56262.54  &  149.81  &       P60  &   r  &     19.42 $\pm$ 0.09 \\
    56265.48  &  152.46  &       P60  &   g  &     19.56 $\pm$ 0.21 \\
    56266.47  &  153.36  &       P60  &   r  &     19.16 $\pm$ 0.04 \\
    56266.47  &  153.36  &       P60  &   g  &     19.62 $\pm$ 0.07 \\
    56268.47  &  155.16  &       P60  &   B  &     20.11 $\pm$ 0.10 \\
    56273.45  &  159.67  &       P60  &   r  &     19.53 $\pm$ 0.05 \\
    56273.45  &  159.67  &       P60  &   g  &     19.86 $\pm$ 0.06 \\
    56281.43  &  166.87  &       P60  &   i  &     19.64 $\pm$ 0.15 \\
    56281.43  &  166.88  &       P60  &   r  &     19.50 $\pm$ 0.09 \\
    56283.43  &  168.68  &       P60  &   r  &     19.55 $\pm$ 0.03 \\
    56283.43  &  168.68  &       P60  &   B  &     20.21 $\pm$ 0.09 \\
    56283.43  &  168.68  &       P60  &   g  &     19.84 $\pm$ 0.05 \\
    56297.39  &  181.29  &       P60  &   i  &     19.66 $\pm$ 0.07 \\
    56297.39  &  181.29  &       P60  &   r  &     19.67 $\pm$ 0.04 \\
    56297.39  &  181.29  &       P60  &   B  &     20.13 $\pm$ 0.10 \\
    56297.39  &  181.29  &       P60  &   g  &     20.00 $\pm$ 0.06 \\
    56323.31  &  204.71  &       P60  &   i  &     20.03 $\pm$ 0.13 \\
    56323.32  &  204.71  &       P60  &   r  &     20.45 $\pm$ 0.15 \\
    56323.32  &  204.71  &       P60  &   g  &     20.39 $\pm$ 0.16 \\
    56324.31  &  205.61  &       P60  &   r  &     20.30 $\pm$ 0.07 \\
    56345.33  &  224.60  &       P60  &   i  &     20.18 $\pm$ 0.13 \\
    56345.34  &  224.60  &       P60  &   r  &     20.41 $\pm$ 0.11 \\
    56345.34  &  224.61  &       P60  &   B  &     20.75 $\pm$ 0.27 \\
    56345.34  &  224.61  &       P60  &   g  &     20.83 $\pm$ 0.16 \\
    56345.34  &  224.61  &       P60  &   i  &     20.22 $\pm$ 0.12 \\
    56345.34  &  224.61  &       P60  &   r  &     20.36 $\pm$ 0.10 \\
    56345.35  &  224.61  &       P60  &   B  &     20.63 $\pm$ 0.24 \\
    56345.35  &  224.61  &       P60  &   g  &     20.61 $\pm$ 0.13 \\
    56346.25  &  225.43  &       P60  &   i  &     20.55 $\pm$ 0.20 \\
    56346.25  &  225.43  &       P60  &   r  &     20.18 $\pm$ 0.12 \\
    56346.26  &  225.44  &       P60  &   g  &     20.43 $\pm$ 0.17 \\
    56347.43  &  226.50  &       P60  &   i  &     20.17 $\pm$ 0.21 \\
    56347.43  &  226.50  &       P60  &   r  &     20.59 $\pm$ 0.22 \\
    56347.44  &  226.50  &       P60  &   g  &     20.62 $\pm$ 0.26 \\
    56349.24  &  228.13  &       P60  &   i  &     19.78 $\pm$ 0.21 \\
    56350.24  &  229.04  &       P60  &   r  &     20.49 $\pm$ 0.25 \\
    56350.25  &  229.04  &       P60  &   g  &     20.55 $\pm$ 0.31 \\
    56352.24  &  230.84  &       P60  &   r  &     20.47 $\pm$ 0.16 \\
    56352.24  &  230.84  &       P60  &   g  &     20.90 $\pm$ 0.28 \\
    56353.23  &  231.74  &       P60  &   r  &     20.46 $\pm$ 0.09 \\
    56353.24  &  231.74  &       P60  &   B  &     21.23 $\pm$ 0.26 \\
    56353.24  &  231.74  &       P60  &   g  &     20.62 $\pm$ 0.10 \\
    56356.23  &  234.44  &       P60  &   i  &     20.44 $\pm$ 0.12 \\
    56362.27  &  239.90  &       P60  &   r  &     20.55 $\pm$ 0.09 \\
    56362.28  &  239.91  &       P60  &   B  &     21.01 $\pm$ 0.21 \\
    56362.28  &  239.91  &       P60  &   g  &     20.82 $\pm$ 0.10 \\
    56362.28  &  239.91  &       P60  &   r  &     20.56 $\pm$ 0.09 \\
    56362.28  &  239.91  &       P60  &   B  &     20.97 $\pm$ 0.19 \\
    56362.28  &  239.91  &       P60  &   g  &     20.82 $\pm$ 0.10 \\
    56364.20  &  241.65  &       P60  &   i  &     20.25 $\pm$ 0.09 \\
    56370.31  &  247.16  &       P60  &   r  &     20.70 $\pm$ 0.13 \\
    56370.31  &  247.16  &       P60  &   B  &     20.84 $\pm$ 0.23 \\
    56371.25  &  248.01  &       P60  &   g  &     21.28 $\pm$ 0.29 \\
    56372.20  &  248.87  &       P60  &   r  &     20.44 $\pm$ 0.22
  \enddata
  \tablenotetext{a}{Calculated using MJD$_{r,\rm peak} = 56096.7$ and $z=0.107$.}
  \tablenotetext{b}{The magnitudes have {\it not} been corrected for
    Galactic extinction.}
\end{deluxetable}
 
\addtocounter{table}{-1}
\begin{deluxetable}{crccc}
  \tablecaption{(continued) Log of observations of \dam.}
  \tablehead{
    \colhead{MJD} &
    \colhead{Phase\a} &
    \colhead{Telescope} &
    \colhead{Filter} &
    \colhead{Magnitude\b} \\
    (days) & (days) & & & AB
  }
  \startdata
    56373.18  &  249.76  &       P60  &   i  &     20.71 $\pm$ 0.25 \\
    56373.18  &  249.76  &       P60  &   r  &     20.51 $\pm$ 0.21 \\
    56374.18  &  250.66  &       P60  &   r  &     20.20 $\pm$ 0.25 \\
    56375.29  &  251.66  &       P60  &   i  &     20.64 $\pm$ 0.14 \\
    56375.29  &  251.66  &       P60  &   r  &     21.33 $\pm$ 0.18 \\
    56375.29  &  251.67  &       P60  &   g  &     21.01 $\pm$ 0.09 \\
    56376.17  &  252.46  &       P60  &   r  &     20.89 $\pm$ 0.25 \\
    56382.17  &  257.88  &       P60  &   B  &     21.04 $\pm$ 0.21 \\
    56386.17  &  261.49  &       P60  &   g  &     21.25 $\pm$ 0.13 \\
    56388.50  &  263.59  &       P60  &   r  &     21.19 $\pm$ 0.24 \\
    56388.50  &  263.59  &       P60  &   g  &     21.37 $\pm$ 0.15 \\
    56390.28  &  265.20  &       P60  &   r  &     20.72 $\pm$ 0.23 \\
    56392.27  &  267.00  &       P60  &   r  &     21.20 $\pm$ 0.26 \\
    56409.27  &  282.36  &       P60  &   i  &     20.78 $\pm$ 0.19 \\
    56412.46  &  285.24  &       P60  &   i  &     21.50 $\pm$ 0.24 \\
    56413.45  &  286.13  &       P60  &   r  &     21.58 $\pm$ 0.26 \\
    56416.39  &  288.79  &       P60  &   r  &     21.91 $\pm$ 0.30 \\
    56416.40  &  288.79  &       P60  &   g  &     21.92 $\pm$ 0.17 \\
    56422.43  &  294.25  &       P60  &   i  &     21.36 $\pm$ 0.20 \\
    56422.44  &  294.25  &       P60  &   g  &     22.35 $\pm$ 0.28 \\
    56425.34  &  296.88  &       P60  &   r  &     21.94 $\pm$ 0.27 \\
    56431.39  &  302.34  &       P60  &   i  &     21.40 $\pm$ 0.25 \\
    56432.32  &  303.18  &       P60  &   i  &     21.57 $\pm$ 0.23 \\
    56452.32  &  321.25  &       P60  &   r  &     21.97 $\pm$ 0.29 \\
    56459.24  &  327.50  &       P60  &   i  &     21.37 $\pm$ 0.18 \\
    56460.29  &  328.45  &       P60  &   i  &     21.45 $\pm$ 0.28 \\
    56470.27  &  337.46  &       P60  &   i  &     21.83 $\pm$ 0.31 \\
    56471.21  &  338.31  &       P60  &   r  &     22.19 $\pm$ 0.25 \\
    56507.22  &  370.84  &       P60  &   r  &     22.10 $\pm$ 0.23 \\
    56070.26  &$-$23.88  &     LCOGT  &   r  &     17.23 $\pm$ 0.02 \\
    56070.26  &$-$23.88  &     LCOGT  &   r  &     17.22 $\pm$ 0.02 \\
    56083.36  &$-$12.05  &     LCOGT  &   r  &     17.00 $\pm$ 0.02 \\
    56086.44  & $-$9.26  &     LCOGT  &   r  &     16.91 $\pm$ 0.02 \\
    56086.45  & $-$9.26  &     LCOGT  &   r  &     16.94 $\pm$ 0.02 \\
    56089.33  & $-$6.65  &     LCOGT  &   r  &     16.89 $\pm$ 0.02 \\
    56089.34  & $-$6.65  &     LCOGT  &   r  &     16.90 $\pm$ 0.02 \\
    56128.38  &   28.62  &     LCOGT  &   r  &     17.20 $\pm$ 0.02 \\
    56128.38  &   28.62  &     LCOGT  &   r  &     17.22 $\pm$ 0.03 \\
    56149.24  &   47.47  &     LCOGT  &   r  &     17.41 $\pm$ 0.02 \\
    56149.25  &   47.47  &     LCOGT  &   r  &     17.42 $\pm$ 0.01 \\
    56152.24  &   50.17  &     LCOGT  &   r  &     17.48 $\pm$ 0.02 \\
    56152.24  &   50.17  &     LCOGT  &   r  &     17.46 $\pm$ 0.03 \\
    56156.24  &   53.79  &     LCOGT  &   r  &     17.55 $\pm$ 0.01 \\
    56156.24  &   53.79  &     LCOGT  &   r  &     17.62 $\pm$ 0.02 \\
    56181.23  &   76.36  &     LCOGT  &   r  &     17.87 $\pm$ 0.04 \\
    56181.23  &   76.36  &     LCOGT  &   r  &     17.91 $\pm$ 0.03 \\
    56187.24  &   81.79  &     LCOGT  &   r  &     17.97 $\pm$ 0.02 \\
    56196.21  &   89.90  &     LCOGT  &   r  &     18.18 $\pm$ 0.02 \\
    56196.22  &   89.90  &     LCOGT  &   r  &     18.03 $\pm$ 0.03 \\
    56197.22  &   90.81  &     LCOGT  &   r  &     18.11 $\pm$ 0.02 \\
    56197.22  &   90.81  &     LCOGT  &   r  &     18.23 $\pm$ 0.02 \\
    56200.21  &   93.51  &     LCOGT  &   r  &     18.19 $\pm$ 0.02 \\
    56200.21  &   93.51  &     LCOGT  &   r  &     18.23 $\pm$ 0.02 \\
    56360.00  &  237.85  &    Keck~I  &   r  &     20.44 $\pm$ 0.30 \\
    56630.00  &  481.75  &    Keck~I  &   r  &     22.46 $\pm$ 0.19
  \enddata
  \tablenotetext{a}{Calculated using MJD$_{r,\rm peak} = 56096.7$ and $z=0.107$.}
  \tablenotetext{b}{The magnitudes have {\it not} been corrected for
    Galactic extinction.}
\end{deluxetable}

\begin{deluxetable}{crccc}
  \tablecaption{Log of observations of \dcc.\label{tab:13dcc_logphotometry}}
  \tablehead{
    \colhead{MJD} &
    \colhead{Phase\a} &
    \colhead{Telescope} &
    \colhead{Filter} &
    \colhead{Magnitude\b} \\
    (days) & (days) & & & AB
  }
  \startdata
    56533.36  &$-$59.36  &       P48  &   R  &     19.68 $\pm$ 0.12 \\
    56533.42  &$-$59.32  &       P48  &   R  &     19.73 $\pm$ 0.06 \\
    56533.45  &$-$59.29  &       P48  &   R  &     19.76 $\pm$ 0.05 \\
    56537.35  &$-$56.57  &       P48  &   R  &     19.66 $\pm$ 0.06 \\
    56537.47  &$-$56.49  &       P48  &   R  &     19.86 $\pm$ 0.04 \\
    56537.50  &$-$56.46  &       P48  &   R  &     19.70 $\pm$ 0.04 \\
    56538.45  &$-$55.80  &       P48  &   R  &     19.80 $\pm$ 0.04 \\
    56539.34  &$-$55.18  &       P48  &   R  &     19.66 $\pm$ 0.07 \\
    56539.38  &$-$55.15  &       P48  &   R  &     19.79 $\pm$ 0.07 \\
    56539.42  &$-$55.12  &       P48  &   R  &     19.80 $\pm$ 0.06 \\
    56540.44  &$-$54.41  &       P48  &   R  &     19.98 $\pm$ 0.19 \\
    56541.34  &$-$53.78  &       P48  &   R  &     19.73 $\pm$ 0.08 \\
    56541.37  &$-$53.76  &       P48  &   R  &     19.55 $\pm$ 0.17 \\
    56541.41  &$-$53.73  &       P48  &   R  &     19.38 $\pm$ 0.17 \\
    56542.34  &$-$53.08  &       P48  &   R  &     19.77 $\pm$ 0.08 \\
    56542.45  &$-$53.00  &       P48  &   R  &     19.69 $\pm$ 0.06 \\
    56542.49  &$-$52.98  &       P48  &   R  &     19.71 $\pm$ 0.06 \\
    56543.39  &$-$52.35  &       P48  &   R  &     19.84 $\pm$ 0.06 \\
    56543.46  &$-$52.30  &       P48  &   R  &     19.91 $\pm$ 0.06 \\
    56543.50  &$-$52.27  &       P48  &   R  &     19.74 $\pm$ 0.05 \\
    56545.33  &$-$51.00  &       P48  &   R  &     19.55 $\pm$ 0.07 \\
    56546.32  &$-$50.30  &       P48  &   R  &     19.76 $\pm$ 0.08 \\
    56546.43  &$-$50.22  &       P48  &   R  &     19.86 $\pm$ 0.05 \\
    56546.46  &$-$50.20  &       P48  &   R  &     19.86 $\pm$ 0.05 \\
    56547.37  &$-$49.57  &       P48  &   R  &     19.79 $\pm$ 0.07 \\
    56547.49  &$-$49.48  &       P48  &   R  &     19.80 $\pm$ 0.06 \\
    56548.32  &$-$48.90  &       P48  &   R  &     19.82 $\pm$ 0.07 \\
    56548.47  &$-$48.80  &       P48  &   R  &     19.76 $\pm$ 0.04 \\
    56548.51  &$-$48.77  &       P48  &   R  &     19.71 $\pm$ 0.06 \\
    56549.41  &$-$48.14  &       P48  &   R  &     19.88 $\pm$ 0.04 \\
    56549.45  &$-$48.11  &       P48  &   R  &     19.82 $\pm$ 0.04 \\
    56549.49  &$-$48.09  &       P48  &   R  &     19.76 $\pm$ 0.03 \\
    56550.39  &$-$47.45  &       P48  &   R  &     19.77 $\pm$ 0.07 \\
    56550.48  &$-$47.39  &       P48  &   R  &     19.83 $\pm$ 0.04 \\
    56550.51  &$-$47.37  &       P48  &   R  &     19.81 $\pm$ 0.03 \\
    56551.42  &$-$46.74  &       P48  &   R  &     19.78 $\pm$ 0.08 \\
    56551.46  &$-$46.71  &       P48  &   R  &     19.86 $\pm$ 0.07 \\
    56551.49  &$-$46.69  &       P48  &   R  &     19.83 $\pm$ 0.04 \\
    56552.40  &$-$46.06  &       P48  &   R  &     19.76 $\pm$ 0.09 \\
    56552.44  &$-$46.02  &       P48  &   R  &     19.88 $\pm$ 0.08 \\
    56552.47  &$-$46.00  &       P48  &   R  &     19.80 $\pm$ 0.09 \\
    56561.28  &$-$39.84  &       P48  &   R  &     19.81 $\pm$ 0.16 \\
    56561.36  &$-$39.79  &       P48  &   R  &     19.87 $\pm$ 0.10 \\
    56561.40  &$-$39.77  &       P48  &   R  &     19.85 $\pm$ 0.09 \\
    56562.45  &$-$39.03  &       P48  &   R  &     19.90 $\pm$ 0.08 \\
    56562.48  &$-$39.01  &       P48  &   R  &     19.81 $\pm$ 0.07 \\
    56562.51  &$-$38.98  &       P48  &   R  &     19.72 $\pm$ 0.08 \\
    56563.42  &$-$38.35  &       P48  &   R  &     19.74 $\pm$ 0.09 \\
    56563.49  &$-$38.30  &       P48  &   R  &     19.75 $\pm$ 0.08 \\
    56563.52  &$-$38.28  &       P48  &   R  &     19.91 $\pm$ 0.12 \\
    56564.43  &$-$37.65  &       P48  &   R  &     19.80 $\pm$ 0.07 \\
    56564.48  &$-$37.61  &       P48  &   R  &     19.95 $\pm$ 0.07 \\
    56564.51  &$-$37.59  &       P48  &   R  &     19.87 $\pm$ 0.08 \\
    56565.41  &$-$36.96  &       P48  &   R  &     19.78 $\pm$ 0.07 \\
    56565.47  &$-$36.92  &       P48  &   R  &     19.88 $\pm$ 0.06 \\
    56565.50  &$-$36.90  &       P48  &   R  &     19.86 $\pm$ 0.08 \\
    56566.40  &$-$36.27  &       P48  &   R  &     19.82 $\pm$ 0.06 \\
    56566.45  &$-$36.23  &       P48  &   R  &     19.79 $\pm$ 0.05 \\
    56566.48  &$-$36.21  &       P48  &   R  &     19.85 $\pm$ 0.07 \\
    56567.39  &$-$35.58  &       P48  &   R  &     19.77 $\pm$ 0.05 \\
    56567.44  &$-$35.54  &       P48  &   R  &     19.76 $\pm$ 0.06 \\
    56567.47  &$-$35.52  &       P48  &   R  &     19.86 $\pm$ 0.06 \\
    56568.38  &$-$34.89  &       P48  &   R  &     19.93 $\pm$ 0.06 \\
    56568.45  &$-$34.84  &       P48  &   R  &     19.83 $\pm$ 0.07 \\
    56568.48  &$-$34.81  &       P48  &   R  &     19.83 $\pm$ 0.07 \\
    56569.39  &$-$34.18  &       P48  &   R  &     19.77 $\pm$ 0.08 \\
    56569.46  &$-$34.13  &       P48  &   R  &     19.84 $\pm$ 0.06 \\
    56569.49  &$-$34.11  &       P48  &   R  &     19.76 $\pm$ 0.07 \\
    56570.39  &$-$33.48  &       P48  &   R  &     19.83 $\pm$ 0.22 \\
    56570.47  &$-$33.43  &       P48  &   R  &     19.39 $\pm$ 0.16 \\
    56570.50  &$-$33.40  &       P48  &   R  &     19.94 $\pm$ 0.22 \\
    56571.40  &$-$32.77  &       P48  &   R  &     20.01 $\pm$ 0.13
  \enddata
  \tablenotetext{a}{Calculated using MJD$_{r,\rm peak} = 56618.3$ and $z=0.431$.}
  \tablenotetext{b}{The magnitudes have {\it not} been corrected for
    Galactic extinction.}
\end{deluxetable}
 
\addtocounter{table}{-1}
\begin{deluxetable}{crccc}
  \tablecaption{(continued) Log of observations of \dcc.}
  \tablehead{
    \colhead{MJD} &
    \colhead{Phase\a} &
    \colhead{Telescope} &
    \colhead{Filter} &
    \colhead{Magnitude\b} \\
    (days) & (days) & & & AB
  }
  \startdata
    56571.46  &$-$32.73  &       P48  &   R  &     19.72 $\pm$ 0.08 \\
    56571.49  &$-$32.71  &       P48  &   R  &     20.08 $\pm$ 0.18 \\
    56572.39  &$-$32.08  &       P48  &   R  &     19.83 $\pm$ 0.08 \\
    56572.46  &$-$32.04  &       P48  &   R  &     19.85 $\pm$ 0.07 \\
    56573.40  &$-$31.38  &       P48  &   R  &     19.83 $\pm$ 0.07 \\
    56573.44  &$-$31.35  &       P48  &   R  &     19.98 $\pm$ 0.07 \\
    56573.47  &$-$31.33  &       P48  &   R  &     19.99 $\pm$ 0.11 \\
    56576.39  &$-$29.29  &       P48  &   R  &     19.83 $\pm$ 0.06 \\
    56576.49  &$-$29.22  &       P48  &   R  &     19.90 $\pm$ 0.08 \\
    56576.53  &$-$29.19  &       P48  &   R  &     19.78 $\pm$ 0.10 \\
    56588.21  &$-$21.03  &       P48  &   R  &     19.76 $\pm$ 0.13 \\
    56588.35  &$-$20.93  &       P48  &   R  &     20.05 $\pm$ 0.16 \\
    56588.38  &$-$20.91  &       P48  &   R  &     19.76 $\pm$ 0.12 \\
    56589.28  &$-$20.28  &       P48  &   R  &     19.62 $\pm$ 0.07 \\
    56589.31  &$-$20.26  &       P48  &   R  &     19.93 $\pm$ 0.10 \\
    56589.33  &$-$20.24  &       P48  &   R  &     19.76 $\pm$ 0.06 \\
    56590.36  &$-$19.53  &       P48  &   R  &     19.74 $\pm$ 0.07 \\
    56590.38  &$-$19.51  &       P48  &   R  &     19.72 $\pm$ 0.08 \\
    56591.28  &$-$18.88  &       P48  &   R  &     19.75 $\pm$ 0.08 \\
    56591.32  &$-$18.85  &       P48  &   R  &     19.64 $\pm$ 0.04 \\
    56591.35  &$-$18.83  &       P48  &   R  &     19.68 $\pm$ 0.04 \\
    56592.30  &$-$18.17  &       P48  &   R  &     19.61 $\pm$ 0.04 \\
    56592.39  &$-$18.11  &       P48  &   R  &     19.67 $\pm$ 0.06 \\
    56592.42  &$-$18.09  &       P48  &   R  &     19.71 $\pm$ 0.04 \\
    56596.19  &$-$15.45  &       P48  &   R  &     19.75 $\pm$ 0.11 \\
    56596.30  &$-$15.37  &       P48  &   R  &     19.66 $\pm$ 0.08 \\
    56596.33  &$-$15.36  &       P48  &   R  &     19.67 $\pm$ 0.07 \\
    56597.23  &$-$14.73  &       P48  &   R  &     19.54 $\pm$ 0.09 \\
    56597.31  &$-$14.67  &       P48  &   R  &     19.70 $\pm$ 0.08 \\
    56597.34  &$-$14.65  &       P48  &   R  &     19.55 $\pm$ 0.07 \\
    56598.24  &$-$14.02  &       P48  &   R  &     19.59 $\pm$ 0.10 \\
    56598.31  &$-$13.97  &       P48  &   R  &     19.62 $\pm$ 0.06 \\
    56598.33  &$-$13.95  &       P48  &   R  &     19.67 $\pm$ 0.07 \\
    56599.24  &$-$13.32  &       P48  &   R  &     19.58 $\pm$ 0.07 \\
    56599.28  &$-$13.29  &       P48  &   R  &     19.71 $\pm$ 0.06 \\
    56599.30  &$-$13.28  &       P48  &   R  &     19.61 $\pm$ 0.06 \\
    56600.25  &$-$12.62  &       P48  &   R  &     19.59 $\pm$ 0.05 \\
    56600.36  &$-$12.54  &       P48  &   R  &     19.54 $\pm$ 0.07 \\
    56600.39  &$-$12.52  &       P48  &   R  &     19.47 $\pm$ 0.16 \\
    56602.17  &$-$11.27  &       P48  &   R  &     19.55 $\pm$ 0.12 \\
    56602.23  &$-$11.23  &       P48  &   R  &     19.57 $\pm$ 0.15 \\
    56602.25  &$-$11.21  &       P48  &   R  &     19.54 $\pm$ 0.11 \\
    56603.17  &$-$10.57  &       P48  &   R  &     19.37 $\pm$ 0.09 \\
    56603.24  &$-$10.53  &       P48  &   R  &     19.44 $\pm$ 0.07 \\
    56603.29  &$-$10.49  &       P48  &   R  &     19.51 $\pm$ 0.06 \\
    56604.29  & $-$9.79  &       P48  &   R  &     19.60 $\pm$ 0.04 \\
    56604.41  & $-$9.71  &       P48  &   R  &     19.57 $\pm$ 0.04 \\
    56604.44  & $-$9.69  &       P48  &   R  &     19.58 $\pm$ 0.06 \\
    56605.34  & $-$9.06  &       P48  &   R  &     19.53 $\pm$ 0.06 \\
    56605.38  & $-$9.03  &       P48  &   R  &     19.62 $\pm$ 0.06 \\
    56605.41  & $-$9.01  &       P48  &   R  &     19.54 $\pm$ 0.05 \\
    56607.16  & $-$7.79  &       P48  &   R  &     19.82 $\pm$ 0.11 \\
    56607.29  & $-$7.69  &       P48  &   R  &     19.54 $\pm$ 0.07 \\
    56607.32  & $-$7.67  &       P48  &   R  &     19.47 $\pm$ 0.07 \\
    56608.22  & $-$7.04  &       P48  &   R  &     19.34 $\pm$ 0.06 \\
    56608.26  & $-$7.02  &       P48  &   R  &     19.60 $\pm$ 0.12 \\
    56608.29  & $-$6.99  &       P48  &   R  &     19.48 $\pm$ 0.10 \\
    56609.19  & $-$6.36  &       P48  &   R  &     19.64 $\pm$ 0.10 \\
    56609.34  & $-$6.26  &       P48  &   R  &     19.45 $\pm$ 0.07 \\
    56537.33  &$-$56.58  &       P60  &   r  &     19.81 $\pm$ 0.04 \\
    56637.12  &   13.15  &       P60  &   i  &     19.95 $\pm$ 0.13 \\
    56637.12  &   13.15  &       P60  &   r  &     19.88 $\pm$ 0.13 \\
    56637.12  &   13.15  &       P60  &   g  &     20.14 $\pm$ 0.11 \\
    56645.12  &   18.74  &       P60  &   r  &     19.93 $\pm$ 0.18 \\
    56645.12  &   18.74  &       P60  &   g  &     20.13 $\pm$ 0.14 \\
    56647.13  &   20.15  &       P60  &   i  &     19.75 $\pm$ 0.16 \\
    56647.13  &   20.15  &       P60  &   r  &     20.83 $\pm$ 0.24 \\
    56647.14  &   20.15  &       P60  &   g  &     20.46 $\pm$ 0.11 \\
    56648.11  &   20.83  &       P60  &   i  &     20.15 $\pm$ 0.10 \\
    56648.12  &   20.84  &       P60  &   r  &     20.02 $\pm$ 0.04 \\
    56648.12  &   20.84  &       P60  &   g  &     20.43 $\pm$ 0.05 \\
    56655.20  &   25.79  &       P60  &   i  &     19.99 $\pm$ 0.12
  \enddata
  \tablenotetext{a}{Calculated using MJD$_{r,\rm peak} = 56618.3$ and $z=0.431$.}
  \tablenotetext{b}{The magnitudes have {\it not} been corrected for
    Galactic extinction.}
\end{deluxetable}
 
\addtocounter{table}{-1}
\begin{deluxetable}{crccc}
  \tablecaption{(continued) Log of observations of \dcc.}
  \tablehead{
    \colhead{MJD} &
    \colhead{Phase\a} &
    \colhead{Telescope} &
    \colhead{Filter} &
    \colhead{Magnitude\b} \\
    (days) & (days) & & & AB
  }
  \startdata
    56655.20  &   25.79  &       P60  &   r  &     20.25 $\pm$ 0.11 \\
    56655.21  &   25.79  &       P60  &   g  &     20.85 $\pm$ 0.15 \\
    56656.17  &   26.46  &       P60  &   i  &     20.17 $\pm$ 0.10 \\
    56656.17  &   26.46  &       P60  &   r  &     20.25 $\pm$ 0.08 \\
    56656.18  &   26.47  &       P60  &   g  &     20.66 $\pm$ 0.09 \\
    56657.13  &   27.14  &       P60  &   i  &     20.14 $\pm$ 0.05 \\
    56657.13  &   27.14  &       P60  &   r  &     20.21 $\pm$ 0.03 \\
    56657.14  &   27.14  &       P60  &   g  &     20.62 $\pm$ 0.04 \\
    56658.13  &   27.83  &       P60  &   i  &     20.15 $\pm$ 0.05 \\
    56658.13  &   27.83  &       P60  &   r  &     20.25 $\pm$ 0.04 \\
    56658.13  &   27.84  &       P60  &   g  &     20.66 $\pm$ 0.05 \\
    56659.20  &   28.58  &       P60  &   i  &     20.21 $\pm$ 0.06 \\
    56660.15  &   29.25  &       P60  &   i  &     20.12 $\pm$ 0.06 \\
    56660.16  &   29.25  &       P60  &   r  &     20.29 $\pm$ 0.05 \\
    56660.16  &   29.25  &       P60  &   g  &     20.65 $\pm$ 0.05 \\
    56662.21  &   30.68  &       P60  &   i  &     20.38 $\pm$ 0.13 \\
    56662.21  &   30.68  &       P60  &   r  &     20.30 $\pm$ 0.08 \\
    56662.22  &   30.69  &       P60  &   g  &     20.82 $\pm$ 0.08 \\
    56671.19  &   36.96  &       P60  &   i  &     20.41 $\pm$ 0.15 \\
    56671.19  &   36.96  &       P60  &   r  &     20.60 $\pm$ 0.14 \\
    56672.19  &   37.66  &       P60  &   i  &     20.88 $\pm$ 0.26 \\
    56672.20  &   37.66  &       P60  &   r  &     20.59 $\pm$ 0.21 \\
    56673.18  &   38.35  &       P60  &   r  &     20.60 $\pm$ 0.19 \\
    56676.13  &   40.41  &       P60  &   i  &     20.70 $\pm$ 0.08 \\
    56676.13  &   40.41  &       P60  &   r  &     21.33 $\pm$ 0.10 \\
    56676.14  &   40.42  &       P60  &   g  &     21.59 $\pm$ 0.09 \\
    56683.25  &   45.39  &       P60  &   i  &     21.13 $\pm$ 0.16 \\
    56683.25  &   45.39  &       P60  &   r  &     21.37 $\pm$ 0.19 \\
    56685.13  &   46.70  &       P60  &   i  &     21.41 $\pm$ 0.14 \\
    56685.13  &   46.70  &       P60  &   r  &     21.65 $\pm$ 0.20 \\
    56697.14  &   55.10  &       P60  &   i  &     21.43 $\pm$ 0.21 \\
    56697.15  &   55.10  &       P60  &   r  &     21.51 $\pm$ 0.19 \\
    56710.18  &   64.20  &       P60  &   i  &     21.25 $\pm$ 0.21 \\
    56701.20  &   57.93  &       DCT  &   r  &     22.20 $\pm$ 0.24 \\
    56701.20  &   57.93  &       DCT  &   i  &     21.74 $\pm$ 0.21 \\
    56908.50  &  202.79  &       DCT  &   r  &     23.20 $\pm$ 0.28 \\
    57070.00  &  315.65  &       HST  & F625W&     25.00 $\pm$ 0.13
  \enddata
  \tablenotetext{a}{Calculated using MJD$_{r,\rm peak} = 56618.3$ and $z=0.431$.}
  \tablenotetext{b}{The magnitudes have {\it not} been corrected for
    Galactic extinction.}
\end{deluxetable}

\begin{acknowledgements}

  It is a pleasure to acknowledge the help of Manos Chatzopoulos with
  the implementation of the semi-analytical light-curve models
  developed by him and his colleagues in our light-curve fitting
  program. We are grateful to Nir Sapir for enlightening discussions,
  and to WeiKang Zheng, Kelsey Clubb and Patrick Kelly for their
  contribution to the 2013 Dec.~3 Keck/LRIS observations of \dcc.  We
  thank the staffs at Palomar and Lick Observatories for their expert
  assistance.  The intermediate Palomar Transient Factory project is a
  scientific collaboration among the California Institute of
  Technology, Los Alamos National Laboratory, the University of
  Wisconsin at Milwaukee, the Oskar Klein Center, the Weizmann
  Institute of Science, the TANGO Program of the University System of
  Taiwan, and the Kavli Institute for the Physics and Mathematics of
  the Universe. This paper is based in part on observations made with
  the NASA/ESA {\it Hubble Space Telescope}.  Some of the data
  presented herein were obtained at the W.M. Keck Observatory, which
  is operated as a scientific partnership among the California
  Institute of Technology, the University of California, and NASA; the
  observatory was made possible by the generous financial support of
  the W.M. Keck Foundation. Research at Lick Observatory is partially
  supported by a generous gift from Google. These results also made
  use of the Discovery Channel Telescope at Lowell Observatory.Lowell
  is a private, non-profit institution dedicated to astrophysical
  research and public appreciation of astronomy and operates the DCT
  in partnership with Boston University, the University of Maryland,
  the University of Toledo, Northern Arizona University and Yale
  University.  LMI construction was supported by a grant AST-1005313
  from the National Science Foundation.

  The Dark Cosmology Centre is funded by the DNRF.  A.G.-Y. is
  supported by the EU/FP7 via ERC grant No. 307260, the Quantum
  Universe I-Core program by the Israeli Committee for planning and
  funding, and the ISF, GIF, Minerva, and ISF grants, WIS-UK ``making
  connections'', and Kimmel and ARCHES awards.  Support for D.A.P. was
  provided by NASA through Hubble Fellowship grant HST-HF-51296.01-A
  awarded by the Space Telescope Science Institute, which is operated
  by the Association of Universities for Research in Astronomy, Inc.,
  for NASA, under contract NAS 5-26555. M.S. acknowledges support from
  EU/FP7-ERC grant 615929.  E.O.O. is incumbent of the Arye
  Dissentshik career development chair and is grateful to support by
  grants from the Willner Family Leadership Institute Ilan Gluzman
  (Secaucus NJ), Israel Science Foundation, Minerva, and the I-CORE
  Program of the Planning and Budgeting Committee and The Israel
  Science Foundation.  Support for I.A. was provided by NASA through
  the Einstein Fellowship Program, grant PF6-170148. A.V.F.'s
  supernova group at UC Berkeley is supported through NSF grant
  AST--1211916, the TABASGO Foundation, and the Christopher R. Redlich
  Fund. The National Energy Research Scientific Computing Center,
  which is supported by the Office of Science of the U.S. Department
  of Energy under Contract No. DE-AC02-05CH11231, provided staff,
  computational resources, and data storage for this project.  Part of
  this research was carried out at the Jet Propulsion Laboratory,
  California Institute of Technology, under a contract with the
  National Aeronautics and Space Administration.

\end{acknowledgements}

\bibliographystyle{apj} 
\bibliography{references}

\end{document}